\documentclass[preprint]{aastex}

\newcommand{\flux}{ergs cm$^{-2}$ s$^{-1}$}

\newcommand{\lumi}{ergs s$^{-1}$}
\newcommand{\col}{cm$^{-2}$}
\newcommand{\nh}{N_{\rm H}}
\newcommand{\einstein}{{\it Einstein}}
\newcommand{\asca}{{ASCA}}
\newcommand{\rosat}{{ROSAT}}
\newcommand{\exosat}{{EXOSAT}}

\title{
Faint X-ray sources resolved in the ASCA Galactic plane survey and
their contribution to the Galactic ridge X-ray emission
}

\author{Mutsumi Sugizaki}
\affil{Tsukuba Space Center, National Space Development Agency of Japan}
\affil{2-1-1 Sengen, Tsukuba, Ibaraki, 305-8505, Japan}
\email{sugizaki@oasis.tksc.nasda.go.jp}
\author{Kazuhisa Mitsuda, Hidehiro Kaneda, Keiichi Matsuzaki,}
\affil{The Institute of Space and Astronautical Science}
\affil{3-1-1 Yoshinodai, Sagamihara, Kanagawa, 229-8510, Japan}
\author{Shigeo Yamauchi}
\affil{Faculty of Humanities and Social Sciences, Iwate University}
\affil{3-18-34 Ueda, Morioka, Iwate, 020-8550, Japan}
\and
\author{Katsuji Koyama}
\affil{Department of Physics, Kyoto University}
\affil{Kyoto 606-8502, Japan}

\begin{abstract}

The X-ray emission from the central region of the Galactic plane, $|l|\lesssim 45\arcdeg$ 
and $|b| \lesssim  0\fdg 4$, was studied in the 0.7--10 keV energy band 
with a spatial resolution of $\sim 3\arcmin$ with the 
ASCA observatory. We developed a new analysis method for the ASCA data 
to resolve discrete sources from the extended Galactic ridge X-ray emission 
(GRXE).  We successfully resolved 163 discrete sources with an X-ray flux down 
to $10^{-12.5}$ {\flux} and determined the intensity variations of the GRXE 
as a function of the Galactic longitude with a spatial resolution 
of about a degree.
The longitudinal intensity variation in the energy band above 4 keV, 
for which there is little absorption in the Galactic plane, 
shows a large enhanced feature within $|l| \lesssim 30\arcdeg$.  
This suggests a strong enhancement of X-ray emissivity 
of the GRXE inside the 4 kpc arm of the Galaxy.
Searches for identifications of the resolved X-ray sources
with cataloged X-ray sources and optical stars show
that the $66\%$ are unidentified. 
Spectral analysis of each source shows that
a large number of the unidentified sources have hard X-ray spectra 
obscured by the Galactic interstellar medium.
We classified the sources into several groups  
by the flux, the hardness and the softness of the spectra,
and performed further detailed analysis for the spectra 
summed within each group.
Possible candidates of X-ray origins of these unidentified sources 
are discussed based on the grouping spectral analysis. 
Also, we derived the LogN-LogS relations of the resolved sources in the energy 
bands below and above 2 keV separately.
The LogN-LogS relation of the Galactic X-ray sources above 2 keV 
was obtained for the first time with this study.  
It is represented by a power-law with an index of $-0.79 \pm 0.07$ after correction 
for the contribution of extragalactic X-ray sources.  This flat power-law 
relation suggests that the spatial distribution of the X-ray sources 
should have an arm-like structure in which the Solar system is included.
The integrated surface brightness of  
the resolved sources is about 10\% of the total GRXE
in both energy bands.  
The approximately 90\% of the emission remaining is still unresolved.

\end{abstract}
\keywords{galaxies: Milky Way --- galaxies: sources --- surveys -- X-rays: general}

\begin{document}

\maketitle


\section{Introduction}

From the first detection of Sco X-1, the brightest X-ray star in the Galaxy,
many bright X-ray sources have been discovered and classified into various
categories.
These categories include young stellar objects, flaring stars, compact
degenerate stars, supernova remnants (SNRs), and diffuse hot plasma in the interstellar medium (ISM).
However, since imaging observations with high sensitivity have always 
been limited to the soft X-ray band below 3 keV,
the comprehensive view of these X-ray sources throughout the Galaxy
is far from understood.
The soft X-rays whose energy is below 3 keV are substantially absorbed 
with the ISM in the Galaxy;
thus the coverage of soft X-ray observations is limited to near the Solar system.
The ISM is almost transparent to X-ray photons above 3 keV
throughout the Galaxy.
However, imaging instruments with good efficiency for hard ($>3$ keV) X-rays 
are difficult to construct because of the limits of the reflection and the diffraction of the hard X-rays.
For this reason, unrecognized high energy phenomena could be hiding in hard X-ray regions.

The Galactic ridge X-ray emission (hereafter GRXE) is one of those open questions.  
The GRXE is unresolved X-ray emission extended along the Galactic plane.
The existence of the excess X-ray emission along the Galactic plane 
was first suggested by the pioneering rocket experiments \cite{Bleach1972}
and clearly confirmed by succeeding observations of X-ray astronomical satellites
(e.g. Worrall et al. 1982; Warwick et al. 1985).
After this, K-shell line emission from He-like iron ions 
was discovered along the Galactic plane \cite{Koyama1986,Yamauchi1993}.
On the other hand, above 10 keV, a power-law component extended  
up to $\gamma$-ray regime with a photon index $\Gamma\simeq 2$, was clearly detected
(Yamasaki et al. 1997; Valinia \& Marshall 1998).

However, the origin of the total emission of the GRXE still remains unresolved.
The X-ray spectrum with the ionized iron K-line in the energy band below 10 keV 
is in good agreement with emission from hot plasma with a temperature of about $10^8$ K.
However, there are some critical problems to be known 
in the purely diffuse plasma emission model.
If such a hot plasma is extended in the Galactic plane, 
it should be free from the gravitational confinement of the Galaxy 
and escape from the Galactic plane (e.g., Koyama et al. 1986).
Also, the plasma pressure should be higher by an order of magnitude than that of the
ISM in the Galaxy (e.g., Kaneda et al. 1997). 
Hence, a great deal of energy must be continuously supplied to maintain this plasma
unless an unidentified plasma confinement mechanism exists.
Tanuma et al. (1999) suggest that the energetics can be explained 
by magnetic reconnection
if the magnetic field in the Galactic plane is locally as high as $\sim 30 \mu$G.
However, any convincing evidence of such a localization of the Galactic magnetic field 
has not been obtained so far.

Another hypothesis is that the GRXE
is a superposition of discrete X-ray sources.
From the similarity of the energy spectra,
Cataclysmic Variable (CV) and RS CVn systems have been considered as candidates 
(e.g., Worrall \& Marshall 1983).
However, a huge number of discrete sources are required
to satisfy the uniformity of the surface brightness observed on 
the Galactic plane \cite{Yamauchi1996}. 
The number density of the CVs estimated near the Solar system is
smaller by two orders of magnitude than that to explain 
the total flux of the GRXE (Worrall et al. 1982).
Also, Ottmann \& Schmitt (1991) estimated the contributions of RS CVn systems
to the GRXE at $\sim$ 27\% based on the optical and the X-ray data
and showed that their X-ray spectra can not completely 
account for that of the GRXE.
However, we have not ever known accurately
how many faint discrete X-ray sources exist in the Galaxy.

{\asca}
has a spatial resolution of $3'$(HPD: half power diameter) 
in a field-of-view (FOV) of $40'$ (diameter) 
in a broad energy range of 0.5--10 keV by utilizing two kinds of detector:
SIS (Solid Imaging Spectrometer) and GIS (Gas Imaging Spectrometer)
combined with XRT (X-ray telescope) 
(Tanaka, Inoue, \& Holt 1994). 
It is capable of imaging observations in the energy band above 3 keV for the first time,
thus making it possible to obtain crucial observational results for understanding
the Galactic X-ray sources and the origin of the GRXE.
Thus, we planned the {\asca} Galactic plane survey and performed it.

In this paper, 
studies of faint Galactic X-ray sources 
and their contribution to the GRXE 
based on the {\asca} Galactic plane survey
are reported. 
We developed a new method of imaging analysis for
data of the {\asca} Galactic plane survey 
and successfully resolved X-ray sources with a flux down to $10^{-12.5}$ {\flux}.
We investigated the identification of the faint unidentified sources 
based on the spectral analysis, in which we classified the sources into several groups
due to the flux, the hardness and the softness,
and analyzed the spectra summed within each group
in order to overcome the poor photon statistics of each source.  
We also derived their LogN-LogS relation and 
the large-scale intensity profile of the unresolved 
GRXE averaged on a scale of a degree. 
In the following,
we describe the observations of the {\asca} Galactic plane survey  
and the data reduction 
in $\S$ \ref{sec:obs_red},
and present the analysis method and the results in $\S$ \ref{sec:ana_res}.
We discuss implications of the obtained results, which include 
longitude intensity variations of the GRXE,
unidentified faint X-ray sources resolved in this ASCA survey,
and their LogN-LogS relations, in $\S$ \ref{sec:discuss}.
Finally we state our conclusions in $\S$ \ref{sec:conclusion}.

\section{Observations and Data reduction}
\label{sec:obs_red}

\subsection{Observations}

The {\asca} Galactic Plane Survey 
is a collaborative project of the ASCA team,
aimed at the systematic study of the Galactic X-ray sources.
It was planned to cover the area of the Galactic inner disk
of $|l|\lesssim 45^\circ$ and $|b|\lesssim 0\fdg 4$ with successive pointing observations,
each of 10 ks exposure time.
The observations were carried out from 1996 March 13 
in the 4-th guest observation cycle (AO-4)
to 1999 April 28 with
a total of 173 pointings
and the total exposure time of 1,370 ks after data screening.
All the observations were performed with the PH-nominal mode for the GIS and 
the 4CCD faint mode at high bit rate and the 4CCD bright mode at medium bit-rate 
for the SIS. 

In addition to the Galactic plane survey,
we utilized data from observations 
of the Galactic ridge edge,
which were proposed 
for the purpose of obtaining information
on the edge of the GRXE.
They cover the area on the Galactic plane from $l=57\fdg 5$ to $l=62\fdg 5$ 
with eight successive pointing observations, each with 10 ks exposure time.
The observations were carried out from 1997 November 5 to 1997 November 9
with the PH-nominal mode for the GIS and 
the 2CCD faint mode at high bit rate and the 2CCD bright mode at medium bit-rate for the SIS. 

\begin{figure}[p]
\begin{center}
\includegraphics[width=17cm]{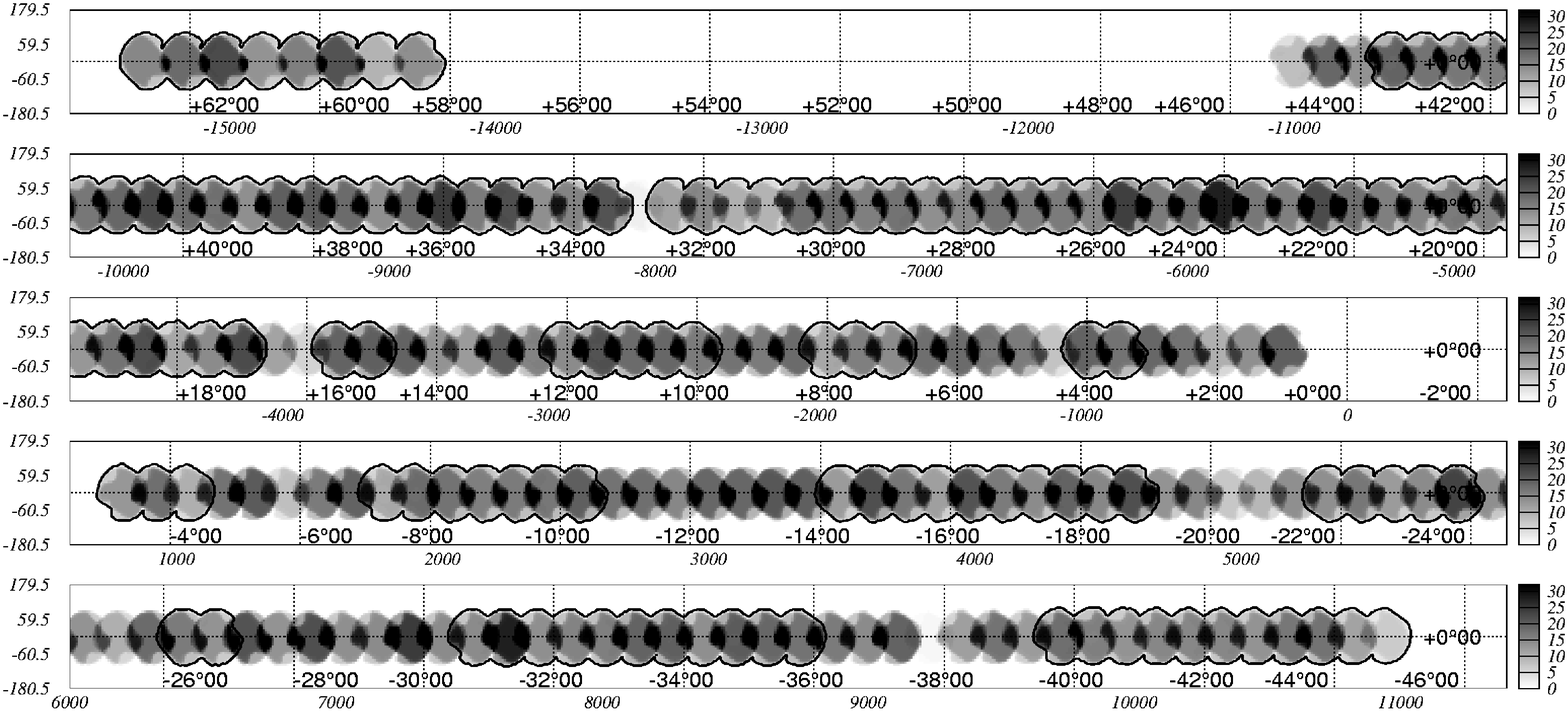}
\caption{Exposure map of the {\asca} Galactic plane survey.
The scale of the color map represents exposure time 
in units of ks (GIS2$+$GIS3). 
The areas enclosed by solid lines are  
the regions used for the source survey excluding stray light
coming from the nearby bright X-ray sources. 
The XY labels represent the GALXY coordinates (see Appendix \ref{app:galxy}).}
\label{fig:obscover}
\end{center}
\end{figure}

Figure \ref{fig:obscover} shows the exposure map with the
GIS FOVs. 
The total area covered by the GIS is 63.04 deg$^2$

\placefigure{fig:obscover}

\subsection{Reduction of the GIS data}

In this Galactic plane survey,
the SIS was not so useful because 
(1) the coverage of the area, which is most important for survey observations, 
was about a half that of the GIS and 
(2) the energy resolution and the efficiency of the SIS were gradually degraded
in orbit due to the radiation damage \cite{Yamashita1997}. 
Thus, the SIS had only a slight advantage over the GIS.
In this paper, 
the analysis of the GIS data is mainly described.
All the data reduction was performed on the {\asca}\_ANL \cite{Kubo1997},
which is a software framework for {\asca} data analysis
developed for event-by-event analysis \cite{Takahashi1995}.

In the analysis 
of searches for unidentified faint sources,
careful event screenings are required in order not to extract fake ones.
For a low background condition and reproducibility of the non X-ray background (NXB),
we adopted the ``flare-cut'' screening criteria for the geographical condition
of the satellite \cite{Ishisaki1997}.
We used standard values for
the other parameters such as elevation from the dark earth rim, 
elevation from the sunlight earth rim and attitude jittering.
The correction of the data such as position linearization, gain
correction and strict rise-time mask were also carried out in this step,
according to the standard procedure. 
In addition, the dead time correction was performed 
based on the hard-wired event counts 
recorded in the GIS monitor data \cite{Makishima1996,Ohashi1996}.

Since the coordinate system employed in the standard data processing
cannot be applied to the analysis of multi-pointing observations,
we adopted a new coordinate system described in Appendix \ref{app:galxy}
to handle all the data of the Galactic plane survey.
The conversion of the coordinates was performed 
for each photon 
according to the position coordinates on the detector and 
the Euler angle of the satellite axis.

\section{Analysis and Results}
\label{sec:ana_res}

\subsection{Basic scheme of the GIS image analysis}

Some of the difficulties of extended emission analysis of the {\asca} data 
come from the complicated image response
which includes vignetting effects and the position dependent PSF (point spread function), 
and from contamination by stray light.
Therefore, we have to carefully interpret 
raw X-ray image obtained with {\asca}.  
Since the effects of these instrumental response functions on the raw image 
are coupled to each other in complicated ways, 
it is impossible to deconvolve the original image from the response functions uniquely. 
We thus utilized model fitting rather than image deconvolution. 

From previous observations of the GRXE (e.g., Kaneda 1997), we know that 
the surface brightness of the GRXE shows substantial intensity variations on 
length scales of a few degrees; however, the variation amplitude decreases 
rapidly on shorter scales.  Thus, as a first approximation, we can consider 
spatial variations on length scales longer and shorter than the size of the FOV 
of the GIS ($\sim$ 1 degree) separately. 

We first constructed a model in which the surface brightness of the GRXE 
is uniform over length scales 
of about a degree, and fitted it to the raw data.
Later, we examined the model parameters and the residuals of the fits to investigate 
the systematic errors of the analysis and the spatial fluctuation of the GRXE 
which were not included in the model.
The image analysis was performed according to the following steps.

\begin{enumerate}

\item[Step-1:] Estimation of large scale intensity profile of the GRXE\\
The brightness of the GRXE averaged over a FOV of the GIS was evaluated,
excluding pixels with high counting rates which are expected to be largely contaminated by
bright X-ray sources.
The average surface brightness obtained for each pointing was utilized to construct 
a uniform brightness model for the GRXE, 
which was required for the source survey in the next step.
The value of the average brightness was reevaluated,
masking all the X-ray sources resolved in the source survey.

\item[Step-2:] Peak finding \\
Peaks of the image which have a significantly higher count-rate 
than the uniform brightness GRXE model
were extracted as source candidates.

\item[Step-3:] Image fitting \\
Image fitting by a surface brightness model that consists of source candidates 
and the uniform brightness GRXE model was carried out to estimate the significance and
the flux of the discrete sources.

\end{enumerate}

In the image fitting, 
precise 2-dimensional response functions must be known.
However, the pattern of stray light, which comes from outside the FOV
through anomalous reflection paths such as single reflection 
and/or backside reflection by the mirror foils, 
is too complicated to be calibrated
sufficiently for the fitting analysis
\cite{Tsusaka1995}.
Thus, we estimated the contribution of the stray light from the nearby bright X-ray sources 
based on the data of the adjacent fields containing the bright sources 
and on previous observations. 
We eliminated data of which more than 5\% are contaminated by stray light
from the bright X-ray sources.
After the data selection, 
the area covered was reduced to 
39.75 deg$^2$ from the total 63.04 deg$^2$ area surveyed.
The resulting coverage is shown in Figure \ref{fig:obscover}.

\subsection{Estimation of large scale intensity profile of GRXE}

The large-scale intensity profile of the GRXE 
was obtained from the background X-ray intensity 
averaged within each FOV of the GIS,
which was determined for all pointings separately.
We employed the fitting method
because the expected number of photons for each pixel in a FOV
is different due to the position dependence of the effective area
even if the observed surface brightness is uniform.
The image fitting in this step 
was performed for the image composed of the GIS2 and the GIS3 data of the 
multi-pointing observation and binned by 12$\times$12 pixels (3$'$$\times$3$'$)
in order to obtain appropriate statistical accuracy
for the background GRXE.

We used the model for the averaged CXB (cosmic X-ray background) image accumulating large blank-sky area
as the standard response for the uniform 
surface emission (Appendix \ref{app:cxbmodel}).
Utilizing the model for the averaged CXB image, $M^{\rm CXB}(\vec{x})$ cnts bin$^{-1}$ s$^{-1}$,
where $\vec{x}$ denotes a position on the GIS image,
the model of the observed image for GRXE which is assumed to have uniform surface brightness, 
$M^{\rm GRXE}(\vec{x})$ cnts bin$^{-1}$, 
is represented by the following equation:
\begin{equation} 
M^{\rm GRXE}_p(\vec{x})=\alpha_p t^{\rm exp}_pM^{\rm CXB}(\vec{x})+M^{\rm NXB}_p(\vec{x}) ~\rm [cnts\, bin^{-1}] ,
\label{equ:backmodel}
\end{equation} 
where $\alpha_p$ is a parameter representing the intensity normalized to that of the CXB, $t^{\rm exp}_p$ is the exposure time
and $M^{\rm NXB}_p(\vec{x})$ is the image of the non X-ray background
not originated from X-rays, defined for each pointing ID, $p$.
The model of the non X-ray background image can be calculated as a function of  
the geographical condition of the satellite with a systematic error of 5\%
\cite{Ishisaki1997}.

We determined the values of $\alpha_p$ 
by fitting the model of $M^{\rm GRXE}_p(\vec{x})$
to the raw observed image $M^{\rm RAW}_p(\vec{x})$.
Here, since the raw image contains a considerable number 
of photons from discrete X-ray sources 
in the same FOV,
we need to eliminate those pixels 
with anomalously high count-rates compared to the average.
Thus, the value of $\alpha_p$ was evaluated by
a fit excluding the pixels with count-rates higher than average$+$2$\sigma$
of the distribution 
$\left\{(M^{\rm RAW}_p(\vec{x})- M^{\rm GRXE}_p(\vec{x}))/t^{\rm exp}_p M^{\rm CXB}(\vec{x}) \right\}$
in the first step.
After the source survey in the following steps, 
it was reevaluated by
a fit masking all the resolved X-ray sources.

\begin{figure}[p]
\begin{center}
\includegraphics[width=15cm, angle=-90]{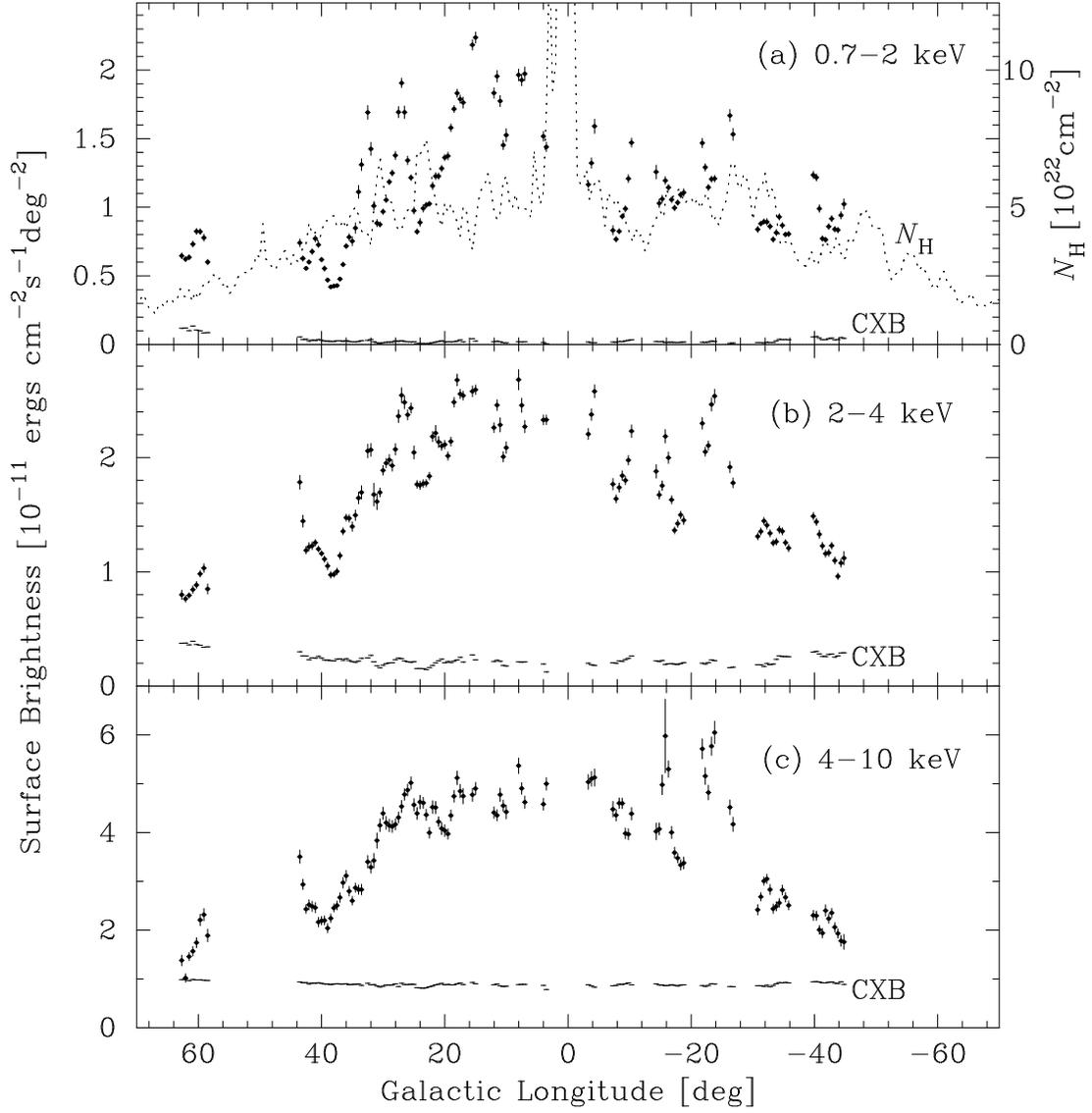}
\caption{
Large-scale intensity variation of the GRXE averaged over each FOV of the GIS
and inclusive contributions of the CXB coming through the Galactic ISM 
in the 0.7--2 keV band (upper),  the 2--4 keV band (middle) 
and the 4--10 keV band (lower).
The Galactic ISM column density estimated from HI and CO-line intensities
is also shown in the upper panel. 
}
\label{fig:ldist}
\end{center}
\end{figure}

These analyses were carried out in each different energy band individually.
Figure \ref{fig:ldist} shows the obtained large scale intensity variations
in the three energy bands of 0.7--2 keV, 2--4 keV and 4--10 keV,
where the intensity is represented by surface brightness in units of 
(ergs cm$^{-2}$s$^{-1}$deg$^{-2}$)
calculated from the photon counting rate
using the ARF file for extended surface emission \cite{Kaneda1997b}.
We also calculated the inclusive contributions of the CXB 
coming through the Galactic ISM 
accounting for the average CXB spectrum 
\cite{Ishisaki1996}
and the ISM column density expected from the HI \cite{Burton1985,Kerr1986} 
and CO-line \cite{Dame1987} intensities
with the CO-H$_2$ conversion factor \cite{Hunter1997},
which are shown together in Figure \ref{fig:ldist}.

\placefigure{fig:ldist}

\subsection{Peak finding}

We searched for peaks in the image, which we used as 
source candidates in the next step.

\begin{figure}[p]
\begin{center}
\includegraphics[width=8cm]{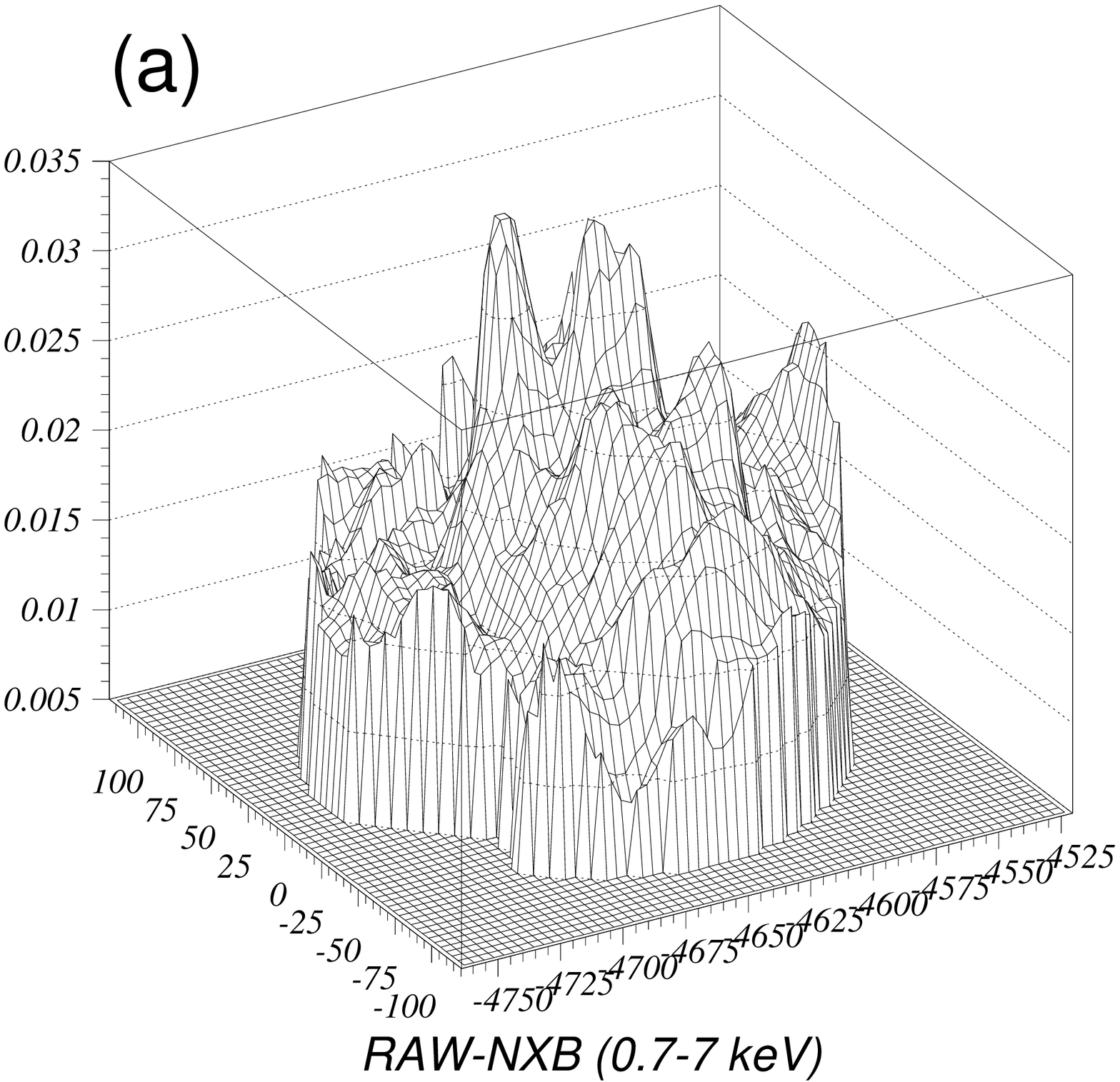}
\includegraphics[width=8cm]{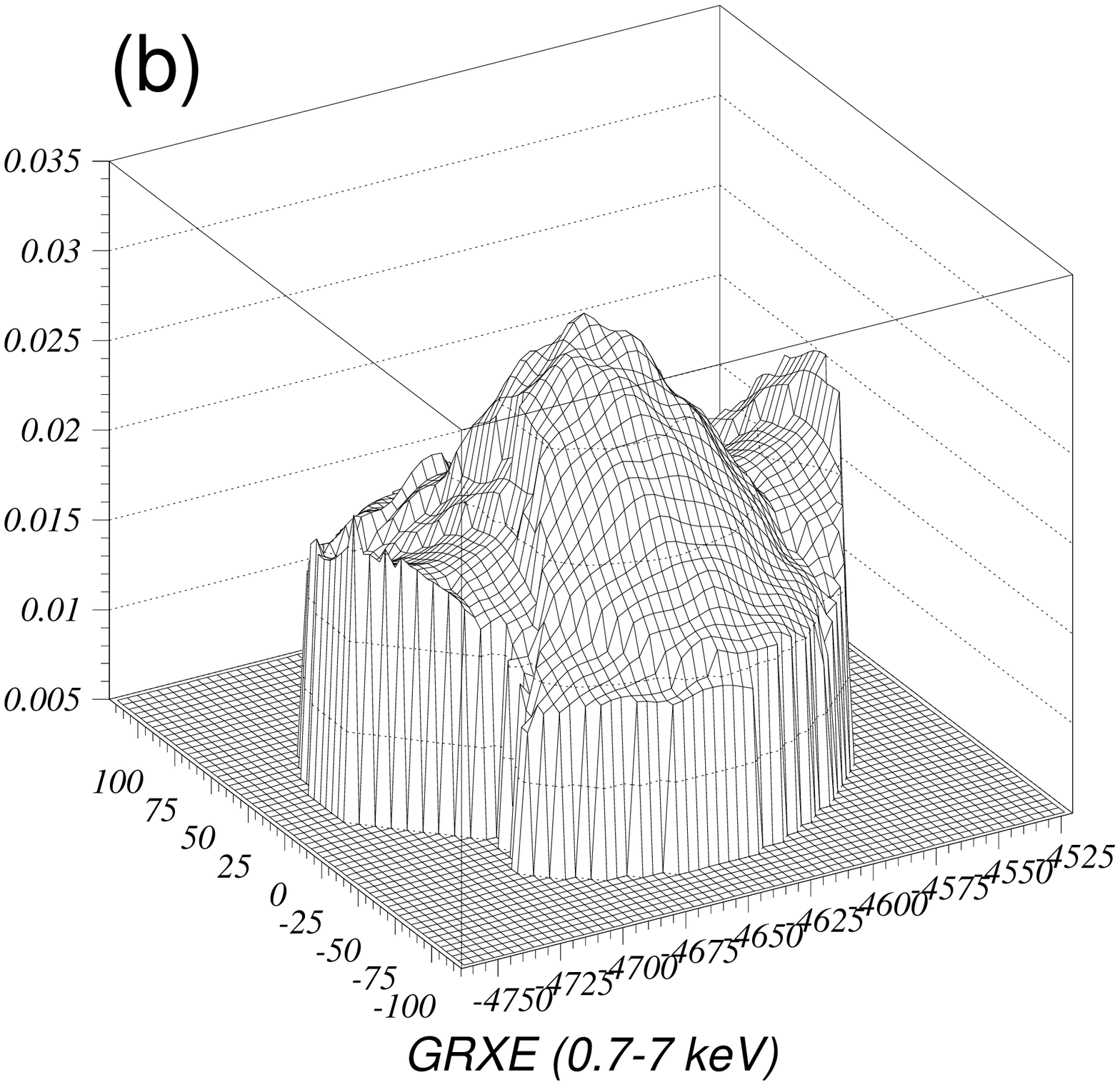}

\includegraphics[width=8cm]{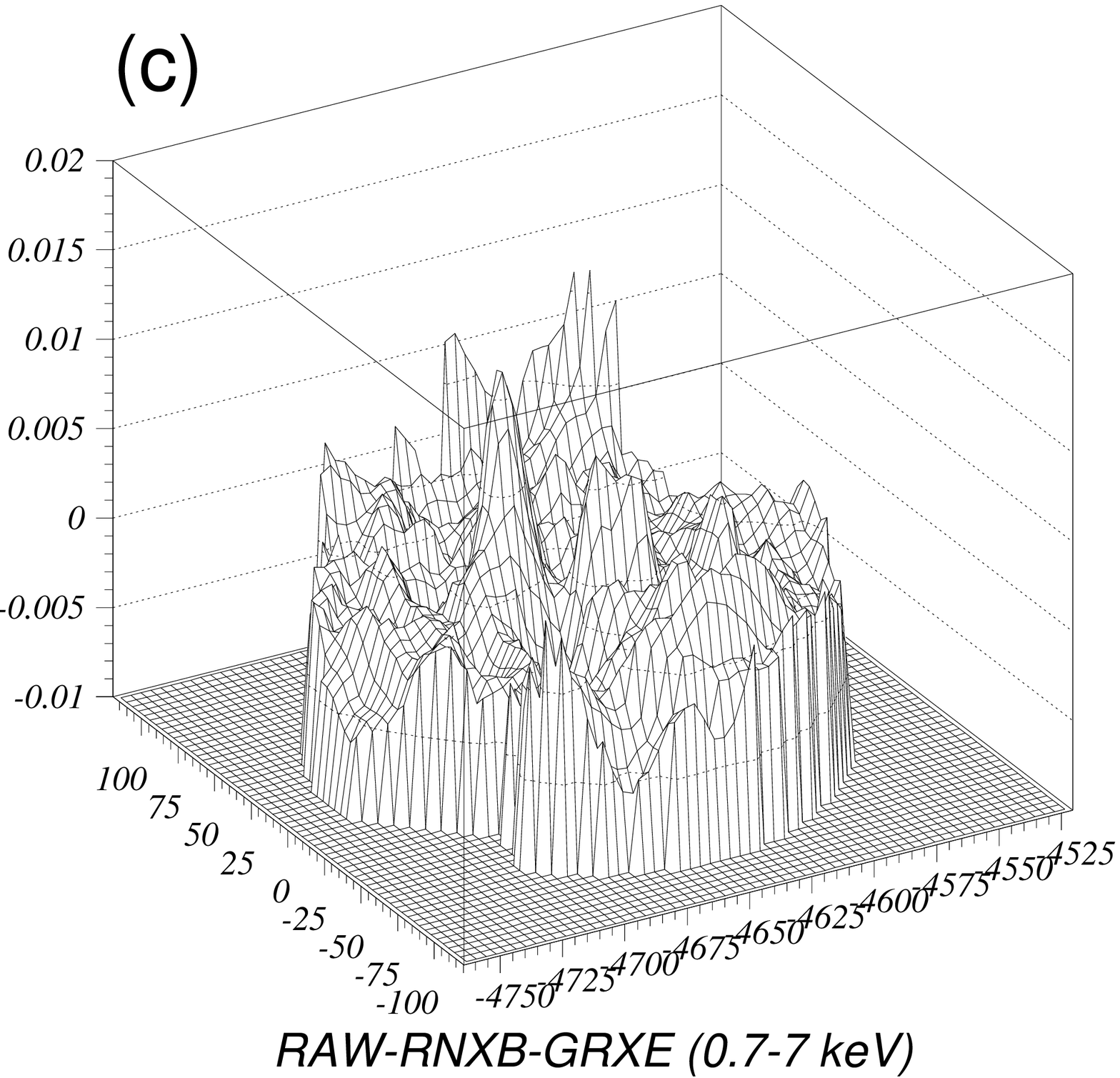}
\includegraphics[width=8cm]{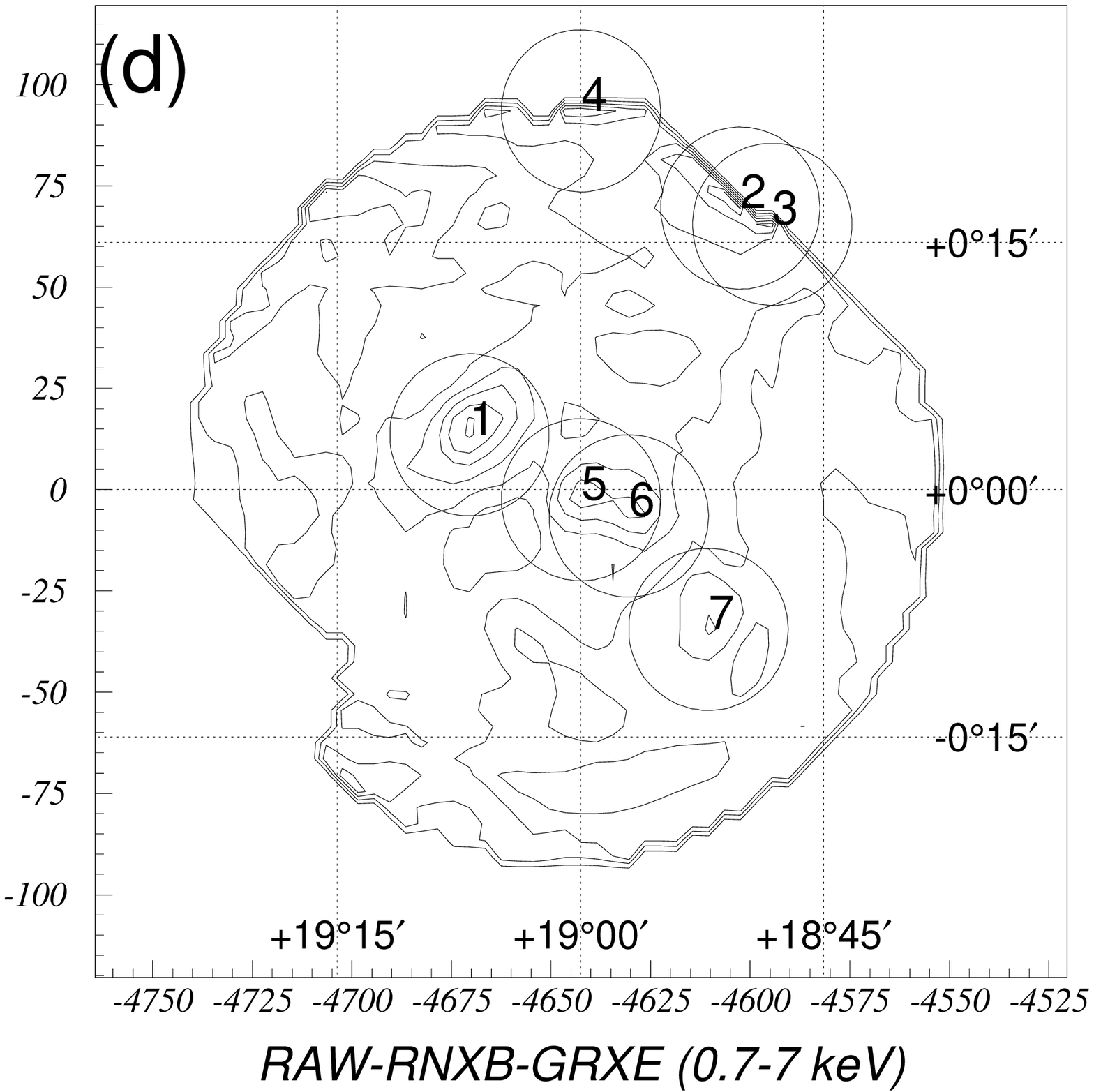}

\caption{
Procedure of image processing and peak finding.
(a): Smoothed image of raw observed image. 
The background GRXE makes a convex structure at the field center
by the vignetting effect.
(b): Smoothed image of the background GRXE model.
(c): Smoothed image after subtracting the background GRXE model.
This is used for peak finding.
The convex profiles at the field center in image (a) are properly removed.
(d): The surface image (c) represented by contours.
Circles represent the peaks recognized as source candidates
in the analysis procedure.
Only the peak labeled with ID=1 in the figure
is recognized as an X-ray source with a significance above 4$\sigma$ 
as the result of image fitting in the next step. 
}
\label{fig:peakfind}
\end{center}
\end{figure}

On the Galactic plane, 
the bright background of the unresolved GRXE
makes a convex structure at the center
of each FOV because of the vignetting of the XRT (see Figure \ref{fig:peakfind}).  
This artificial structure needs to be subtracted 
before peak finding.
To subtract the background component properly, 
the 2-dimensional profile of the unresolved GRXE
must be known. 
Here, we assumed as a first approximation that the surface brightness of the GRXE is 
uniform on the scale of a FOV of the GIS.
The profile for the uniform GRXE model can be obtained from
equation (\ref{equ:backmodel}).
Even if the model is not statistically acceptable, the artificial structure of the 
vignetting and the stray light 
should be appropriately removed in this procedure.
Therefore, it is thought to be an appropriate method.

To avoid picking false peaks as much as possible,
after subtracting the background, 
the following processes were performed on the raw image beforehand.

\begin{enumerate}
 \item Smoothing by cross-correlation with the PSF.\\
  To improve photon statistics without degrading position resolution,
  the image was smoothed by cross-correlation with the PSF.
 \item Compositing multi-pointing observation.\\
  Overlapping regions of the multi-pointing observation with GIS2 and GIS3 were
  summed to improve photon statistics.
 \item Correction of effective exposure time.\\\label{item:expcorr}
  The correction was required 
  not only for the exposure time of multi-pointing overlapping observations
  but also for the vignetting effect, 
  which causes differences in the effective area within a FOV. 
\end{enumerate}

The smoothed image $M^{\rm SMO}(\vec{y})$ on which peak finding was carried out 
is represented by the following equation,
\begin{equation}
 M^{\rm SMO}(\vec{y})=\frac{\sum_p \int_{\vec{x}\in p}{\it PSF}_p(\vec{x},\vec{y})\left\{ M_p^{\rm RAW}(\vec{x})-M^{\rm GRXE}(\vec{x})\right\}d\vec{x} }{\sum_p \int_{\vec{x}\in p}{t^{\rm exp}_p}{\it PSF}_p(\vec{x},\vec{y})M_p^{\rm Mask}(\vec{x}) d\vec{x}}, \label{equ:smo}
\end{equation}
where ${\it PSF}_p(\vec{x},\vec{y})$ represents the point spread function
for the X-rays coming from the direction $\vec{y}$ on the sky and
$M_p^{\rm Mask}(\vec{x})$ represents the selected region on the detector.
The region near the edge of the GIS FOV, more than 20$'$ from the detector center,
is usually not used because of 
the large background ratio and the large calibration uncertainty.
Here, considering the following step,  
the region within a radius of 22$'$ from the detector center was 
selected for the peak finding.

We chose peaks exceeding a certain threshold as source candidates.
The proper threshold value depends on the averaged intensity of the trial field.
We found after some trials that 
3$\sigma$ of the distribution of the smoothed image, 
$\left\{M^{\rm SMO}(\vec{y})\right\}$, 
is appropriate for the threshold,
where the $\sigma$ is the best-fit value
obtained from fitting the distribution to a Gaussian function. 
Since the average intensity of a FOV is different from
pointing to pointing, the threshold was determined for each pointing 
by performing the Gaussian fitting separately.
These procedures are illustrated in Figure \ref{fig:peakfind}.

The procedures of peak finding were carried out
for the three energy bands of 0.7--2 keV, 2--10 keV and 0.7--7 keV,
each of which the CXB image model of the 0.7--2 keV, 2--7 keV and 0.7--7 keV band 
(Appendix \ref{app:cxbmodel})
were respectively 
used.

\placefigure{fig:peakfind}

\subsection{Image fitting}
 
To estimate the significance of the source detection and the flux,
we employed image fitting.
A model which consists of source candidates and background
was fitted to the raw data.
The detection significance of the source
was estimated from the fitting error of the flux.

The peaks detected in the previous step were used as 
the source candidates. 
As the background, the uniform model $M^{\rm GRXE}(\vec{x})$
represented by equation (\ref{equ:backmodel}) was employed.
The model is represented by
\begin{eqnarray}
M^{\rm Model}(\vec{x}) 
&=& \sum_p \left\{ \sum_{i}{t^{\rm exp}} f_i A_p(\vec{y}_i){\it PSF}_p(\vec{x},\vec{y}_i)+M^{\rm GRXE}(\vec{x})) \right\} \\
&=& \sum_p \left\{ \sum_{i}{t^{\rm exp}} f_i A_p(\vec{y}_i){\it PSF}_p(\vec{x},\vec{y}_i) +\alpha_p {t^{\rm exp}} M_p^{\rm CXB}(\vec{x}) +M_p^{\rm RNXB}(\vec{x})\right\} \label{equ:imagemo}
\end{eqnarray}
where $\vec{f_i}$, $\vec{y_i}$ are the flux and the sky coordinates, respectively, 
of the $i$-th source candidate
and $A_p(\vec{y_i})$ is the effective area for the sky coordinates, $\vec{y_i}$.
In this model, parameters to be fitted are source fluxes $\vec{f_i}$ and 
a normalization factor of GRXE, $\alpha_p$,
which was reevaluated here.
The fits were performed for the image composed of the GIS2 and the GIS3 data from the 
multi-pointing observation and binned by 4$\times$4 pixels (1$'$$\times$1$'$)
to improve statistics and speed up the calculation time.
Since the HPD of the PSF is $3'$, 
this bin size is reasonable.

The flux of the source was defined with the count rate measured 
as if the source was at the GIS nominal position.
Its units are (cnts GIS$^{-1}$ ks$^{-1}$).
Then, $A(\vec{y})$ is a correction factor of the effective area against that of the GIS nominal position.
It is slightly dependent on the energy spectrum of the source; however, 
the dependence is not important 
as far as the spectrum is that of a typical X-ray source.
The $A(\vec{y})$ was derived by {\it jbldarf}, which is a program to build an ARF file
in the standard analysis procedure.
The energy flux ({\flux}) is easily calculated from the 
the GIS count rate utilizing the ARF file at the GIS nominal position 
and assuming an appropriate spectrum for the target source.
As an example, the conversion factors for typical power-law spectra
with the typical ISM absorption densities in each energy band are
shown in Table \ref{tab:fluxconv}.

To cover a survey area as large as possible,
we extended the image fitting region to a 20$'$ radius
from the detector center.
Since the PSF of the XRT+GIS system was not calibrated 
outside of a 17$'$ radius \cite{Takahashi1995},
we approximated the PSF outside 17$'$ radius by that at the 17$'$ radius.
Thus, the reproducibility of the fitting model was not good for these regions
compared to the inner regions.
Also, we included source candidates within a 22$'$ radius 
in the model of the image fitting 
because X-ray photons from those sources 
can contribute to the flux at a 20$'$ radius through the extended PSF.

The fits were performed by
the maximum likelihood method based on the
Poisson distribution of photon counts.
This method is valid even when counting statistics are too poor to use 
the $\chi^2$ method.
The likelihood ${\cal L}$ is defined as
\begin{equation}
\ln({\cal L}) = \sum_{\vec{x}} \left\{ -m_{\vec{x}} +d_{\vec{x}}\ln\left(\frac{ m_{\vec{x}}}{ d_{\vec{x}} }\right) \right\},
\end{equation}
where $d_{\vec{x}} = M^{\rm RAW}(\vec{x})$ [cnts] represents observed photon counts and 
$m_{\vec{x}}= M^{\rm Model}(\vec{x})$ [cnts]
represents predicted counts from the model
at the position on the detector, $\vec{x}$.

The errors of the flux were estimated
by the $\chi^2$ statistic.
Since the bins around the detected sources 
have enough photon statistics,
the $\chi^2$ statistic is appropriate for the error estimation.

The significance of the source detection is defined as
\begin{equation}
\rm significance \, [\sigma]= \frac{\rm best~fit~flux}{\rm 1\sigma~error~of~flux}, \label{equ:sensi}
\end{equation}
where the 1$\sigma$ error is a fitting error of 1$\sigma$(68\%) confidence level
based on the $\chi^2$ statistics.
Here, $\chi^2$ is defined by
\begin{equation}
\chi^2 
= \sum_{\vec{x}}\left( \frac{ M^{\rm Data}(\vec{x}) - M^{\rm Model}(\vec{x}) }{M^{\rm Error}(\vec{x})} \right)^2,
\label{equ:defechi}
\end{equation}
\begin{equation}
M^{\rm Error}(\vec{x})
= \sqrt{ M^{\rm Model}(\vec{x}) + \left\{ M^{\rm Syserr}(\vec{x})\right\}^2},
\label{equ:defechierr}
\end{equation}
where $M^{\rm Syserr}(\vec{x})$ represents the systematic error of the model.
In this analysis, 5\% of $M^{\rm Model}(\vec{x})$ was set to $M^{\rm Syserr}(\vec{x})$ 
which represents the systematic errors of the PSF, CXB and NXB models 
and the small-scale fluctuation of GRXE. 
The adequacy of this systematic error was verified later in $\S$ \ref{sec:verify}.

From the definition of the $\chi^2$ statistic,
a total of 25 photons gives a
significance of 5$\sigma$ in an ideal observation
where there is no background and the PSF is represented only by a single pixel peak.
In the practical situation of the extended PSF and the background emission 
on the Galactic plane,
the threshold flux of 4$\sigma$ detection was typically about
50 cnts GIS$^{-1}$ (10ks)$^{-1}$.

\placetable{tab:fluxconv}

\subsection{Source list}

\begin{figure}[p]
\begin{center}
\includegraphics[width=17cm]{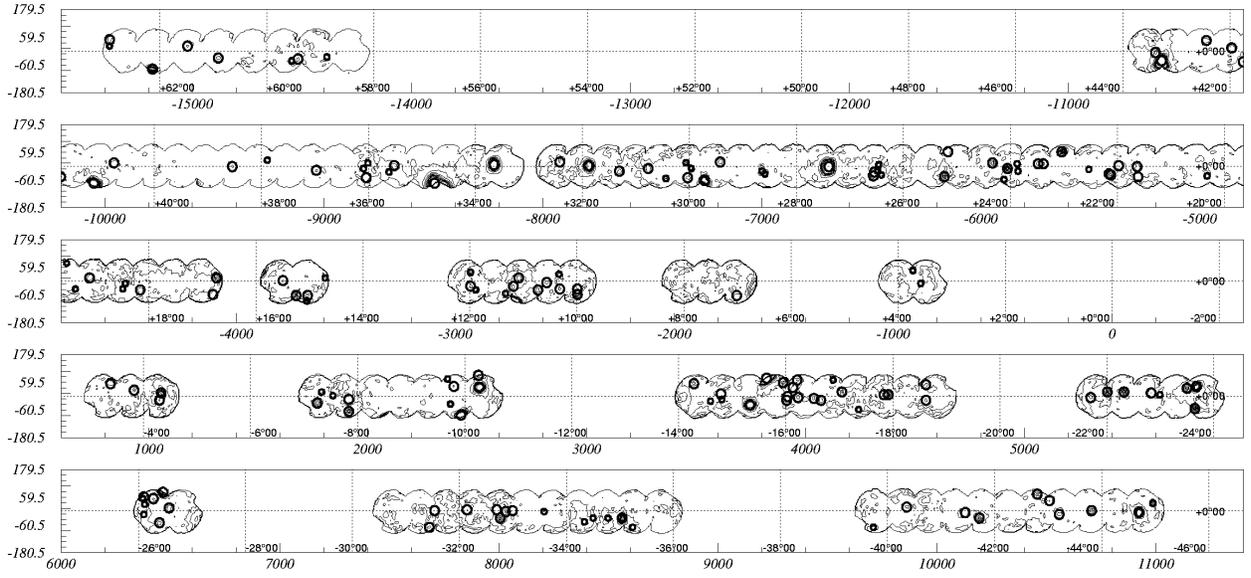}
\caption{
ASCA Galactic sources detected in the source survey with more than
4$\sigma$ significance. Large circles and small circles represent
those with significance above 5$\sigma$ and below 5$\sigma$, respectively.
The contour map is the mosaic image of the area used for the source survey 
in the 0.7--7 keV band, which is smoothed by the PSF and corrected  
for the exposure time and the vignetting of the XRT.
}
\label{fig:plotsrc}
\end{center}
\end{figure}

As a result of the image analysis,
we detected a total of 207 sources  
with a significance above 4$\sigma$
in at least one of the 0.7--2, 2--10 and 0.7--7 keV energy bands.
We recognized sources which are detected in more than two energy bands 
and less than 1$'$ apart to be identical.
After that, we investigated the position and the profile of the image 
around all the detected sources and excluded apparent fake sources such 
as peaks of stray light patterns located at the edge of the survey fields
near to the bright X-ray sources
and fractional parts of diffuse emission.
Subtracting the fake ones, the total number of the detected sources 
with a significance above 4$\sigma$
becomes 163.
The numbers of sources detected in each energy band 
are summarized in 
Table \ref{tab:numsrc}. 
The obtained source list is presented in Table \ref{tab:srclist}
with the parameters of position, flux and detection significance,
excluding the sources recognized as fake ones.
Figure \ref{fig:plotsrc} shows the position of all the sources 
in the list on the 0.7--7 keV image.

\placetable{tab:numsrc}
\placetable{tab:srclist}
\placefigure{fig:plotsrc}

We compared the obtained {\asca} Galactic source catalog
with the known X-ray source catalogs,
which are {\einstein} IPC Catalog 
and {\rosat} Bright Source Catalog \cite{Voges1999},
in order to search for the past X-ray identifications.
We next investigated identifications with the Galactic SNRs 
\cite{Green1998}, which are most promising candidates 
as the origins of diffuse X-ray emission.
The number of these cataloged sources involved in the survey area 
and that of the detected sources 
in the ASCA survey are summarized in Table \ref{tab:catid}.
We also examined the position coincidence with optical stars
in SIMBAD database.
Here, note that their position coincidence does not necessarily 
indicate the identification of the origins, 
since the visibility limit in optical band ($\sim$ 1 kpc) 
is much smaller than that in the ASCA-GIS 0.7--10 keV band ($\sim$ 5 kpc) 
in the Galactic plane
and the ASCA position determination error inaccuracy 
($\sim 1\arcmin$ in radius) is relatively large against the density of optical stars.
The number of the ASCA sources coincident with the SIMBAD stars 
is 22, which is shown together in Table \ref{tab:catid}.

After all,  
$107/163\simeq 66\%$ of the resolved sources were unidentified
with any of these source catalogs.

\placetable{tab:catid}

\subsection{Spectral analysis of the resolved sources and their grouping analysis}
\label{sec:anaspec}

In order to investigate the origins of X-ray emission of the resolved sources,
we performed spectral analysis.
However, photon statistics of faint X-ray sources with a significance 
as low as $\sim$4--5$\sigma$ 
are too poor to obtain any spectral information other than the position and the total flux.
Then, we took a method to select samples for the spectral analysis
and classify them into several groups according to the flux and the simple spectral parameters 
such as the hardness and the softness.
As for samples with photon statistics unsatisfactory for a detail analysis,
we inquired further spectral information for source groups and 
analyzed the spectra summed within each group.
Here, we utilized a photon index and an absorption column density obtained from 
a spectral fitting with a power-law model as the hardness and the softness
because simple photon counts in each energy band reflect 
not only the shapes of original X-ray spectra 
but also the effective area whose energy dependence varies within a FOV of the GIS.

All the spectral analysis was performed according to the standard analysis 
procedure supported by the ASCA Guest Observer Facility of NASA/GSFC.
The X-ray spectrum of each source was extracted within $3'$ from the source center
and the background spectrum is extracted from the the same off-axis region
within the FOV of the GIS masking the detected sources.
To gain the statistical accuracy,
we utilized the spectra summed of the GIS-2 and the GIS-3 for the model fitting.
Also, we eliminated the sources more than 18$'$ apart from the optical center on the GIS
beforehand,
since the response function near the edge of the FOV is not calibrated.

\begin{figure}
\begin{center}
\includegraphics[width=12cm, angle=-90]{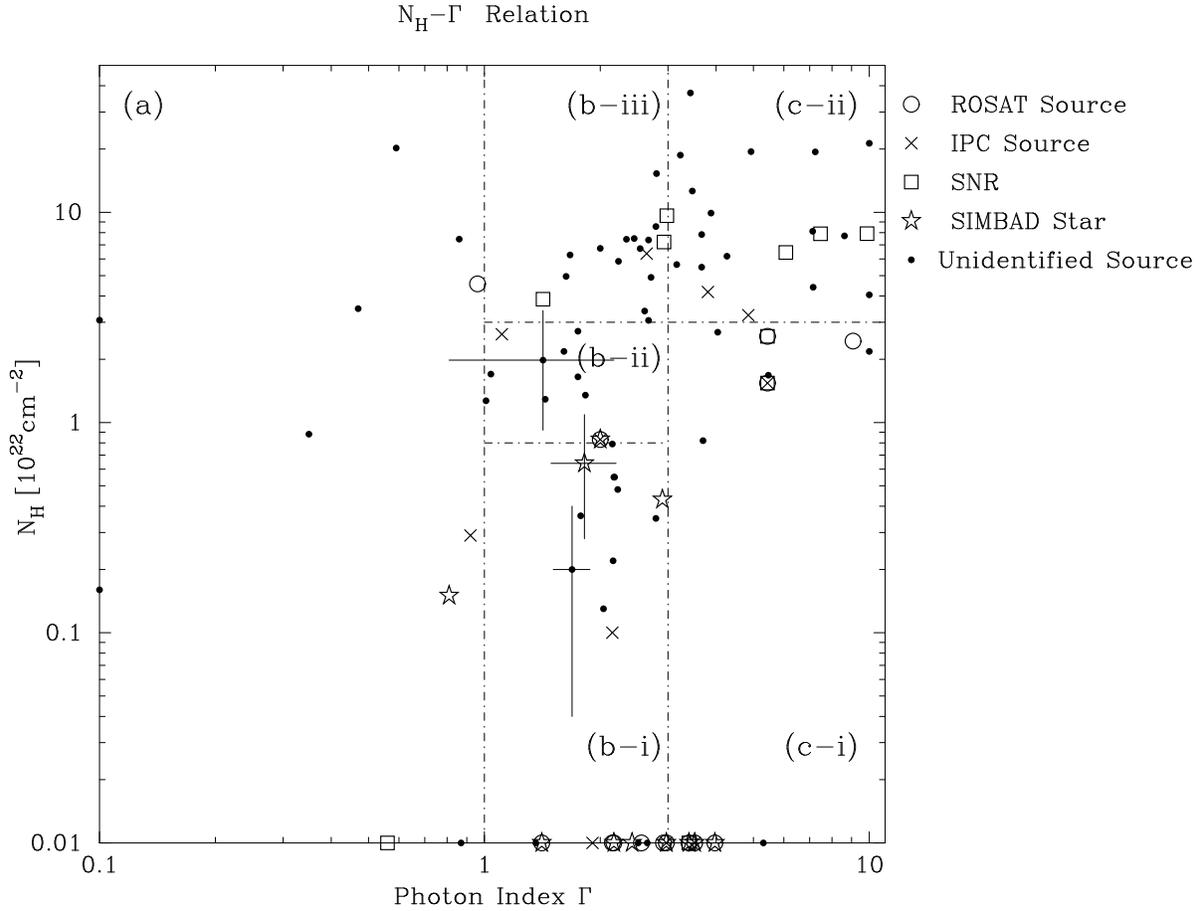}
\caption{$\nh - \Gamma$ relation of the faint X-ray sources resolved in the ASCA Galactic plane survey.
The 1-$\sigma$ error bars are shown in three sources typically.
The dot-dashed lines represent the boundary of the source groups in the spectral analysis. 
The sources identified with the Einstein, the ROSAT and the Galactic SNR source catalogs
are plotted by different marks.
Marks on the lower-boundaries of the plotting area mean that the best-fit values
are lower than the limit, however these fitting errors are relatively large.
}
\label{fig:nh_pow}
\end{center}
\end{figure}

We first collected the sources with a significance larger than 5$\sigma$ 
as samples for the spectral analysis
and fitted these spectra with a simple power-law with an absorption model.
The fluxes of 5$\sigma$ sources are approximately  $\sim 0.5\times 10^{-12}$ {\flux} in this survey condition.
The fits were accepted in most sources within a 90\% confidence level 
of the relatively large statistical errors
although there are a few sources not to accept that simple model.
The parameters of the best-fit power-law models are shown for each sample sources 
in Table \ref{tab:srclist}. 
Also, relation between the absorption column densities and the photon indices 
of all the sources is shown in Figure \ref{fig:nh_pow},
where the sources identified with the cataloged sources are distinguished by 
different marks. 
We see in Figure \ref{fig:nh_pow} that the sources identified with the ROSAT 
and the SIMBAD-star catalogs are distributed near the lower boundary 
of the absorption column density, $N_{\rm H}\simeq 10^{20}$ {\col}.
It is naturally expected from the fact that the optical and the soft X-ray photons
are easily absorbed by the Galactic ISM. 

\placefigure{fig:nh_pow}

We next extracted ten relatively-bright sources with a flux above $1.0\times 10^{-11}$ {\flux}
based on the flux of the best-fit power-law model.
These sources were excluded from the grouping spectral analysis in the below,
since such bright sources dominate the summed spectra within each group 
if they are added to the group members.
These sources have enough photon statistics to investigate each source properties individually.
We also examined flux variabilities for these sources by fitting a constant flux model 
to the light curves summed of the GIS-2 and the GIS-3 with 256-s and 1024-s bins.
We recognized a source to be variable if the constant model was unacceptable 
within a 99\% confidence limit. 
Obtained spectral parameters, flux variabilities, and known identifications of these sources 
are summarized in Table \ref{tab:10brtsrc}.
We can easily see in Table \ref{tab:10brtsrc} that these sources can be distinguished into two groups:
Some with soft X-ray spectra, exhibiting a large power-law index of $\Gamma>2$,
are identified with SNRs, and the others with hard X-ray spectra, 
exhibiting a small power-law index of $\Gamma<2$,
show flux variabilities. 

\placetable{tab:10brtsrc}

\begin{figure}
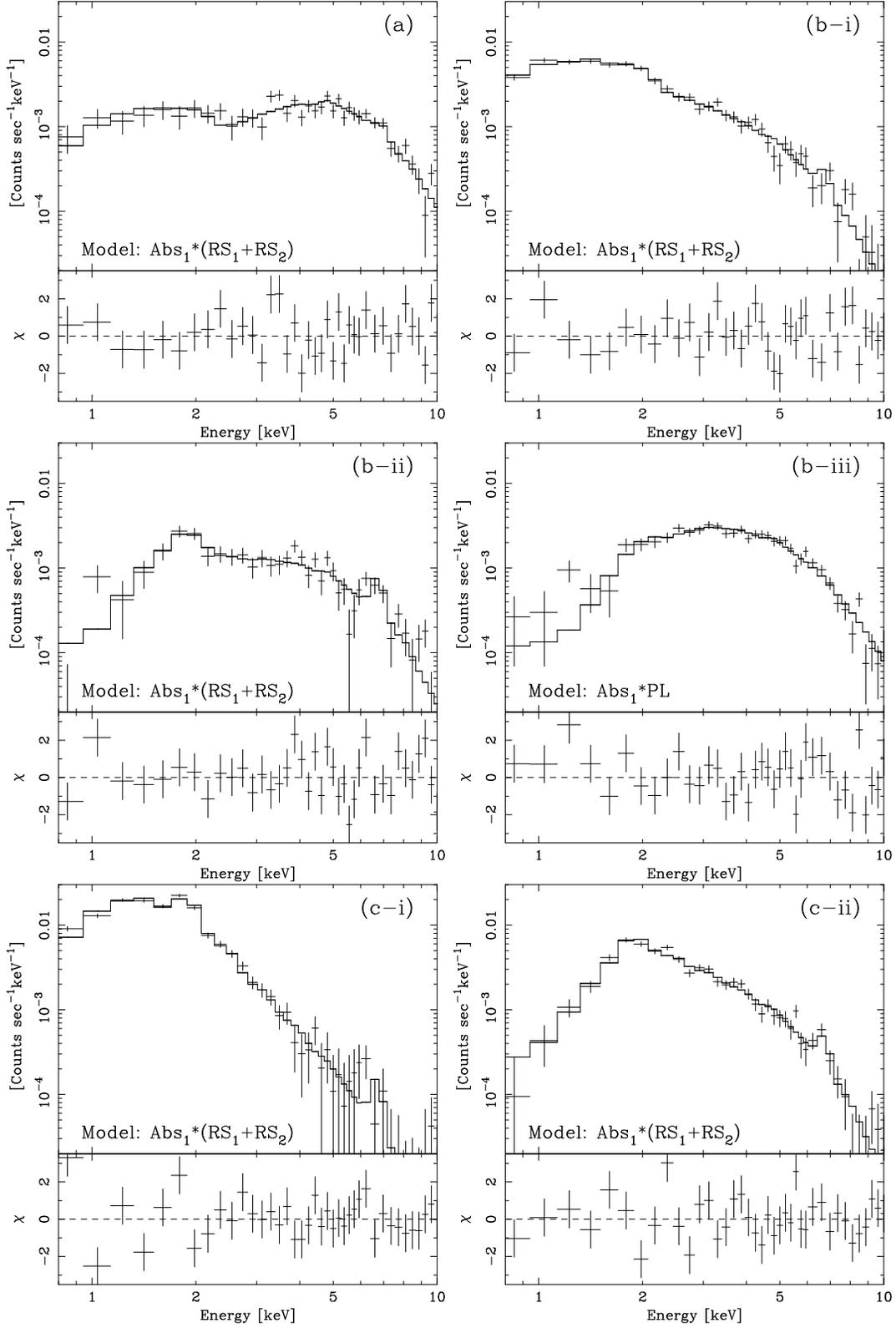

\begin{center}

\includegraphics[width=7cm, angle=-90]{fig6a.ps}
\includegraphics[width=7cm, angle=-90]{fig6bi.ps}

\includegraphics[width=7cm, angle=-90]{fig6bii.ps}
\includegraphics[width=7cm, angle=-90]{fig6biii.ps}

\includegraphics[width=7cm, angle=-90]{fig6ci.ps}
\includegraphics[width=7cm, angle=-90]{fig6cii.ps}

\caption{Summed spectra of the GIS2$+$GIS3 within each group (1$\sigma$ cross error bars) and 
these best-fit models (step lines) described in Table \ref{tab:specfit}.
}
\label{fig:groupspec}
\end{center}
\end{figure}

After that, we classified the residual 93 faint sources
into six groups by the best-fit power-law index, $\Gamma$, 
and the absorption column density, $N_{\rm H}$,
and analyzed spectra summed within each group.
The fluxes of the sources utilized here are within  $(0.5-10) \times 10^{-12}$ {\flux}.
Considering possible shapes of X-ray spectra 
expected from those of known X-ray sources
and the scatter of the $N_{\rm H} - \Gamma$ relation in Figure \ref{fig:nh_pow}, 
we adopted the grouping criteria described in Table \ref{tab:classification}.
The boundaries of the groups on $N_{\rm H} - \Gamma$ relation are illustrated in Figure \ref{fig:nh_pow}.
The number of sources belonging to each group 
and the numbers of the members identified with SNRs and SIMBAD optical stars 
are shown in Table \ref{tab:classification}.
Figure \ref{fig:groupspec} shows
the summed spectra within each group. 
It is remarkable that iron K-shell line emission 
is apparently seen at the energy $\sim$6.5 keV 
in the spectra of the groups (b-ii) and (c-ii), however 
the other group have no apparent iron-line features.
Also, the spectra of the groups (b-ii), (c-i), and (c-ii) show silicon K-shell line
at the energy $\sim$1.8 keV. 
The coexistence of the silicon and the iron lines in the groups (b-ii) and (c-ii) 
implies that the plasma of the emission origin does not reach the thermal equilibrium state.
Then, we tried to fit those spectra to some spectral models including power-law,
thin-thermal plasma emission coded by Raymond\& Smith (1977), and 
their two-component combinations.
Here, we used the sum of the response functions for all the group members
as the response for the summed spectra.
Thus, the flux obtained from the fitting model 
represents the averaged flux within the group.
We first assumed the plasma metal abundance, $Z$, as 1 solar 
and next floated it if any 1-solar plasma models are unacceptable.
Results of these spectral fits are summarized in Table \ref{tab:specfit}
and the best-fit models are shown in Figure \ref{fig:groupspec}
together with the data.

\placetable{tab:classification}
\placefigure{fig:groupspec}
\placetable{tab:specfit}

\subsection{Completeness map and LogN-LogS relation}

Sensitivity of the source survey depends on the image response functions of
the instruments,
the background and the exposure time.
These factors vary from pixel to pixel in the area surveyed. 
To study the population of the discrete sources 
(for example, the LogN-LogS relation),  
correction for the difference in sensitivity is required.

\begin{figure}[p]
\begin{center}
\includegraphics[width=12cm, angle=-90]{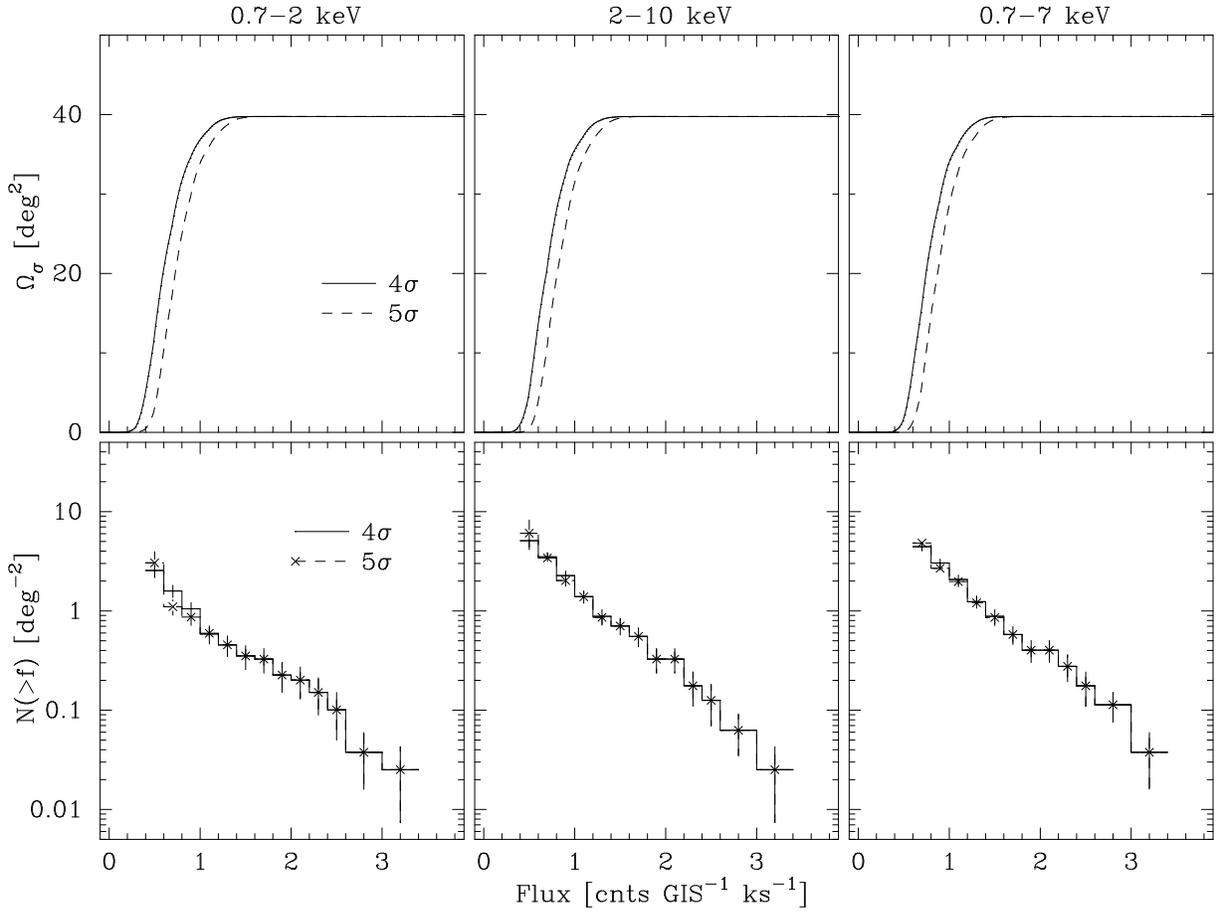}
\caption{
Completeness areas $\Omega_\sigma(f)$ and LogN-LogS relations $N(>f)$
measured with the detection thresholds of 4 and 5 $\sigma$
in each energy band of 0.7--2, 2--10 and 0.7--7 keV.
}
\label{fig:comparea_lnls}
\end{center}
\end{figure}

From the definitions of $\chi^2$ in equation (\ref{equ:defechi})
and the significance in equation (\ref{equ:sensi}),
we can calculate the significance of a trial point source 
with any flux at any position in the area surveyed
if we know the profile of the background GRXE image, $M^{\rm GRXE}(\vec{x})$.
The completeness area, $\Omega_\sigma(f)$, 
which is the area surveyed with a certain significance limit of $\sigma$ for a given flux $f$, 
can be easily obtained from the significance map.
Here, we employed the best-fit model for the GRXE image in the image fitting of the source survey
and calculated the completeness area for the significance limits of 4$\sigma$ and 5$\sigma$.
Figure \ref{fig:comparea_lnls} shows
the obtained completeness area 
in each energy band of 0.7--2, 2--10 and 0.7--7 keV.
The adequacy of the background model is discussed in the next section.

The number density of X-ray sources with a flux larger than a given value $N(>f)$, 
so called 'LogN-LogS relation', gives
important clues about the spatial distribution and the luminosity function.
The LogN-LogS relation can be derived 
from the raw number of resolved X-ray sources and the completeness area.
We calculated the LogN-LogS relation 
from the differential number density,
which is represented by
\begin{equation}
N(f)df = \frac{ n_\sigma(>f+df)-n_\sigma(>f) }{\Omega_\sigma(f)},
\end{equation}
where $n_\sigma(>f)$ is the raw number of detected sources with a flux larger than $f$ 
and a significance above $\sigma$.
The LogN-LogS relation $N(>f)$ was derived by integrating the above equation.
Figure \ref{fig:comparea_lnls} shows the obtained LogN-LogS relations 
for significance limits of 4$\sigma$ and 5$\sigma$ and 
in the 0.7--2, 2--10 and 0.7--7 keV energy bands.

\placefigure{fig:comparea_lnls}

\subsection{Verification of the image fitting analysis}\label{sec:verify}

In this section, we verify the adequacy of 
the methods of the image fitting analysis
and the error estimation.

In this image fitting analysis,
we assumed that the surface brightness of the GRXE can be approximated
by the composition of two components: discrete point sources and 
uniform background emission.

For the model of the discrete sources, 
the PSF and the effective area are calibrated well by observational data
within a systematic error smaller than 5\% \cite{Takahashi1995}.
Those errors were included in estimating the fitting errors
as the 5\% systematic errors. 
If a source is spatially extended more than the HPD of the PSF, 
the discrepancy of the real response from the PSF should be large.
However, extended X-ray sources are fairly restricted to 
only a few kinds of sources such as SNRs, which are expected to be 
known from the catalog of the Galactic SNRs observed in radio 
\cite{Green1998}.
Also, the fraction of SNRs in the resolved X-ray sources was as small as 10\%
(Table \ref{tab:catid}). 
Thus, systematic error in the model for discrete sources 
is not severe.

The adequacy of the uniform background emission model 
had not been confirmed for the GRXE.
To verify the background model, 
we fitted the raw observed image to the background model
without systematic errors,
masking the resolved sources.
The residuals of the fit should imply the systematic errors of the model and 
the fluctuation of the background emission
in a scale smaller than that of the GIS FOV ($\sim 30\arcmin$).
The fit was performed for the image composed of the GIS2 and the GIS3 data
and binned by 12$\times$12 pixels (3$'$$\times$3$'$)
in order to obtain good enough photon statistics for the background GRXE image.
In the 12$\times$12-bin image, 40--50 photons 
were typically expected in each bin under the conditions of this survey observation.
Also, since the bin size of 12 pixel $=3'$ is similar to the HPD of the PSF,
event counts of each bin can be considered to be independent
throughout the observed field. 
The fit resulted in a significant detection of the residual fluctuation.
However, the amplitude of the residual fluctuation 
was evaluated to be at most 11\% of the Poisson errors 
in the 12$\times$12-bin image
\cite{Sugizaki1999}.
If we assume that the residuals of the fluctuation 
is scaled by a bin size,
the ratio of the residual fluctuation to the Poisson error 
is reduced to 3.5\% in the 4$\times$4-bin image.
That assumption is reasonable because the fluctuation on scales smaller than
the HPD of the PSF should be smoothed by the PSF in the observed image.
The obtained value of 3.5\% is in reasonable agreement with the 5\% systematic errors included 
in the model fitting of source survey.
Thus, the discrepancy of the background model from the real surface brightness 
was appropriately accounted for in the source survey.

From these verifications, the results of the image fitting analysis
are confidently confirmed.
The origin of the residual fluctuation in the background GRXE image
detected in this verification procedure 
will be discussed in a separate paper.

\section{Discussion}
\label{sec:discuss}

As a result of image analysis,
we succeeded in resolving 163 faint X-ray sources with 
a flux down to $10^{-12.5}$ {\flux} and 
determined large-scale longitude intensity variations of the GRXE,
eliminating a contamination of the resolved discrete X-ray sources.
We analyzed spectra of the resolved sources and 
classified due to the spectral properties
in order to search for their identifications.
We also derived the LogN-LogS relations of these sources.
Particularly, in the energy band above 2 keV, 
the LogN-LogS relation of the Galactic X-ray sources down to $10^{-12.5}$ {\flux}
was obtained for the first time in this {\asca} survey.

In the follwing, we consider the implications of these results;
(1) longitude intensity profiles of the GRXE,
(2) spectral properties and identifications of the resolved sources, and
(3) LogN-LogS relations of the resolved sources.
We discuss in the soft and the hard band separately
since the influence of the obscuration by the ISM on the Galactic plane
is quite different between the X-ray bands below and above 2 keV.

\subsection{Longitude profile of the Galactic ridge X-ray emission}

\subsubsection{Soft band}

We can see from Figure \ref{fig:ldist} that 
the large-scale longitude profile of the GRXE in the soft X-ray band below 2 keV 
is fairly variable and asymmetric across the Galactic center
with an enhancement on the $l>0$ side.
The large-scale profile of soft X-ray on the Galactic plane 
has been observed in the past X-ray sky surveys
such as the HEAO-1 \cite{Garmire1992} and the {\rosat} \cite{Snowden1997} surveys.
The result of the {\asca} survey is consistent with the past observations
although there are slight differences in detail 
because of the differences of the spatial resolution and 
the energy dependence of effective area
among those surveys.

The observed large variability of the profile is reasonably expected 
from the fact that the soft X-ray band below 2 keV
is heavily obscured by the ISM on the Galactic plane. 
Thus, the profile of the surface brightness 
reflects the both distributions of emission sources and absorbing matters 
located less than $\sim 1$ kpc from the Solar system. 
From the all-sky map obtained in the past sky survey \cite{Garmire1992,Snowden1997},
it is suggested the the large-area emission comes from hot plasma 
with a temperature of about $10^{6.5}$ K 
associated with local hot bubbles such as the Loop I and the Local Hot Bubble (LHB) 
(e.g., Egger \& Ashenbach 1995).  
The longitude profile obtained in the {\asca} survey 
also confirmed this picture approximately
although the profile on the Galactic plane is rather complicated 
because of the coexistence of the emission region
and the thick obscuring ISM.
The large enhancement of the emission in the direction towards $l\simeq 15^\circ$ corresponds
to the direction to the Loop I and the peaks in the direction towards $l\simeq 28^\circ$ 
and $l\simeq -28^\circ$ correspond to the interacting rims of the Loop I with the LHB.

Also, it is suggested from
the energy spectra of the scutum arm obtained by {\asca} 
that the origin of the soft X-ray component in the GRXE can be
explained by hot plasma produced in SNRs \cite{Kaneda1997b}.
These spectral properties are also consistent with the picture
of local hot bubbles.

\subsubsection{Hard band}

The large-scale profile of the GRXE in the X-ray band above 2 keV
has previously estimated by the {\exosat} survey in the 2--6 keV band \cite{Warwick1985},
by the Ginga survey in the 1--20 keV band \cite{Yamauchi1993}
and by the RXTE survey in the 2--60 keV band \cite{Valinia1998}.
However, since the previous observations had no imaging capability, 
their sensitivities for discrete sources were seriously affected
by the confusion of bright Galactic X-ray sources.
Thus, inclusive contribution of faint discrete sources has been 
difficult to be estimated.
In this {\asca} survey, we successfully determined the profile,
eliminating the contamination of resolved discrete sources 
with a flux down to $\sim 10^{-12.5}$ {\flux}.

If the energy band is limited to above 4 keV, 
the obscuration by the ISM is negligible
and also confusion of the soft X-ray component of the GRXE
may be neglected \cite{Kaneda1997b}.
We can see from Figure \ref{fig:ldist} that the obtained longitude profile 
in the energy band above 4 keV is relatively symmetrical and that
the emission is strongly enhanced within the area of $|l|\lesssim 30^\circ$
if the CXB component is subtracted.
The area of $|l|\lesssim 30^\circ$ corresponds to the Galactic inner disk
with a radius of 4 kpc.
Thus, it reveals that the emissivity of the GRXE is considerably 
concentrated within the Galactic inner disk.

Also, in addition to the symmetrical component,
large peaks are apparently seen in the directions to $l\simeq 23^\circ$,  
which correspond to the tangential direction of the 4 kpc arm.
Then, we next compared the fine longitude variation of the GRXE in the 4--10 keV band
with the profiles of radio emission at 408-MHz \cite{Haslam1982} and 115-GHz CO-line \cite{Dame1987}.
These radio bands are almost transparent in the Galactic ISM 
and the both spatial resolutions are $\lesssim 0.5\arcdeg$,
which is sufficient for comparison with the obtained GRXE profile.
The 408-MHz radio emission is mainly originated by synchrotron radiation of 
relativistic cosmic-ray electrons with an energy of $\sim 10$ GeV 
under the $\sim 3\mu$G magnetic field in the Galaxy,
thus the profile 
is expected to be associated with high-energy phenomena such as SN explosions in the Galaxy.
On the other hand, the CO-line emission is associated with dense molecular clouds,
which should be good tracer of the Galactic arm structures.
 
\begin{figure}[p]
\begin{center}
\includegraphics[width=12.5cm, angle=-90]{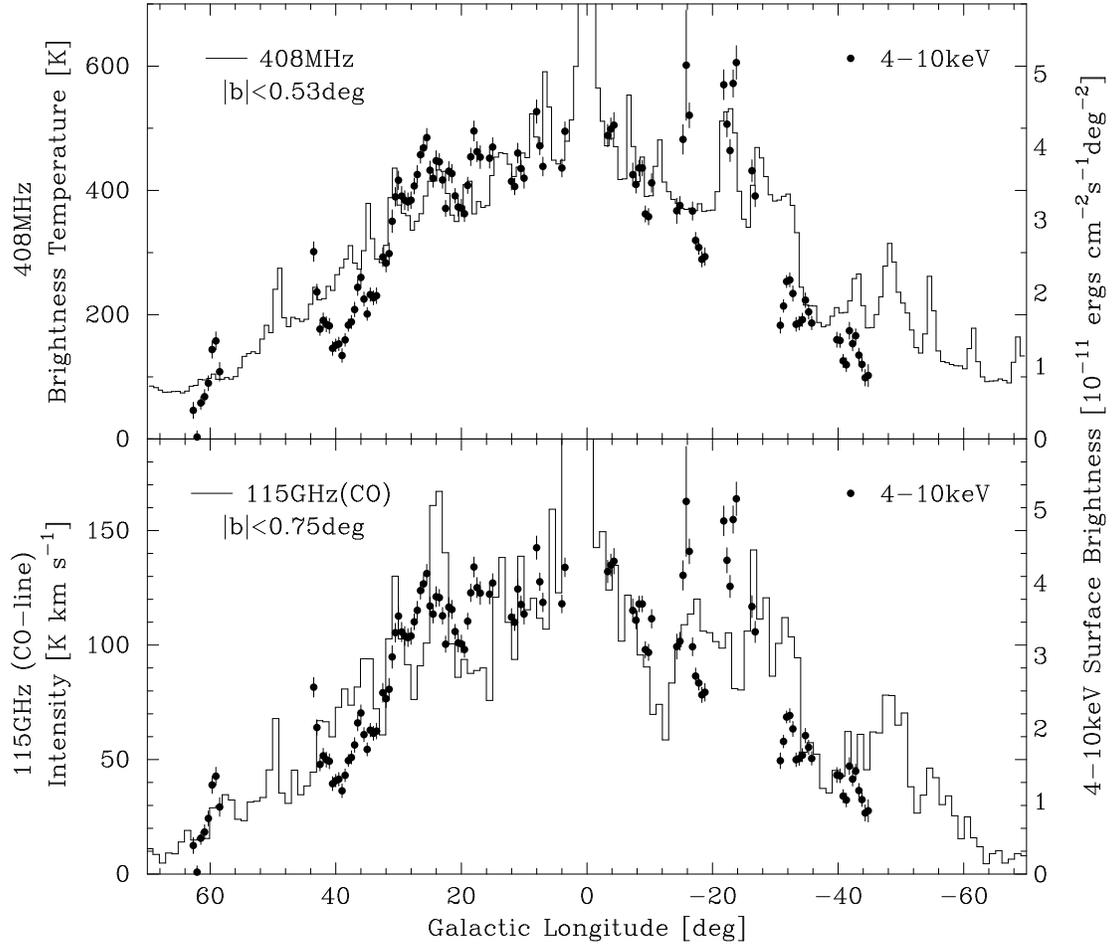}
\caption{
Correlation of longitude intensity profiles of the Galactic radio emission 
at 408-MHz (upper) and at 115-GHz (CO-line) (lower) 
with the profile of the Galactic X-ray emission in the 4--10 keV band.
Inclusive contributions of the CXB in the 4--10 keV band 
is subtracted
(see Figure \ref{fig:ldist}). 
}
\label{fig:radio_xray_dist}
\end{center}
\end{figure}

Figure \ref{fig:radio_xray_dist} shows 
the longitude intensity profiles of radio emission at 408-MHz and CO-line, 
superposing the profile of the GRXE in the 4--10 keV band, where
the profile at 408-MHz is averaged within $|b|<0\fdg 53$ and 
the profile at CO-line is averaged within $|b|<0\fdg 75$ 
in the direction to the Galactic latitude.
Also, the inclusive CXB component is subtracted in the 4--10 keV profie.
We see in Figure \ref{fig:radio_xray_dist} that 
the correlation between the 4--10 keV X-ray and the 408-MHz radio
is fairly good and apparently better than the correlation with the CO-line.
These relationships might imply the connections of the GRXE
with the cosmic-ray electrons suggested by 
Valinia et al. (2000), 
and/or with the magnetic field suggested by Tanuma et al. (1999), 
although the electrons contributing to the GRXE 
should be different from those responsible 
for the 408-MHz synchrotron radiation (Skibo, Ramaty \& Purcell 1996).

\placefigure{fig:radio_xray_dist}

\subsection{Identification of the resolved X-ray sources}
\label{sec:disid}

The $107/163\simeq 66\%$ of the resolved X-ray sources in this ASCA survey
have been unidentified with cataloged X-ray sources, SNRs, and optical stars
in the SIMBAD database.
Thus, they might represent a new aspect of X-ray sources in the Galaxy.
Here, we consider the origins of the unidentified sources
resolved in the ASCA survey based on the results of the spectral analysis
in $\S$ \ref{sec:anaspec}.

\subsubsection{Relatively bright X-ray sources with $F>10^{-11}$ {\flux}}

As for the ten relatively bright sources with a flux above $1.0\times 10^{-11}$ {\flux},
we analyzed the spectra individually and also investigated the flux variabilities.
These results are clearly distinguished into two groups.
Some with soft X-ray spectra showing a power-law index of $\Gamma>2$,
are identified with SNRs.
These soft X-ray spectra are agreed with the typical SNR spectra, which show hot plasma emission
with a temperature of $\sim$0.1--2 keV.
Results of these detailed spectral analyses have been already reported (see references in Table \ref{tab:10brtsrc}).
On the other hand, the others with hard X-ray spectra showing a power-law index of $\Gamma<2$,
are unidentified with the cataloged X-ray sources and found to show flux variabilities.
These variable sources are considered to be binary X-ray systems 
embodying compact degenerate stars such as white dwarfs, neutron stars, and black holes.
Since the absorption column density is larger than $3\times 10^{22}$ {\col} for any of these sources, 
it is reasonable that no optical counterparts are identified for these sources.
Comparing with the spectra of the known Galactic X-ray sources,
one of these, AX J153818$-$5541, which shows the power-law index of $\Gamma\simeq 2$,
is likely to be a low-mass X-ray binary (LMXB) (e.g. Mitsuda et al. 1984)
and the others with the indices of $\Gamma < 1.0 $ is likely to be
high-mass X-ray binaries (HMXBs) (e.g. Nagase 1989).
Magnetic CVs are also considered as anther possible candidates
for the latter harder X-ray sources
since the spectra can be explained by a multi-absorption hot plasma 
emission model, too.
As one example, AX J183221$-$0840, which shows a 1549-s coherent pulsation,
are considered to be probably a magnetic CV 
because a significant iron K-line emission with a line center energy of 6.7 keV 
are detected with a follow-up observation \cite{Sugizaki2000}. 

These identifications are summarized in Table \ref{tab:10brtsrc}.

\subsubsection{Faint X-ray sources with $F<10^{-11}$ {\flux}}

As for the faint sources with a flux below $1.0\times 10^{-11}$ {\flux},
we collected source samples with a significance above 5$\sigma$,
and classified them into the six groups
due to the power-law index and the absorption column density
obtained from the fitting with a power-law model.
We analyzed the spectra summed within each source group.
Since the sources belonging to the same group share common spectral properties,
they are considered to be a similar kind of X-ray sources.
In the following, we estimate the typical luminosity and 
the space number density for each source group taking account 
of the grouping criteria and the observation condition in this ASCA survey,
and discuss about the origins.

Assuming the limit of the source visibility from the solar system 
as $d$ pc for all the area surveyed,
the number density of the detected sources, $\rho$ pc$^{-3}$, 
is represented by 
\begin{equation}
\rho = N/(\Omega d^3/3) \label{equ:density}
\end{equation}
where $N$ is the number of detected sources
and $\Omega$ is the solid angle of the area surveyed, here 39.8 deg$^2 = 1.21\times 10^{-2}$ str.
We can estimate the number density from equation (\ref{equ:density})
if we know the limit of the visibility.
Here, we estimate the limit of the visibility to each source group
from the constraint to the ISM column density,
assuming the averaged ISM density on the Galactic plane to be $\rho_{\rm H}\simeq 1$ cm$^{-3}$
as a canonical value.
Since the area surveyed is fairly restricted to the low Galactic latitude area of $|b|\lesssim 0\fdg 4$,
the visibility determined by the ISM absorption
should be almost same within the area surveyed.

The group (a) is classified with a hard spectral index of $\Gamma <1.0$.
Assuming that all the sources are belonging to the Galaxy, 
$d$ is approximated by the distance to the edge of the Galaxy, $\sim$18 kpc.
Thus, the number density of this group
is estimated at $\sim 4\times 10^{-10}$ pc$^{-3}$ as the lower limit.
Also, assuming that the distance to the sources is 8 kpc as the typical Galactic sources, 
the luminosity is estimated at $1.9 \times 10^{34} d_{\rm 8 kpc}^2${\lumi} 
from the averaged flux in table \ref{tab:specfit}.
Comparing the summed spectrum with those of the known X-ray sources, 
the property of the flat spectrum is good agreed with that of HMXB pulsars (e.g. Nagase 1989).
As one evidence, a 715-s periodic pulsation is discovered from 
one of these sources, AX J170006$-$4157, 
using data taken in the observation for another target \citep{Torii1999}.
In this survey observation, pulsation is hard to be recognized because
the exposure time of 10 ks in each pointing is not enough for pulse search analysis.
The obtained luminosity of $\sim 10^{34} d_{\rm 8 kpc}^2${\lumi}
is by one order of magnitude smaller than that of the known faintest X-ray pulsar.
Thus, unidentified sources in this group are considered to be HMXB pulsars 
which have ever been unrecognized under the sensitivity limit.    
Another possibility to explain the flat spectrum is a multi-component spectrum 
with different absorption column densities.
In fact, the summed spectrum is also fitted with a multi-absorption thin-thermal
plasma model with a temperature of $\gtrsim 15$ keV. 
Such spectra are observed in some magnetic CVs \citep{Ezuka1999}
and the obtained average luminosity in this group is 
also agreed with that of the typical magnetic CVs.

The group (b-i) is classified with a photon index of $1<\Gamma<3$ 
and an absorption column density of $N_{\rm H}< 0.8 \times 10^{22}$ {\col}.
The summed spectrum is represented by a single power-law and shows no significant emission line.
From the upper-limit of $0.8\times 10^{22}$ {\col} to the ISM column density,
the distance to the sources is restricted to be less than $\sim$ 2.6 kpc.
Then, the luminosity is estimated to be below $8.9 \times 10^{32} d_{\rm 2.6kpc}^2$ {\lumi}
from the observed flux
and the number density is estimated at $\sim 4\times 10^{-7}$ pc$^{-3}$.
Six of a total of 23 sources in this group are identified with optical stars in the SIMBAD database. 
However, the summed spectrum in this group, 
represented by a single power-law of $\Gamma \simeq 2$,
is not agreed with typical X-ray spectra of stars, 
which show thin-thermal plasma emission with a temperature of $\sim$0.1--3 keV 
even if they are late-type \cite{Singh1999} 
or early-type (Cohen, Cassinelli \& Waldron 1997). 
If the observed spectrum is fitted with thermal models,
it is derived that the plasma temperature should be as high as $\sim 5$ keV 
and the plasma abundance should be as low as $\sim 0.2$.
Although such a low abundance plasma is observed in some active coronae \cite{Singh1999},
the plasma temperature of $\sim 5$ keV seems to be too high.
Thus, emission from optical stars should not occupy a dominant fraction of this group.
The spectral property of a power-law of $\Gamma \simeq 2$ rather
resembles that of bright Low Mass X-ray Binaries (LMXBs) (e.g. Mitsuda et al. 1984).
However, the luminosities of bright LMXBs are within $10^{36}$ -- $10^{38}$ {\lumi}.
When the LMXBs are in quiescence with a luminosity as low as $10^{33}$ {\lumi},
they likely show very soft spectra represented by a thermal bremsstrahlung 
with a temperature of $\sim 0.5$ keV \citep{Asai1998},
which should be classified in the group (c-i).
However, some of quiescent LMXBs exhibit hard tails with a photon index of $\sim 2$ \citep{Asai1996}.
They are agreed with this group.
Thus, this type of LMXBs in quiescent state is one of possible candidates.
Another possibility is a rotation powered pulsar such as the Crab pulsar.
Some of rotation powered pulsars and their pulsar nebula are found to have non-thermal hard spectra
represented by a power-law of $\Gamma \simeq 2$ (Kawai, Tamura \& Shibata 1998). 
Since the photon statistics of each source is very poor and the time resolution is not so good
in this survey observation, 
evidence of pulsation is hard to be recognized even if they show periodic pulsation.
Considering those facts, 
the sources in this group might be quiescent LMXBs or faint Crab-like pulsars 
which come to be observed in this survey.

The group (b-ii) is classified with a photon index of $1<\Gamma<3$ and 
an absorption column density of $N_{\rm H}=(0.8-3)\times 10^{22}$ {\col}.
Assuming that the absorption column density is responsible for
the ISM, the distances to the sources
are in $\sim 2.6-9.7$ kpc range.
Thus, the typical luminosity and the number density of the sources in this group are 
estimated at $\sim 7.5 \times 10^{33} d_{\rm 8ks}^2$ {\lumi} and $\sim 3\times 10^{-9}$ pc$^{-3}$, 
respectively.
The summed spectrum shows strong iron and silicon K-line emissions.
The coexistence of the iron and the silicon lines implies that the emission
comes from hot plasma in multi-temperature and/or ionization nonequilibrium state.
The spectrum can be fitted with a two-temperature thin-thermal plasma model
with best-fit temperatures of 6.7 keV and 0.28 keV.
The most promising candidates of the origins of this group are considered to be magnetic CVs
from the high plasma-temperature of $\sim$ 6.7 keV and the strong iron K-line.
Magnetic CVs exhibit spectra of multi-temperature plasma due to 
temperature gradient in the accretion column \cite{Fujimoto1997}.
X-ray emission from hot plasma at the bottom of the accretion column 
is expected to be absorbed by the outer cool part.
Thus, it is possible that the observed $N_{\rm H}$ value includes intrinsic absorption to some content,
and that we choose magnetic CVs by source selection with a medium absorption column density
in this group.

The group (b-iii) is classified with a photon index of $1<\Gamma<3$ and 
an absorption column density of $\nh >3\times 10^{22}$  {\col}.
The column density of $3\times 10^{22}$ {\col} is near to that of the interstellar medium 
integrated all over the Galaxy.
Thus, if the emission source has no intrinsic absorption,
it should be outside of the Galaxy.
The X-ray spectrum of the power-law with the photon index $\Gamma \simeq 2$
is agreed with that of typical active galactic nuclei (AGNs).
Estimating the number of extragalactic sources included in this source sample 
from the LogN-LogS relation of the extragalactic sources \citep{Ueda1999} and this survey condition, 
it is derived to be 12, which occupy $12/17\simeq70$\% in this group.
Thus, the major part of this group is considered to be extragalactic sources 
observed through the Galactic ISM.
On the other hand, three sources in this group 
are identified with SNRs.
The two of these are classified into center-filled type SNRs, including
G10.0$-$0.3 which is also identified as soft gamma-ray repeater, SGR1806$-$20 \cite{Sonobe1994, Kouveliotou1998}.
Thus, some of this group might be compact X-ray sources involved in SNRs.

The group (c-i) is classified 
with a photon index of $\Gamma>3$ and an absorption column density of $\nh <3\times 10^{22}$  {\col}.
From the upper-limit of $3\times 10^{22}$ {\col} to the ISM column density,
the distance to the sources is constrained to be less than $\sim$ 9.7 kpc in this group.
Thus, the typical luminosity is estimated at $1.4\times 10^{34} d_{\rm 8 kpc}^2$ {\lumi}
and the number density of the sources is estimated at $\sim 4\times 10^{-7}$ pc$^{-3}$.
The X-ray spectrum summed in this group shows strong silicon K-line emission
and is represented by two-component thin-thermal plasma model with temperatures of 
$\sim$0.4 keV and $\gtrsim 2$ keV.
The model of the hard component is not determined here
since the relative intensity of the hard component is very small 
and an iron K-line emission is not significantly observed.
Six of a total of 12 sources in this group are cross-identified with the ROSAT all-sky survey
and three of these are also identified with SNRs.
Also, other three sources are identified with optical stars in the SIMBAD database.
The most of the ROSAT sources on the Galactic plane are identified with stellar active coronae and 
the rest are identified with OB stars and CVs \citep{Motch1997}.
The obtained summed X-ray spectrum, exhibiting plasma emission with a temperature of $\sim 0.4$ keV,
are good agreed with these sources.
Thus, the majority of the sources in this group is considered to be active coronae
and some of OB stars, CVs, and SNRs are expected to be contained.
LMXBs in quiescence are also considered as another minor candidate for this group,
which is described in the group (b-i).

The group (c-ii) is classified with a photon index of $\Gamma>3$
and an absorption column density of $\nh >3\times 10^{22}$ {\col}.
We cannot estimate the distance to the sources from the ISM absorption.
Thus, we approximate it by the distance to the edge of the Galaxy, $\sim$18 kpc, 
assuming that all the sources in this group are in the Galaxy. 
The number density of this group is estimated at $\sim 9\times 10^{-10}$ pc$^{-3}$ as the lower limit.
Also, assuming the distance to the typical Galactic sources as 10 kpc,
the typical luminosity is estimated at $1.6 \times 10^{34} d_{\rm 10 kpc}^2${\lumi}
from the averaged flux.
The X-ray spectrum summed within this group, exhibiting strong silicon and iron K-line emissions
and fitted with a two-temperature thin-thermal plasma model, 
resembles that of the group (b-ii). 
The difference is in the temperature of the hard component: $kT_1\sim 2$ keV in this group 
is lower than  $kT_1\sim 7$ keV in the the group (b-ii).
Thus, these identifications can be discussed similarly
and magnetic CVs are considered to be the most promising candidates.
However, since the plasma temperature of $\sim 2$ keV is relatively low, 
active coronae with a highest plasma temperature are considered as another possible candidates.
Also, the three sources are identified with the SNRs.
However, since the plasma temperature of the typical SNR is not so high to 
exhibit strong iron K-shell line,
the fraction of SNRs in this group should be little
if unidentified SNRs are contained.

The estimated number densities, the typical luminosities, and the identifications 
of the grouped faint X-ray sources 
are summarized in Table \ref{tab:discussrc}.

\subsection{LogN-LogS relations of the faint Galactic X-ray sources}

The LogN-LogS relation is one of the most important sources of information
about the spatial distribution and the luminosity function.
However, in the case of Galactic sources,
simple approximation of an infinite Euclidean distribution is not appropriate
because of the finite scale of the Galaxy, local irregularities such as arm structures, 
obscuration of the thick ISM on the Galactic plane, and
contamination from the extragalactic sources coming through the ISM.
Moreover, we need to take account of the limited area coverage to the low latitude region of $|b|<0\fdg 4$ 
in this {\asca} survey.
We discuss the implications of the LogN-LogS relations considering those conditions.

\subsection{Soft band}

\subsubsection{Contamination by extragalactic sources
and the LogN-LogS relation of the Galactic sources}

The obtained LogN-LogS relation might contain some extragalactic sources.
Here, we estimate the LogN-LogS relation of the extragalactic sources
observed through the Galactic plane.

The LogN-LogS relation of the extragalactic X-ray sources below 2keV
have already been measured by {\einstein} (e.g., Gioia et al. 1990), 
{\rosat} (e.g., Hasinger et al. 1993) 
and {\asca} \cite{Ueda1999} and
the consistency of these results was confirmed. 
We utilized the LogN-LogS relation obtained in the {\einstein} survey here,
which is expressed by
\begin{equation}
N(>f)=2.68\times 10^{-19} \left(f{\rm [ergs~cm^{-2}s^{-1}]}\right)^{-1.48} ~{\rm [deg^{-2}]},
\end{equation} 
where $f$ is the flux of the 0.3--3.5 keV band.

On the Galactic plane, soft X-rays from extragalactic sources are
heavily absorbed by the ISM.
These fluxes need to be corrected for the absorption of the ISM in order to
estimate the contamination from extragalactic sources.  
We estimated the averaged column density for the extragalactic sources in the area surveyed
at 5$\times$10$^{22}$ {\col} from Figure \ref{fig:ldist}.
Assuming a power-law photon index of $\Gamma=2$ as a typical spectrum of extragalactic sources, 
photons coming through the Galactic plane with $N_{\rm H} \simeq 5\times 10^{22}$ {\col} 
are reduced by a factor of $8.1\times 10^{-4}$ in the GIS 0.7--2 keV band.  
Converting the energy flux of the 0.3--3.5 keV band to the photon flux of the GIS 0.7--2 keV band
and taking the Galactic absorption into account for the typical power-law
spectrum, 
the LogN-LogS relation of the extragalactic sources observed
through the Galactic plane is estimated to be
\begin{equation} 
N(>f)=1.05\times 10^{-3} \left(f{\rm [cnts~GIS^{-2}s^{-1}]}\right)^{-1.48} ~{\rm [deg^{-2}]}.
\label{equ:lnls_cont_s}
\end{equation}
This is shown in Figure \ref{fig:lnls_s_etc}
together with the LogN-LogS relation of the Galactic X-ray sources resolved in this
survey.  
We see that the contamination from extragalactic sources is negligible in the
soft X-ray band.
Thus, all the sources detected in the soft X-ray band 
are considered to be Galactic sources located near to the Solar system
without large interstellar absorption.
 
\subsubsection{Comparison with previous work and the implications of the LogN-LogS relation}

\begin{figure}[p]
\begin{center}
\includegraphics[width=10cm, angle=-90]{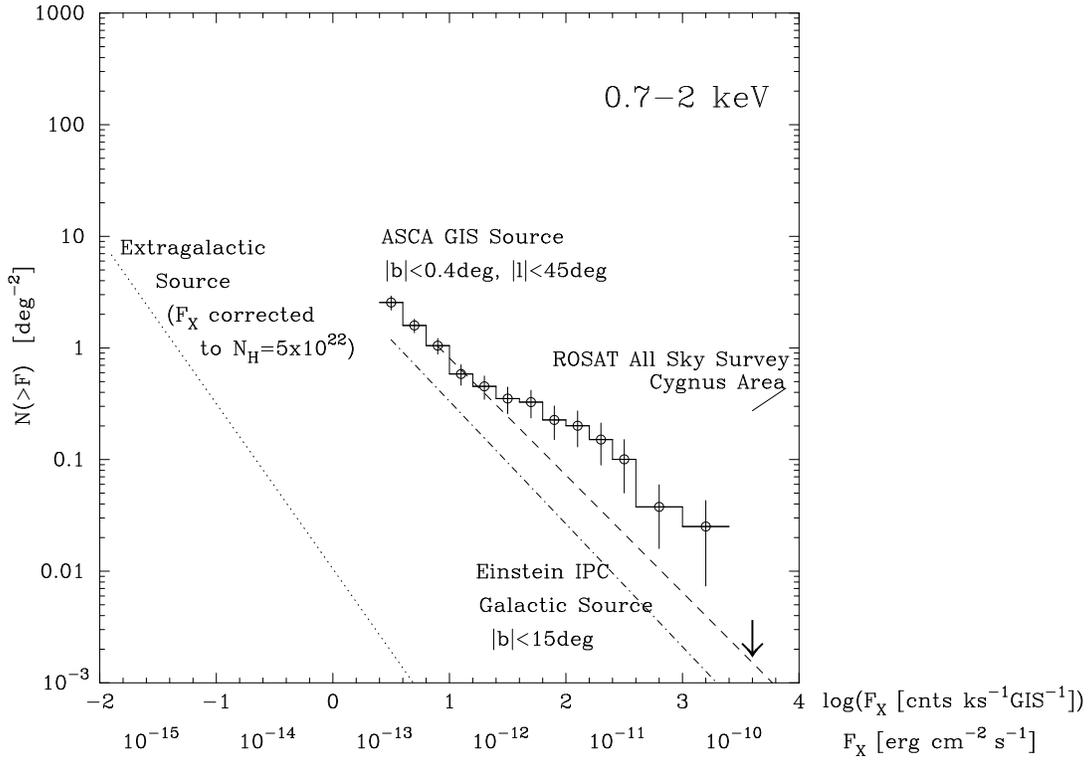}
\caption{
LogN-LogS relation of the Galactic X-ray sources with the {\asca} (open
circle with error bar), {\rosat} (dashed line), and {\einstein} (dot-dashed line)
surveys and the expected contamination from the extragalactic sources in
the 0.7--2 keV band (dotted line).}
\label{fig:lnls_s_etc}
\end{center}
\end{figure}

The obtained LogN-LogS relation of the Galactic X-ray sources 
in the soft X-ray band is shown in Figure \ref{fig:lnls_s_etc}.
The profile is rather flat and far from a simple single power-law 
which is expected for a uniform infinite spatial source distribution.
In the soft X-ray band, the LogN-LogS relation of Galactic X-ray sources 
has previously been measured with {\einstein} \cite{Hertz1984} 
and {\rosat} \cite{Motch1997} surveys although
the survey areas are different from each other.  
The LogN-LogS relations of the {\einstein} and the {\rosat} surveys are shown 
in Figure \ref{fig:lnls_s_etc} together with the {\asca} result.
Here, notice that the fluxes measured with different instruments should depend slightly on the
assumed spectral model.  We assume a power-law spectrum with a photon index of $\Gamma=2$ and 
an absorption column density of $N_{\rm H} = 1.0\times 10^{22}$ {\col} for the Galactic X-ray sources
and use the PIMMS software \cite{Mukai1993} to convert energy flux.

We can see that the LogN-LogS relations in the soft X-ray band 
obtained by the {\einstein}, the {\rosat}, and {\asca} surveys 
are slightly different.
It can be reasonably expected from the fact 
that the kinds and the populations of X-ray sources in these source samples 
are quite different from each other
due to the differences of the energy dependence of effective area and
the direction of the area surveyed.
The {\einstein} survey in Hertz \& Grindlay (1984) 
is based on a serendipitous source survey within $|b|<15^\circ$ in
the IPC field, eliminating a primary target.  
The total covering area of 275.7 deg$^2$ takes 2.6\% of the $|b|<15^\circ$ region.  
The $\sim$46\% of the {\einstein} sources are identified with stellar coronae
and the contamination of extragalactic sources is estimated at $\sim$23\%.
Such a significant contamination of extragalactic sources 
is quite different from the other surveys.
It is owing to the relatively large scale height of $b\simeq 15^\circ$ of the area surveyed, 
where the column density of the ISM is $\sim 10^{20}$ {\col}, smaller by two orders of magnitude
than that on $b\simeq 0^\circ$.
Also, the {\rosat} survey in Motch et al. (1997)
used the data of the low diffuse-background Cygnus area 
observed in the {\rosat} all sky survey, 
which covers 64.5 deg$^2$ of a $10\times 10^\circ$ rectangular region 
centered at $(l,b)=(90\fdg 0, 0\fdg 0)$.
The majority about 85\% of the ROSAT sources are identified with stellar active coronae. 
On the other hand, we know from the spectral analysis of the resolved sources in $\S$ \ref{sec:anaspec}
and the discussion in $\S$ \ref{sec:disid}
that the ASCA sources include several kinds of X-ray sources with various X-ray spectra 
and the ratio of candidates of active coronae with soft X-ray spectra
is not so large.
It is considered to be due to the energy dependence of the effective area of the ASCA-GIS 
increasing toward higher energy in the 0.7--2 keV band,
which is contrary to the ROSAT-PSPC.
Thus, 
the LogN-LogS relation of the ASCA Galactic X-ray sources obtained here
in the soft X-ray band
is considered to reflect the complex mixtures 
of various X-ray sources located near the Solar system.

\placefigure{fig:lnls_s_etc}

\subsubsection{Contribution to the GRXE}

We estimate the contribution of the resolved faint X-ray sources
to the GRXE based on the LogN-LogS relation obtained.  
The average surface brightness within the central emission region 
of $|l|<35\arcdeg$ in the 0.7--2 keV band is derived to be
$S_{\rm GRXE} = 1.3\times 10^{-11}$ ergs cm$^{-2}$ s$^{-1}$ deg$^{-2}$ 
from the large-scale profile in Figure \ref{fig:ldist}. 
The total flux of the resolved X-ray sources, $S_{\rm p}$, 
is calculated from the LogN-LogS relation by 
\begin{equation}
S_{\rm p} = \int_{f_{\rm min}}^{f_{\rm max}} N(f)f df = \int_{f_{\rm min}}^{f_{\rm max}} \left\{ -\frac{dN(>f)}{df}f \right\} df,
\label{equ:lnls_cont}
\end{equation}
where $f_{\rm min}$ and $f_{\rm max}$ are the minimum and the maximum fluxes
of the contributing X-ray sources.  Here, we adopt $f_{\rm
min}=3.2$ cnts GIS$^{-1}$ ks$^{-1}$ from the minimum flux of the
resolved X-ray sources and $f_{\rm max}=285$ cnts GIS$^{-1}$ ks$^{-1}$
corresponding to $10^{-11}$ {\flux}, the lower-limit of bright sources excluded from the analysis.
Calculating equation (\ref{equ:lnls_cont}) and
converting the unit of flux to ergs cm$^{-2}$ s$^{-1}$ deg$^{-2}$, 
the accumulated surface brightness, $S_{\rm p}$, is derived to be $1.5\times 10^{-12}$ ergs
cm$^{-2}$ s$^{-1}$ deg$^{-2}$.  Therefore, the contribution of the
resolved faint X-ray sources to GRXE is $S_{\rm p}/S_{\rm GRXE} = 11\%$.
Although we do not know the number density of the sources 
with a flux smaller than the detection limit,
such a small value for the contribution of the resolved sources to the GRXE 
in this soft X-ray band is reasonable
based on the hypothesis that the origin of the Galactic soft X-ray emission is 
diffuse hot plasma in local hot bubbles (e.g., Kaneda at al. 1997).


\subsection{Hard band}

\subsubsection{Contamination by extragalactic sources and the LogN-LogS relation of the Galactic sources}

\begin{figure}[p]
\begin{center}
\includegraphics[width=10cm, angle=-90]{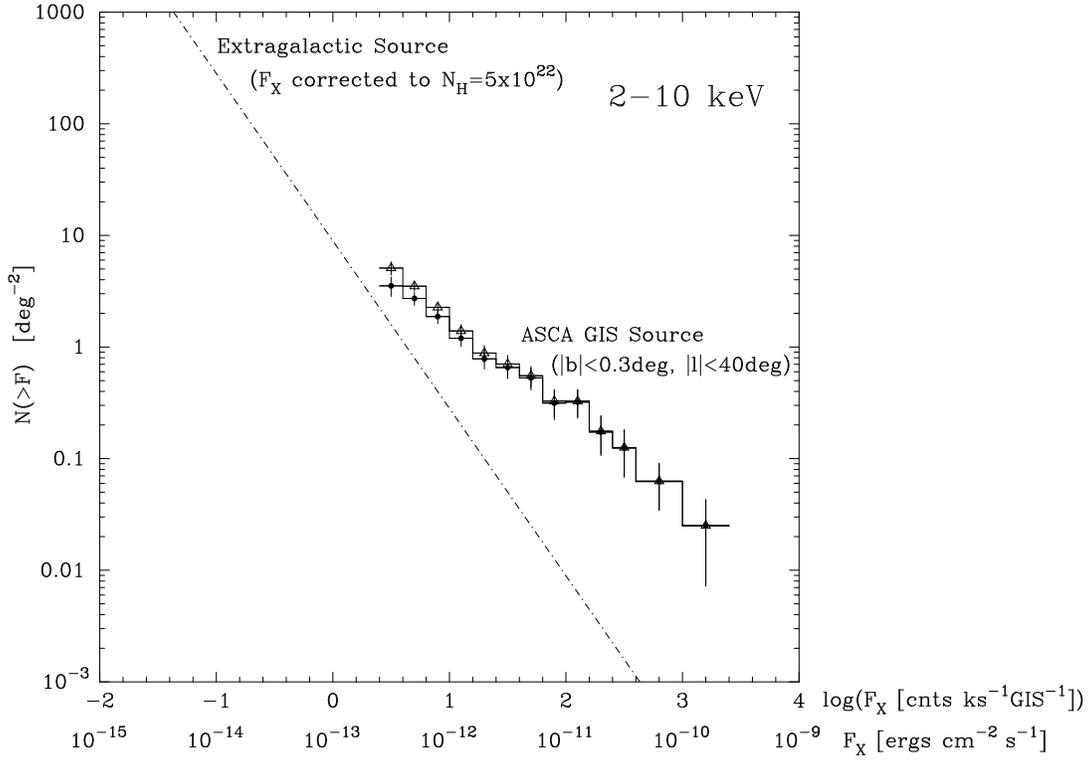}
\caption{
LogN-LogS relation of the {\asca} Galactic X-ray sources (step functions with open triangles and solid circles) 
and the expected contribution of the extragalactic sources (dot-dashed) in the 2--10 keV band.
Open triangles are the LogN-LogS relation before subtracting the
contamination from the extragalactic sources and filled circles are
those after subtracting the extragalactic sources.}
\label{fig:lnls_h_etc}
\end{center}
\end{figure}

Since the ISM on the Galactic plane is almost transparent
in the X-ray band above 3 keV,
the extragalactic X-ray sources should 
contribute to the LogN-LogS relation obtained in the 2--10 keV band.
The LogN-LogS relation of the extragalactic sources in the band above 2 keV 
has been measured with HEAO-1 A2 \cite{Piccinotti1982}, 
Ginga (Hayashida, Inoue \& Kii 1990) and {\asca} \cite{Ueda1999}.
These results are consistent within the errors.  Here, we adopt the ASCA result,
which is represented by
\begin{equation}
N(>f)=4.67\times 10^{-19} \left(f{\rm [ergs~cm^{-2}s^{-1}]}\right)^{-1.5} ~{\rm [deg^{-2}]}.
\end{equation}
The flux in the 2--10 keV band 
is reduced by a factor of 0.78 for a typical power-law spectrum 
with a photon index of $\Gamma=2$
due to the absorption of the ISM with $N_{\rm H}\simeq 5.0\times 10^{22}$ {\col}.  
Correcting this effect of the absorption and converting the energy flux in the
units of {\flux} to the count flux in the units of cnts GIS$^{-1}$ks$^{-1}$, the
LogN-LogS relation of the extragalactic sources observed through the
Galactic plane is estimated to be
\begin{equation}
N(>f)=8.80\times \left(f{\rm [cnts~GIS^{-1} ks^{-1}]}\right)^{-1.5} ~{\rm [deg^{-2}]}.
\label{equ:lnls_cnt_exgal}
\end{equation}
This relation is shown in Figure \ref{fig:lnls_h_etc},
together with that of the Galactic X-ray sources resolved in this
survey.  
We see that the LogN-LogS relation must be corrected for extragalactic sources.

The LogN-LogS relation of the Galactic X-ray sources in the 2--10 keV band, 
subtracting the contribution of extragalactic sources, is 
shown with the solid circles in Figure \ref{fig:lnls_h_etc}.  
Fitting a power-law model to the LogN-LogS relation of the Galactic sources, 
it is represented by
\begin{equation}
N(>f)= 9.2 \times \left( f {\rm [cnts~GIS^{-1}ks^{-1}]}\right)^{-0.79\pm 0.07}~~~{\rm [deg^{-2}]},
\label{equ:asca_h_lnls} 
\end{equation}
where the error of the power-law index represents the 90\% confidence
level.  
The slope of the power-law model is significantly flatter than $-1$, 
the value expected for a uniform infinite-plane source distribution.
Also, note that it is consistent with the slope of the LogN-LogS relation 
of the bright Galactic X-ray sources 
measured by the previous non-miller instrument on Ariel-V \cite{Warwick1981}
although these flux ranges are not overlapped.

\placefigure{fig:lnls_h_etc}

\subsubsection{Comparison with simulated LogN-LogS relations for typical source distributions}
\label{sec:spatialdis}

\begin{figure}[p]
\begin{center}
\includegraphics[width=10.5cm]{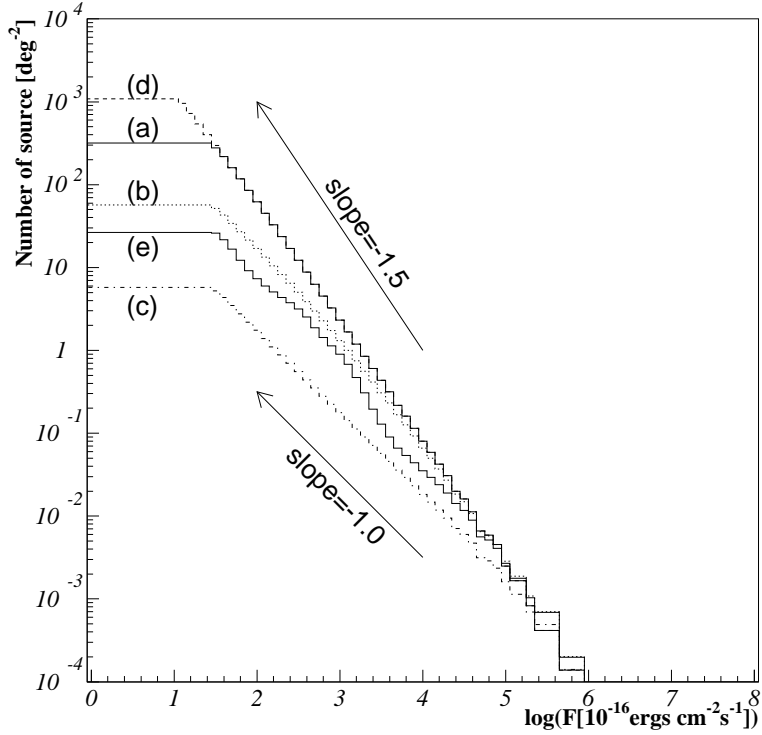}
\caption{Calculated LogN-LogS relation 
under the observational conditions of the {\asca} Galactic plane survey
for some typical source distributions
(see text).
}
\label{fig:lnls_sim}
\end{center}
\end{figure}

In the 2--10 keV band, we can observe almost throughout the Galaxy,
escaping from the obscuration of the ISM.
Thus, we can obtain information about the spatial source distribution in the Galaxy 
from the LogN-LogS relation.
However, since the observed LogN-LogS relation is biased 
by the limited area coverage in this survey,
it is not so easy to derive the implications.
Thus, we take a method to calculate LogN-LogS relations to be observed for typical source distributions 
taking account of the area surveyed,
and compare those with the observed relation. 
We examine the following simple axisymmetrical source distributions
to confirm effects of scale height, Galactocentric radius and 
an arm structure:
\begin{enumerate}
\item[(a)] uniform finite disk within $R_{\rm Gal}\leq10$ kpc and with an exponential scale height of $H=$100 pc
\item[(b)] uniform finite disk within $R_{\rm Gal}\leq10$ kpc and with an exponential scale height of $H=$10 pc
\item[(c)] uniform finite disk within $R_{\rm Gal}\leq10$ kpc and with an exponential scale height of $H=$1 pc
\item[(d)] uniform finite disk within $R_{\rm Gal}\leq20$ kpc and with an exponential scale height of $H=$100 pc
\item[(e)] circular arms at $R_{\rm Gal}=$ 4, 6, 8 kpc with an exponential scale height of $H=10$ pc,
\end{enumerate}
where $R_{\rm Gal}$ represents the Galactocentric radius.
The distance from the Galactic center to the Solar system is assumed to be 8 kpc
and the absorption of the ISM is neglected, which is adequate for the 2--10 keV band.
The luminosity and the number density of the sources are assumed to be $10^{32}$ {\lumi} and $10^{-6}$ pc$^{-3}$,
however those do not affect the slope of the LogN-LogS relation. 
We here concentrate only on the slope of the LogN-LogS relation and do not care for 
the normalization.
The scale height of $H=$100 pc agrees with that of the GRXE (e.g., Warwick et al. 1985).
Figure \ref{fig:lnls_sim} shows
results of calculation for spatial source distributions (a)--(e), together.

\begin{figure}[p]
\begin{center}
\includegraphics[width=8cm]{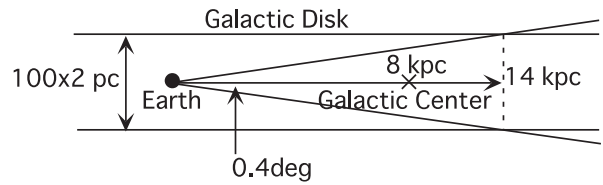}
\caption{
Schematic cross section of the Galactic disk and volume surveyed by
the {\asca} FOV in the Galactic plane survey.
}
\label{fig:diskconfig}
\end{center}
\end{figure}

We can see from the profile of (a)
that the slope of the LogN-LogS relation becomes $-1.5$ for the simple finite disk distribution 
with an exponential scale height of 100 pc. 
The slope of $-1.5$ is consistent with that for an infinite uniform 3-dimensional Euclidean distribution. 
This is because the area of the {\asca} Galactic plane survey is limited to the low latitude of 
$|b|\lesssim 0\fdg 4$.
The scale height subtending a FOV of $0\fdg 4$ radius 
reaches 100 pc at a distance of 14 kpc 
from the Solar system, which is comparable with the size of the Galaxy.
Thus, the effect of the 2-dimensional distribution in the Galactic disk does not appear
and the volume surveyed increases 3-dimensionally until the distance reaches 14 kpc.
This geometrical configuration is illustrated in Figure \ref{fig:diskconfig}.
Comparing the profiles of (a), (b) and (c), 
it is found that the slope of the LogN-LogS relation becomes about $-1$
if the scale height of the source distribution is as low as 10 pc.

The difference between models of (a) and (d) is in the radius of the distribution.
The difference of the LogN-LogS relations 
is only in the lower limit of the power-law relation 
and has little influence on the slope of the LogN-LogS relation.
Therefore, we need not care about the Galactocentric radius of 
the uniform finite disk distribution
when we consider the slope of the LogN-LogS relation.

From comparison of the profiles of (b) and (e),
we find that the LogN-LogS relation becomes flat for the arm distribution. 
This is because the spatial distribution of each arm is 1-dimensional.

\placefigure{fig:lnls_sim}
\placefigure{fig:diskconfig}

\subsubsection{Implications of the observed LogN-LogS relation}

We consider possible source distributions in the Galaxy 
to satisfy the observed LogN-LogS relation,
comparing with the calculated LogN-LogS relations for 
the typical source distributions taking account of the area surveyed.

First, we consider the situation where the luminosity of major X-ray sources 
contributing to the LogN-LogS relation is approximately the same and uniform spatially.
The uniform finite disk model with a scale height of 100 pc
cannot explain the slope of the observed LogN-LogS relation.
If the scale height of the source distribution is restricted to be as low as 10 pc,
the slope of the LogN-LogS relation comes near to $-1$.
However the observed LogN-LogS relation in the 2--10 keV is even flatter.

One possible explanation is that 1-dimensional distributions of the sources 
in the Galactic plane, 
i.e. arm structures, are contributing.
The calculated profile for model (e) approximately shows a flatter LogN-LogS relation
than without arm structures. 
In order to make the LogN-LogS relation still flatter,
the source density needs to decrease with increasing distance.
This implies that the Solar system belongs to a spiral arm 
where the source density is high \cite{Matilsky1979} and/or
that the spiral arm structure has a small pitch angle \cite{Johnson1978}.
If an appropriate model of the Galactic arm is assumed, 
a flat LogN-LogS relation consistent with the observation
is possibly produced.
However, this is beyond the scope at present discussion.

Another possibility is that the observed LogN-LogS relation reflects the luminosity function
rather than the spatial distribution, like the soft X-ray band.
However it is very unlikely 
considering that the obscuration of the ISM is almost negligible in this hard X-ray band.
If the LogN-LogS relation represents the luminosity function of some dominant Galactic X-ray sources,
all of the dominant sources in the Galaxy must be detected.
For this, their luminosity should be higher than $\sim 10^{34}d_{20\rm kpc}$ {\lumi}.
It differs from the obtained properties of the resolved faint sources, of which a half
shows little absorption column density, thus should have a luminosity much smaller than $\sim 10^{34}$ {\lumi}.
Thus, the observed LogN-LogS relation should truly represent the spatial distribution.


\subsubsection{Contribution of the resolved sources to the GRXE}

We found from the grouping spectral analysis in $\S$\ref{sec:anaspec}
that there are groups of X-ray sources showing strong iron K-line emission,
which have a possibility to satisfy the spectrum of the unresolved GRXE.
These group members occupy $31/81=38\%$ of the total groups of the Galactic X-ray sources   
in a flux range of $(0.5-10)\times 10^{-12}$ {\flux}.
If such kinds of X-ray sources still remain unresolved in the flux range smaller than the detection limit,
they should contribute to the GRXE in some degree.

Then, we estimate the contribution of the resolved faint X-ray sources to
GRXE based on the obtained LogN-LogS relation.  
The calculation is the same as that for the 0.7--2 keV band.  
The average surface brightness within the central emission region 
of $|l|<35\arcdeg$ in the 2--10 keV band is derived to be 
$S_{\rm GRXE} = 5.2\times 10^{-11}$ ergs cm$^{-2}$ s$^{-1}$ deg$^{-2}$
from the large-scale profile in Figure \ref{fig:ldist}. 
The minimum flux of the resolved X-ray sources is $f_{\rm min}=3.2$
cnts GIS$^{-1}$ ks$^{-1}$.  The maximum flux of the integration is
adopted as $f_{\rm max}=91$ cnts GIS$^{-1}$ ks$^{-1}$ corresponding
to $10^{-11}$ {\flux}, the lower limit of the X-ray sources excluded 
from the faint source analysis.  
The integrated flux of the resolved X-ray sources is
calculated by equation (\ref{equ:lnls_cont}), which becomes $4.9\times
10^{-12}$ ergs cm$^{-2}$ s$^{-1}$ deg$^{-2}$.  Therefore, the
contribution of the resolved faint X-ray sources to the GRXE is
$S_{\rm p}/S_{\rm GRXE} =9.4\%$.
Moreover, even if we extrapolate the LogN-LogS relation 
in the flux range lower than the detection limit from the power-law relation of the resolved sources,
the integrated flux increases at most by a factor of 1.3 
because of the flat slope of the relation.
Thus, we can explain only approximately 10\% of the GRXE in the hard band
by accumulation of discrete X-ray sources.
The main origin of the GRXE is still unresolved.

\section{Conclusion}
\label{sec:conclusion}

We analyzed all the data of the {\asca} Galactic plane survey 
covering the area of $|l|\lesssim 45^\circ$ and $|b|\lesssim 0\fdg 4$ almost completely,
taking careful account of the complicated image response function. 
We derived the following observational results
about the Galactic ridge X-ray emission (GRXE) 
and the resolved X-ray sources:

\begin{enumerate}

\item  The longitude intensity profile of the GRXE in a scale of a degree
is determined in the energy bands of 0.7--2 keV, 2--4 keV, and 4--10 keV
eliminating the resolved discrete sources.
It is revealed from the profile in the energy band above 4 keV that 
the emissivity of the GRXE is strongly enhanced within the 4 kpc arm.
The combined analysis of multi-wavelength data 
such as the correlation between the X-ray and the radio emission profiles 
might give a important clue to solve the unidentified origin of the GRXE.

\item 
We resolved 163 X-ray sources by image analysis 
and investigated these identifications by catalog searches and spectral analyses. 
In order to overcome too poor photon counts of each faint source,
we classified the sources into some groups due to the fluxes, the spectral indices
and the absorption column densities,
and analyzed the spectra summed within each group.
We found the groups possibly identified with
Cataclysmic Variables, high-mass X-ray binaries, quiescent low-mass X-ray binaries, 
and Crab-like X-ray pulsars which have not ever been recognized 
because of the low flux and the large absorption column density.
However, since the observation condition is not enough to clarify
the properties of each source individually, 
confident evidences such as periodic variations 
have not been obtained.
Thus, we highly recommend following observations 
to confirm those facts by next generation X-ray satellites with a high sensitivity
such as XMM-Newton.

\item  
The LogN-LogS relations of the Galactic X-ray sources 
are obtained in the energy bands below and above 2 keV 
as the results of the source survey.
In the band below 2 keV, 
the LogN-LogS relation has a flat slope with a power-law index 
larger than $-1$
and shows a complex profile far from a single power-law relation.
The relation is considered to reflect the complex mixtures 
of various X-ray sources located near the Solar system, 
taking account of the various spectra of the resolved X-ray sources 
and the visibility limit below 1 kpc in the 0.7--2 keV band.
On the other hand, in the energy band above 2 keV,
the LogN-LogS relation was obtained for the first time 
in the flux down to $10^{-12.5}$ {\flux}.
It is well-approximated by a single power-law
with a slope of $-0.79\pm 0.07$, which is also flatter than a slope of $-1$.
Taking account of the good visibility in the energy band above 2 keV and 
the low-latitude area coverage of the ASCA survey,
it implies that the spatial source distribution has a 1-dimensional arm structure
in which the Solar system is included.

\item 
The integrated fluxes of resolved X-ray sources
contribute to the surface brightness of GRXE only by $\sim$10\%
in both energy bands below and above 2 keV.
The flux of the unresolved GRXE in the below 2 keV 
is considered to come from hot bubbles near the Solar system.
However, as for the origin of the GRXE in the energy band above 2 keV, 
no proper candidate can be recognized in this {\asca} survey.
This problem is left to future missions.
As to resolving discrete sources, 
much more progress is expected from Chandra
with a spatial resolution higher by three orders of magnitude 
than that of the ASCA.

\end{enumerate}

\acknowledgments

We thank all the members 
of the {\asca} Galactic plane survey team for their support in observation planning, 
satellite operation, and data acquisition
and the {\asca}\_ANL software developers for supporting this good analysis platform.
We would like to thank the referee for
his help in making this paper much more complete and clear.
We are also grateful to Damian Audley
for his careful review of the manuscript.
This research has made use of the HEASARC EOSCAT {\einstein} IPC Source List
provided by NASA/GSFC and the SIMBAD database operated at CDS, Strasboutg, France.


\appendix

\section{Coordinate system}\label{app:galxy}

The coordinate system employed in the standard data processing
cannot be applied to the analysis of multi-pointing observations,
because of large distortion when the image is larger than 
one GIS FOV.
Also, the 'SKYXY' coordinates adopted in the analysis of the LSS (Large
Sky Survey, e.g., Ishisaki 1996, Ueda et al. 1999) are not appropriate for the analysis 
of the Galactic plane survey.
Because it contains the tangential projection of the spherical image,
the discrepancies of the celestial areas per pixel on the projected plane 
become severe for the long area map over $\sim 10^\circ$. 
Thus, a new coordinate system, namely 'GALXY', is utilized
to handle all the data of the Galactic plane survey.
'GALXY' is a coordinate system based on the Galactic coordinate system
adopting a Cartesian projection defined in FITS format (e.g., Harten et al. 1988).
In this coordinate,
the celestial area per pixel is kept constant on the Galactic plane.
The size of the pixel agrees with that of the other coordinates at the center of the focal plane.
The origin is assigned to the Galactic center.
The value of (GALX, GALY) is easily calculated from the Galactic coordinates by 
the equation:
\begin{equation}
GALX=-l/d_{\rm pixel} ~~~~~~ GALY=b/d_{\rm pixel}, 
\end{equation}
where $d_{\rm pixel}$ is the pixel size of the DETXY coordinates at the detector center;
$ d_{\rm pixel}= 0.2455'$ for the GIS. 
The values of the GALXY coordinates are labeled in Figure \ref{fig:obscover} throughout all the 
observed fields of the Galactic plane survey.

\section{Response functions for extended uniform surface emission}\label{app:cxbmodel}

In the case where the emission region is 
uniformly extended over more than the FOV of the GIS, 
45\% of all the detected photons in the 0.7--2 keV band and 30\% in the 2--10 keV band 
are stray light coming from outside the FOV \cite{Ishisaki1996}.  
Although the probability that a photon is reflected as stray light is very low,
we have a large amount of stray light, as mentioned above,
because the solid angle of the sky which can contribute is large. 

We derive an image response to a uniform emission from observation.
For that purpose, we can use CXB (cosmic X-ray background) images
obtained from large blank sky observations \cite{Ueda1999}.
However, since the counting statistics of the CXB image are still not good enough
even if we sum up all blank sky observations (8 cnts pixel$^{-1}$)
and since some CXB data contain spatial structure due to discrete sources,
we decide to utilize day-earth images.
The day-earth is the bright earth surface exposed to sunlight
and can be regarded as having uniform surface brightness. 
The day-earth data accumulated from various observations with 4,120-ks exposure time 
have very high statistics of about 50 cnts pixel$^{-1}$, 
therefore their statistical error is very small.
The spectrum of the day-earth is extremely soft,
so the energy dependence of the response needs to be corrected for.
Since the energy dependence of the response function 
approximately depends only on the off-axis angle, 
the difference of the energy spectrum is corrected by the radial profile.
The radial profile of the CXB image averaged over azimuthal angle 
has sufficient statistics.
The detailed procedure to construct the template CXB model
is described by Ishisaki (1996).

The CXB image models for the uniform emission are constructed 
in the three energy bands of 0.7--2 keV (soft), 2--7 keV (hard) and 0.7--7 keV (total).
Then, the CXB image model of the appropriate energy band is adopted in the energy-sorted analysis.
The image response depends on the energy spectrum;
however, the differences of the integrated image for uniform surface emission are expected to be small
within each energy band.
The differences of the 0.7--7 keV image from both the image of the 0.7--2 keV and the 2--7 keV   
are less than 5\% in RMS.

\clearpage

\clearpage

\begin{table}[p]
\begin{center}
\caption{Conversion factors$^a$ from the GIS count rate to energy flux 
for typical power-law spectra with interstellar absorption.}
\label{tab:fluxconv}
\begin{tabular}{l@{~~~~~}ccc@{~~~~~}ccc@{~~~~~}ccc}
\tableline
\tableline
 Energy band                     & \multicolumn{3}{c@{\hspace{1cm}}}{0.7--2 keV} & \multicolumn{3}{c@{\hspace{1cm}}}{2--10 keV} & \multicolumn{3}{c}{0.7--7 keV} \\
\hline
$N_{\rm H}{}^b$ $\backslash$ $\Gamma$ & 1.0 & 2.0 & 4.0 &  1.0 & 2.0 & 4.0 &  1.0 & 2.0 & 4.0 \\
\hline
0.0  &0.430 & 0.499 & 0.729 & 1.430 & 1.021 & 0.638 & 0.776 & 0.639 & 0.716\\
1.0  &0.334 & 0.352 & 0.394 & 1.521 & 1.096 & 0.682 & 0.875 & 0.680 & 0.476\\
5.0  &0.235 & 0.277 & 0.312 & 1.825 & 1.368 & 0.857 & 1.126 & 0.971 & 0.692\\
\hline
\end{tabular}
\tablenotetext{a}{Units of conversion factors are [$10^{-13} {\rm erg\, cm^{-2} s^{-1}/(cnts\, GIS^{-1} ks^{-1})}$].}
\tablenotetext{b}{Units of absorption column densities are [$10^{22} {\rm cm^{-2}}$].}
\end{center}
\end{table}

\begin{table}[p]
\renewcommand{\baselinestretch}{1}
\scriptsize
\tiny
\begin{center}
\caption{Source list of the ASCA Galactic plane survey.}
\label{tab:srclist}
\begin{tabular}{cccccccl}
\hline
\hline
ASCA name$^a$      & \multicolumn{2}{c}{Position$^b$[$^\circ$]} & \multicolumn{2}{c}{Flux/Significance$^c$} & \multicolumn{2}{c}{PL parameters$^d$ } & Identification$^e$\\
               & $l$ & $\alpha_{2000}$ &  0.7--2 keV & 0.7--7 keV & $\Gamma$    & Flux       & \\
               & $b$ & $\delta_{2000}$ &   2--10 keV &            & $N_{\rm H}$ & $\chi_\nu$ & \\
\hline
AX J143148$-$6021 &  $-$44.958 &    217.954 &    ---/--- &   10.5/4.2 &                        &      & \\
                  &      0.137 &  $-$60.353 &    9.5/4.1 &            &                        &      & \\
AX J143416$-$6024 &  $-$44.696 &    218.570 & 116.0/26.8 & 168.0/31.8 & $2.56_{-}^{-}$ & 3.78 & 1RXSJ143415.0$-$602438\\
                  &   $-$0.027 &  $-$60.404 &  53.8/17.5 &            & $0.00_{-}^{-}$ & 2.76 & SIM: HD127535 -- Variable of RS CVn type(K1IV/Ve)   \\
AX J144042$-$6001 &  $-$43.812 &    220.177 &  15.4/15.2 &  20.2/16.1 & $3.40_{-0.24}^{+0.27}$ & 0.22 & 1RXSJ144037.9$-$600138\\
                  &      0.006 &  $-$60.023 &    4.5/6.1 &            & $0.00_{-0.00}^{+0.05}$ & 1.46 & G316.3$-$0.0, SIM: HD128696 -- Star(G5) \\
AX J144519$-$5949 &  $-$43.207 &    221.330 &    0.6/1.1 &  12.4/10.3 & $2.67_{-0.49}^{+0.54}$ & 0.90 & \\
                  &   $-$0.059 &  $-$59.830 &  10.5/10.0 &            & $3.06_{-0.93}^{+1.21}$ & 0.35 & \\
AX J144547$-$5931 &  $-$43.026 &    221.446 &    6.1/4.0 &   15.6/6.3 & $7.13_{-3.71}^{+1.74}$ & 0.34 & \\
                  &      0.186 &  $-$59.532 &    9.7/4.8 &            & $8.12_{-2.30}^{+2.89}$ & 0.89 & \\
AX J144701$-$5919 &  $-$42.797 &    221.757 &    8.2/4.5 &   19.7/7.3 & $4.04_{-1.65}^{+5.20}$ & 0.54 & \\
                  &      0.301 &  $-$59.330 &   12.9/5.8 &            & $2.69_{-1.45}^{+3.98}$ & 0.89 &    \\
AX J145605$-$5913 &  $-$41.717 &    224.022 &    ---/--- &  34.3/17.3 & $0.96_{-}^{-}$ & 5.61 & 1RXSJ145540.4$-$591320\\
                  &   $-$0.125 &  $-$59.230 &  31.6/17.5 &            & $4.57_{-}^{-}$ & 3.14 &    \\
AX J145732$-$5901 &  $-$41.455 &    224.386 &    ---/--- &    6.0/5.5 & $4.93_{-3.36}^{+3.21}$ & 0.69 & \\
                  &   $-$0.027 &  $-$59.022 &    4.8/4.7 &            & $19.44_{-17.03}^{+19.23}$ & 1.89 &    \\
AX J150436$-$5824 &  $-$40.358 &    226.152 &    ---/--- &    6.3/5.3 & $1.44_{-0.62}^{+0.74}$ & 0.58 & \\
                  &      0.072 &  $-$58.412 &    5.7/5.9 &            & $1.29_{-0.81}^{+1.30}$ & 1.25 &    \\
AX J151005$-$5824 &  $-$39.736 &    227.524 &    ---/--- &    7.4/2.4 &                        &      & \\
                  &   $-$0.288 &  $-$58.415 &    8.5/4.8 &            &                        &      & \\
AX J153751$-$5556 &  $-$35.234 &    234.466 &    ---/--- &    8.5/4.9 &                        &      & \\
                  &   $-$0.288 &  $-$55.938 &    7.4/4.2 &            &                        &      & \\
AX J153818$-$5541 &  $-$35.038 &    234.576 &    4.5/4.6 & 124.8/38.9 & $1.96_{-0.11}^{+0.10}$ & 21.92 & \\
                  &   $-$0.125 &  $-$55.690 & 130.8/40.6 &            & $6.95_{-0.80}^{+1.66}$ & 0.91 &    \\
AX J153947$-$5532 &  $-$34.776 &    234.949 &    1.8/3.4 &    3.9/4.6 &                        &      & \\
                  &   $-$0.125 &  $-$55.534 &    ---/--- &            &                        &      & \\
AX J154122$-$5522 &  $-$34.498 &    235.342 &    4.2/4.3 &    ---/--- &                        &      & \\
                  &   $-$0.125 &  $-$55.367 &    ---/--- &            &                        &      & \\
AX J154233$-$5519 &  $-$34.334 &    235.641 &    ---/--- &    5.5/4.2 &                        &      & \\
                  &   $-$0.190 &  $-$55.321 &    4.3/3.9 &            &                        &      & \\
AX J154557$-$5443 &  $-$33.581 &    236.491 &    ---/--- &    ---/--- &                        &      & \\
                  &   $-$0.010 &  $-$54.719 &    3.1/4.2 &            &                        &      & \\
AX J154905$-$5420 &  $-$32.992 &    237.272 &    ---/--- &    5.1/5.0 &                        &      & \\
                  &      0.006 &  $-$54.342 &    4.1/4.6 &            &                        &      & \\
AX J154951$-$5416 &  $-$32.861 &    237.465 &    3.7/5.2 &  13.8/10.7 & $0.92_{-0.30}^{+0.34}$ & 1.51 & 2E1545.9$-$5407\\
                  &   $-$0.010 &  $-$54.273 &  11.4/10.0 &            & $0.29_{-0.29}^{+0.48}$ & 0.70 &    \\
AX J155035$-$5408 &  $-$32.697 &    237.648 &    2.0/3.0 &    5.1/5.0 & $2.17_{-0.81}^{+1.21}$ & 0.24 & \\
                  &      0.022 &  $-$54.145 &    3.3/3.9 &            & $0.55_{-0.55}^{+1.29}$ & 0.73 &    \\
AX J155052$-$5418 &  $-$32.763 &    237.719 &    9.4/9.3 &  35.3/19.1 & $3.81_{-0.37}^{+0.42}$ & 2.11 & 2E1547.0$-$5408(9)\\
                  &   $-$0.125 &  $-$54.301 &  25.9/16.6 &            & $4.18_{-0.59}^{+0.69}$ & 0.80 &    \\
AX J155331$-$5347 &  $-$32.140 &    238.382 &    2.5/3.6 &    6.8/5.9 & $2.04_{-0.43}^{+1.26}$ & 0.27 & \\
                  &      0.022 &  $-$53.794 &    3.9/4.2 &            & $0.13_{-0.13}^{+1.11}$ & 0.80 &    \\
AX J155644$-$5325 &  $-$31.535 &    239.185 &    3.8/5.5 &    5.0/5.2 & $5.47_{-2.63}^{+2.03}$ & 0.06 & 1RXSJ155644.7$-$532441\\
                  &      0.006 &  $-$53.419 &    ---/--- &            & $1.68_{-1.56}^{+3.37}$ & 1.05 & 2E1552.8$-$5316   \\
\hline
\end{tabular}
\tablenotetext{a}{The mark $^{*}$ on a source name 
denotes reduced accuracy because the source location is near the edge of the field of view.}
\tablenotetext{b}{Uncertainty of the position coordinates is $\sim 1\farcm 0 =0\fdg 016$ (Ueda et al. 1999)}
\tablenotetext{c}{Flux obtained from image fitting analysis are shown 
in a unit of [cnts ks$^{-1}$ GIS$^{-1}$], a count-rate observed at the GIS nominal position, for each energy band. 
Dash lines ('---') mean no source candidates in the energy band.}
\tablenotetext{d}{Best fit parameters of a power-law model 
(photon index $\Gamma$, absorption column density $N_{\rm H}$ [10$^{22}$ cm$^{-2}$], 
flux [10$^{-12}$ ergs cm$^{-2}$ s$^{-1}$] in 0.7--10 keV and reduced chi-squared $\chi_\nu$ for 10 degrees of freedom)
are shown for the sources with the significance above 5$\sigma$.
All errors represent 90\% confidence limits. Dash lines ('$-$') in the errors mean that the best-fit 
power-law model is unacceptable within a 97\% confidence limit ($\chi_\nu<2$).}
\tablenotetext{e}{Identified catalog sources within the position detemination accuracy of 1' 
in the {\it Einstein} IPC X-ray Source Catalog (2E), the ROSAT All-Sky Survey Bright Source Catalog (1RXS), the Green Galactic SNR Catalogue (G), and catalogues of optical stars in SIMBAD database (SIM:).}
\end{center}
\end{table}

\begin{table}
\renewcommand{\baselinestretch}{1}
\tiny
\addtocounter{table}{-1}
\begin{center}
\caption{--- Continued.}
\begin{tabular}{cccccccl}
\hline
\hline
ASCA name$^a$      & \multicolumn{2}{c}{Position$^b$[$^\circ$]} & \multicolumn{2}{c}{Flux/Significance$^c$} & \multicolumn{2}{c}{PL parameters$^d$ } & Identification$^e$\\
               & $l$ & $\alpha_{2000}$ &  0.7--2 keV & 0.7--7 keV & $\Gamma$    & Flux       & \\
               & $b$ & $\delta_{2000}$ &   2--10 keV &            & $N_{\rm H}$ & $\chi_\nu$ & \\
\hline
AX J155831$-$5334 &  $-$31.436 &    239.631 &    5.9/5.5 &   13.4/7.7 & $2.00_{-0.69}^{+0.93}$ & 1.11 & SIM: GSC08697-01180 -- Star\\
                  &   $-$0.288 &  $-$53.581 &    8.2/5.8 &            & $0.83_{-0.70}^{+1.12}$ & 1.41 & \\
AX J161929$-$4945 &  $-$26.460 &    244.871 &    ---/--- &  125.7/8.2 &                        &      & SIM: HD146628 -- Star(B1/B2Ia)\\
                  &      0.334 &  $-$49.758 &  136.9/8.8 &            &                        &      & \\
AX J162011$-$5002 &  $-$26.575 &    245.048 &    ---/--- &    7.9/5.0 & $2.51_{-}^{-}$ & 0.23 & \\
                  &      0.055 &  $-$50.036 &    5.8/5.0 &            & $0.00_{-}^{-}$ & 2.58 &    \\
AX J162046$-$4942 &  $-$26.280 &    245.194 &    5.3/4.6 &    9.4/5.8 & $7.15_{-1.42}^{+0.82}$ & 0.27 & SIM: EM* VRMF 1 -- Emission-line Star\\
                  &      0.219 &  $-$49.713 &    4.1/3.6 &            & $4.40_{-}^{-}$ & 0.91 &    \\
AX J162125$-$4933 &  $-$26.100 &    245.356 &    7.2/3.4 &   19.1/6.1 & $10.00_{-3.32}^{+0.00}$ & 0.02 & \\
                  &      0.252 &  $-$49.562 &    ---/--- &            & $2.18_{-1.99}^{+0.43}$ & 1.69 &    \\
AX J162138$-$4934 &  $-$26.084 &    245.409 &    ---/--- &    ---/--- &                        &      & SIM: CD-49 10572 -- Star in double system(B) \\
                  &      0.219 &  $-$49.574 &    9.5/4.2 &            &                        &      & \\
AX J162155$-$4939 &  $-$26.116 &    245.480 &    ---/--- &    9.0/4.9 &                        &      & SIM: HD147070 -- Star(K3III)\\
                  &      0.121 &  $-$49.667 &    5.5/4.0 &            &                        &      & \\
AX J162208$-$5005 &  $-$26.395 &    245.535 &    ---/--- &   15.0/7.3 & $4.27_{-1.56}^{+2.31}$ & 0.79 & \\
                  &   $-$0.207 &  $-$50.095 &   13.7/6.5 &            & $6.18_{-3.04}^{+4.57}$ & 0.96 &    \\
AX J162246$-$4946 &  $-$26.100 &    245.696 &    ---/--- &    9.3/4.9 &                        &      & \\
                  &   $-$0.059 &  $-$49.782 &    6.1/4.1 &            &                        &      & \\
AX J163159$-$4752 &  $-$23.677 &    248.000 &    ---/--- & 124.6/37.3 & $0.24_{-0.16}^{+0.18}$ & 48.52 & \\
                  &      0.170 &  $-$47.879 & 149.2/42.1 &            & $8.77_{-1.10}^{+1.21}$ & 1.31 &    \\
AX J163252$-$4746 &  $-$23.497 &    248.218 &    4.5/4.9 &  41.6/18.9 & $2.64_{-0.64}^{+0.76}$ & 5.05 & 2E1629.1$-$4738\\
                  &      0.137 &  $-$47.769 &  43.8/19.2 &            & $6.36_{-1.73}^{+2.11}$ & 1.61 &    \\
AX J163351$-$4807 &  $-$23.644 &    248.463 &  43.5/17.0 &  60.1/18.3 & $2.97_{-0.25}^{+0.40}$ & 1.12 & 1RXSJ163352.2$-$480643, 2E1630.1$-$4800   \\
                  &   $-$0.223 &  $-$48.122 &   18.4/8.8 &            & $0.07_{-0.07}^{+0.19}$ & 1.19 & SIM: HD148937 -- Variable Star(O+)\\
AX J163524$-$4728 &  $-$22.990 &    248.854 &    ---/--- &    5.8/4.2 &                        &      & \\
                  &      0.022 &  $-$47.474 &    5.4/4.5 &            &                        &      & \\
AX J163555$-$4719 &  $-$22.826 &    248.981 &    ---/--- &    8.8/6.7 & $2.80_{-1.02}^{+1.36}$ & 1.21 & \\
                  &      0.055 &  $-$47.331 &    9.5/7.6 &            & $15.29_{-5.88}^{+8.13}$ & 0.91 &    \\
AX J163751$-$4656 &  $-$22.318 &    249.464 &  14.2/13.3 &  35.3/20.5 & $1.69_{-0.18}^{+0.19}$ & 2.02 & \\
                  &      0.072 &  $-$46.944 &  21.2/15.7 &            & $0.20_{-0.16}^{+0.20}$ & 0.99 &    \\
AX J163904$-$4642 &  $-$22.007 &    249.768 &    ---/--- &  38.9/18.2 & $-0.01_{-0.60}^{+0.66}$ & 19.18 & \\
                  &      0.072 &  $-$46.713 &  47.8/21.7 &            & $12.82_{-6.88}^{+8.58}$ & 0.86 &    \\
AX J164042$-$4632 &  $-$21.696 &    250.176 &    ---/--- &   10.4/6.9 & $2.98_{-0.89}^{+1.13}$ & 1.21 & G338.3$-$0.0\\
                  &   $-$0.027 &  $-$46.545 &   10.1/7.3 &            & $9.63_{-3.30}^{+4.72}$ & 0.60 &    \\
AX J165105$-$4403 &  $-$18.619 &    252.774 &    5.8/5.4 &   13.3/7.5 & $1.75_{-0.60}^{+0.75}$ & 1.45 & \\
                  &      0.203 &  $-$44.058 &    9.2/6.1 &            & $1.65_{-0.60}^{+1.66}$ & 1.92 &    \\
AX J165217$-$4414 &  $-$18.619 &    253.073 &    3.7/4.7 &    9.0/7.6 & $1.78_{-0.62}^{+0.86}$ & 0.50 & \\
                  &   $-$0.076 &  $-$44.235 &    5.5/5.7 &            & $0.36_{-0.36}^{+0.81}$ & 0.49 &    \\
AX J165420$-$4337 &  $-$17.915 &    253.585 &  11.9/12.0 &  30.2/19.3 & $1.41_{-0.11}^{+0.19}$ & 1.59 & 1RXSJ165424.6$-$433758\\
                  &      0.022 &  $-$43.628 &  17.6/14.6 &            & $0.02_{-0.02}^{+0.19}$ & 1.54 & SIM: HD326426 -- Star(K5)   \\
AX J165437$-$4333 &  $-$17.833 &    253.657 &    6.9/9.4 &    ---/--- & $3.97_{-0.53}^{+0.62}$ & 0.06 & 1RXSJ165442.5$-$433346\\
                  &      0.022 &  $-$43.564 &    ---/--- &            & $0.00_{-0.00}^{+0.07}$ & 1.34 & SIM: HD152335 -- Double or multiple star(F7V)   \\
AX J165646$-$4239 &  $-$16.884 &    254.193 &    9.0/4.5 &    9.7/4.6 &                        &      & 1RXSJ165650.0$-$423906\\
                  &      0.284 &  $-$42.660 &    ---/--- &            &                        &      & \\
AX J165707$-$4255 &  $-$17.047 &    254.280 &   10.2/5.9 &   16.9/6.9 & $2.42_{-0.20}^{+0.34}$ & 0.90 & \\
                  &      0.072 &  $-$42.921 &    7.2/5.9 &            & $0.00_{-0.00}^{+0.17}$ & 0.38 &    \\
AX J165723$-$4321 &  $-$17.358 &    254.348 &    ---/--- &    3.5/2.0 &                        &      & \\
                  &   $-$0.239 &  $-$43.358 &    6.9/4.2 &            &                        &      & \\
AX J165901$-$4208 &  $-$16.212 &    254.758 &    ---/--- &   18.4/8.4 & $0.86_{-0.59}^{+0.70}$ & 6.40 & SIM: V* V921 Sco -- Emission-line Star(Bpe) \\
                  &      0.284 &  $-$42.135 &   20.3/9.3 &            & $7.46_{-3.09}^{+4.14}$ & 1.05 &    \\
AX J165904$-$4242 &  $-$16.654 &    254.770 &    ---/--- &   12.7/8.3 & $3.67_{-0.82}^{+1.12}$ & 1.13 & \\
                  &   $-$0.076 &  $-$42.705 &   12.4/8.6 &            & $7.84_{-2.49}^{+4.01}$ & 1.33 &    \\
AX J165922$-$4234 &  $-$16.523 &    254.846 &    ---/--- &    6.0/5.4 & $10.00_{-5.42}^{+0.00}$ & 0.20 & \\
                  &   $-$0.043 &  $-$42.582 &    5.5/5.6 &            & $21.30_{-17.13}^{+11.39}$ & 1.96 &    \\
AX J165951$-$4209 &  $-$16.131 &    254.965 &    ---/--- &    4.4/2.7 & $0.47_{-0.59}^{+0.68}$ & 4.04 & \\
                  &      0.153 &  $-$42.152 &    8.6/5.3 &            & $3.48_{-2.08}^{+3.14}$ & 0.28 &    \\
\hline
\end{tabular}
\end{center}
\end{table}

\begin{table}
\renewcommand{\baselinestretch}{1}
\tiny
\addtocounter{table}{-1}
\begin{center}
\caption{--- Continued.}
\begin{tabular}{cccccccl}
\hline
\hline
ASCA name$^a$      & \multicolumn{2}{c}{Position$^b$[$^\circ$]} & \multicolumn{2}{c}{Flux/Significance$^c$} & \multicolumn{2}{c}{PL parameters$^d$ } & Identification$^e$\\
               & $l$ & $\alpha_{2000}$ &  0.7--2 keV & 0.7--7 keV & $\Gamma$    & Flux       & \\
               & $b$ & $\delta_{2000}$ &   2--10 keV &            & $N_{\rm H}$ & $\chi_\nu$ & \\
\hline
AX J170006$-$4157 &  $-$15.950 &    255.028 &    5.8/3.0 &   26.9/7.4 & $0.35_{-}^{-}$ & 5.06 & \\
                  &      0.235 &  $-$41.959 &   35.8/8.1 &            & $0.88_{-}^{-}$ & 3.98 &    \\
AX J170017$-$4220 &  $-$16.229 &    255.075 &    ---/--- &  13.2/10.3 & $-0.34_{-0.23}^{+0.31}$ & 2.69 & \\
                  &   $-$0.027 &  $-$42.340 &  14.5/11.8 &            & $0.16_{-0.16}^{+1.06}$ & 1.70 &    \\
$^{*}$AX J170047$-$4139 &  $-$15.639 &    255.199 &   11.1/1.9 & 1427.3/54.7 &                        &      & 2E1657.2$-$4134\\
                  &      0.317 &  $-$41.664 & 1751.0/62.2 &            &                        &      & SIM: EXO 1657-419 -- High Mass X-ray Binary \\
AX J170052$-$4210 &  $-$16.032 &    255.221 &    4.6/5.2 &    ---/--- & $1.91_{-0.67}^{+1.76}$ & 0.24 & 2E1657.3$-$4204\\
                  &   $-$0.010 &  $-$42.175 &    ---/--- &            & $0.00_{-0.00}^{+10.24}$ & 0.60 &    \\
AX J170112$-$4212 &  $-$16.016 &    255.304 &    ---/--- &    ---/--- & $0.59_{-0.91}^{+0.47}$ & 5.15 & \\
                  &   $-$0.076 &  $-$42.203 &   10.3/6.9 &            & $20.23_{-13.27}^{+17.46}$ & 1.01 &    \\
AX J170349$-$4142 &  $-$15.329 &    255.956 &  61.6/23.7 & 175.9/39.9 & $7.47_{-}^{-}$ & 4.83 & G344.7$-$0.1\\
                  &   $-$0.158 &  $-$41.709 & 110.1/31.0 &            & $7.91_{-}^{-}$ & 6.57 &    \\
AX J170444$-$4109 &  $-$14.788 &    256.186 &    ---/--- &    4.4/5.6 & $1.01_{-0.46}^{+0.51}$ & 0.79 & \\
                  &      0.039 &  $-$41.160 &    4.2/6.0 &            & $1.27_{-0.79}^{+1.11}$ & 0.46 &    \\
AX J170506$-$4113 &  $-$14.805 &    256.276 &    2.8/4.9 &    3.1/4.1 &                        &      & \\
                  &   $-$0.059 &  $-$41.233 &    ---/--- &            &                        &      & \\
AX J170536$-$4038 &  $-$14.281 &    256.403 &    8.0/6.1 &   12.8/7.0 & $2.92_{-0.49}^{+0.71}$ & 0.23 & 1RXSJ170541.0$-$403850\\
                  &      0.219 &  $-$40.647 &    4.8/3.7 &            & $0.00_{-0.00}^{+1.97}$ & 1.91 &    \\
AX J170555$-$4104 &  $-$14.592 &    256.482 &    3.8/4.3 &    ---/--- &                        &      & \\
                  &   $-$0.092 &  $-$41.083 &    ---/--- &            &                        &      & \\
$^{*}$AX J171715$-$3718 &  $-$10.254 &    259.312 &    ---/--- &   32.5/4.9 &                        &      & \\
                  &      0.366 &  $-$37.312 &   41.4/6.2 &            &                        &      & \\
AX J171804$-$3726 &  $-$10.270 &    259.519 &  38.9/22.4 & 189.3/52.7 & $4.45_{-}^{-}$ & 13.32 & G349.7$+$0.2\\
                  &      0.153 &  $-$37.448 & 150.1/47.6 &            & $5.56_{-}^{-}$ & 4.13 &    \\
AX J171910$-$3652 &   $-$9.681 &    259.793 &    7.2/3.9 &   10.4/4.7 &                        &      & 1RXSJ171913.1$-$365144\\
                  &      0.301 &  $-$36.881 &    ---/--- &            &                        &      & \\
AX J171922$-$3703 &   $-$9.795 &    259.846 &    ---/--- &    4.0/3.8 & $4.95_{-}^{-}$ & 0.98 & \\
                  &      0.170 &  $-$37.050 &    5.4/5.3 &            & $76.89_{-}^{-}$ & 2.49 &    \\
AX J172050$-$3710 &   $-$9.730 &    260.212 &    4.0/4.4 &    5.8/4.3 &                        &      & \\
                  &   $-$0.141 &  $-$37.174 &    ---/--- &            &                        &      & \\
AX J172105$-$3726 &   $-$9.926 &    260.274 & 306.4/13.5 & 366.8/16.7 &                        &      & 1RXSJ172106.9$-$372639\\
                  &   $-$0.338 &  $-$37.448 & 161.6/10.5 &            &                        &      & \\
AX J172550$-$3533 &   $-$7.831 &    261.461 &    ---/--- &    5.5/6.0 & $1.04_{-}^{-}$ & 0.65 & \\
                  &   $-$0.059 &  $-$35.561 &    4.6/5.7 &            & $1.70_{-}^{-}$ & 2.21 &    \\
AX J172623$-$3516 &   $-$7.536 &    261.597 &    4.0/4.9 &    ---/--- &                        &      & SIM: HD319810 -- Star(G0)  \\
                  &      0.006 &  $-$35.280 &    ---/--- &            &                        &      & \\
AX J172642$-$3504 &   $-$7.324 &    261.676 &    ---/--- &    5.2/4.1 &                        &      & \\
                  &      0.072 &  $-$35.067 &    4.9/4.3 &            &                        &      & \\
AX J172642$-$3540 &   $-$7.831 &    261.678 &  22.7/12.3 &  35.8/14.1 & $2.15_{-0.17}^{+0.23}$ & 1.69 & 1RXSJ172641.7$-$354030\\
                  &   $-$0.272 &  $-$35.680 &   14.3/7.9 &            & $0.00_{-0.00}^{+0.12}$ & 0.88 &    \\
AX J172743$-$3506 &   $-$7.242 &    261.931 &  18.0/10.8 &  34.8/13.7 & $9.87_{-}^{-}$ & 0.80 & G352.7$-$0.1\\
                  &   $-$0.125 &  $-$35.109 &   16.7/8.8 &            & $7.92_{-}^{-}$ & 2.39 &    \\
AX J173441$-$3234 &   $-$4.328 &    263.675 &  49.0/20.1 &  58.6/20.3 & $3.52_{-0.17}^{+0.24}$ & 0.60 & 1RXSJ173442.6$-$323449, 2E1731.4$-$3232\\
                  &      0.055 &  $-$32.574 &   10.6/7.0 &            & $0.00_{-0.00}^{+0.07}$ & 1.00 & SIM: NGC 6383 -- Cluster of Stars \\
AX J173518$-$3237 &   $-$4.295 &    263.826 &    ---/--- &   11.3/6.4 & $6.07_{-1.71}^{+0.62}$ & 0.57 & G355.6$-$0.0\\
                  &   $-$0.076 &  $-$32.617 &    3.9/2.9 &            & $6.44_{-2.41}^{+3.78}$ & 0.34 &    \\
AX J173548$-$3207 &   $-$3.820 &    263.950 &    4.0/5.4 &   12.2/9.4 & $1.83_{-0.64}^{+0.85}$ & 0.69 & \\
                  &      0.104 &  $-$32.120 &    8.5/7.7 &            & $1.35_{-0.78}^{+1.09}$ & 0.63 &    \\
AX J173628$-$3141 &   $-$3.378 &    264.117 &    ---/--- &   10.3/4.7 & $3.43_{-2.34}^{+1.29}$ & 1.64 & \\
                  &      0.219 &  $-$31.686 &   10.8/5.2 &            & $36.98_{-21.92}^{+41.30}$ & 0.35 &    \\
\hline
\end{tabular}
\end{center}
\end{table}

\begin{table}
\renewcommand{\baselinestretch}{1}
\tiny
\addtocounter{table}{-1}
\begin{center}
\caption{--- Continued.}
\begin{tabular}{cccccccl}
\hline
\hline
ASCA name$^a$      & \multicolumn{2}{c}{Position$^b$[$^\circ$]} & \multicolumn{2}{c}{Flux/Significance$^c$} & \multicolumn{2}{c}{PL parameters$^d$ } & Identification$^e$\\
               & $l$ & $\alpha_{2000}$ &  0.7--2 keV & 0.7--7 keV & $\Gamma$    & Flux       & \\
               & $b$ & $\delta_{2000}$ &   2--10 keV &            & $N_{\rm H}$ & $\chi_\nu$ & \\
\hline
AX J175331$-$2538 &      3.726 &    268.381 &    7.0/2.9 &   12.6/4.7 &                        &      & \\
                  &      0.186 &  $-$25.644 &    ---/--- &            &                        &      & \\
AX J175404$-$2553 &      3.579 &    268.518 &    ---/--- &    5.6/4.6 &                        &      & \\
                  &   $-$0.043 &  $-$25.887 &    3.9/3.8 &            &                        &      & \\
AX J180225$-$2300 &      7.017 &    270.608 &    8.2/5.0 &   13.2/5.8 & $3.70_{-0.88}^{+1.26}$ & 0.44 & \\
                  &   $-$0.256 &  $-$23.016 &    4.2/2.8 &            & $0.82_{-0.50}^{+0.68}$ & 0.81 &    \\
AX J180800$-$1956 &     10.340 &    272.004 &    4.3/4.0 &    ---/--- &                        &      & \\
                  &      0.121 &  $-$19.934 &    ---/--- &            &                        &      & \\
AX J180816$-$2021 &      9.996 &    272.070 &    ---/--- &    7.1/5.8 & $2.71_{-}^{-}$ & 0.81 & \\
                  &   $-$0.141 &  $-$20.362 &    7.8/6.8 &            & $4.90_{-}^{-}$ & 2.38 &    \\
AX J180838$-$2024 &      9.996 &    272.161 &    ---/--- &  37.6/14.8 & $1.42_{-0.32}^{+0.22}$ & 6.90 & G10.0$-$0.3\\
                  &   $-$0.239 &  $-$20.410 &  39.6/16.2 &            & $3.86_{-0.99}^{+1.13}$ & 1.02 &    \\
AX J180857$-$2004 &     10.323 &    272.239 &    ---/--- &    ---/--- & $7.24_{-4.35}^{+2.76}$ & 0.00 & \\
                  &   $-$0.141 &  $-$20.076 &    7.1/5.3 &            & $19.38_{-10.95}^{+5.69}$ & 0.67 &    \\
AX J180902$-$1948 &     10.569 &    272.259 &    ---/--- &    4.3/4.8 & $1.61_{-0.59}^{+0.69}$ & 0.56 & \\
                  &   $-$0.027 &  $-$19.805 &    5.0/6.0 &            & $2.18_{-1.09}^{+1.64}$ & 1.13 &    \\
AX J180948$-$1918 &     11.093 &    272.452 &    6.9/7.2 &  17.0/11.2 & $1.82_{-0.33}^{+0.38}$ & 1.05 & SIM: HD166077 -- Star(A2/A3II/III)\\
                  &      0.055 &  $-$19.308 &   11.4/9.3 &            & $0.64_{-0.36}^{+0.45}$ & 0.47 &    \\
AX J180951$-$1943 &     10.733 &    272.465 &  18.2/12.2 &  20.6/11.2 & $9.08_{-1.73}^{+0.92}$ & 0.10 & 1RXSJ180951.5$-$194345\\
                  &   $-$0.158 &  $-$19.726 &    ---/--- &            & $2.44_{-0.74}^{+0.53}$ & 0.80 &    \\
AX J181033$-$1917 &     11.191 &    272.639 &  12.3/10.4 &  20.9/12.9 & $2.16_{-0.40}^{+0.44}$ & 0.99 & \\
                  &   $-$0.092 &  $-$19.293 &    8.3/6.8 &            & $0.22_{-0.22}^{+0.31}$ & 0.62 &    \\
AX J181116$-$1828 &     11.993 &    272.820 &    ---/--- &    ---/--- &                        &      & \\
                  &      0.153 &  $-$18.471 &    6.3/4.7 &            &                        &      & \\
AX J181120$-$1913 &     11.338 &    272.836 &    ---/--- &    ---/--- &                        &      & SIM: HD166441 -- Star(G8/K0)\\
                  &   $-$0.223 &  $-$19.227 &    7.5/4.5 &            &                        &      & \\
AX J181211$-$1835 &     11.993 &    273.048 &    ---/--- &    6.7/5.4 & $0.56_{-0.37}^{+1.71}$ & 0.89 & G12.0$-$0.1\\
                  &   $-$0.092 &  $-$18.590 &    4.5/4.1 &            & $0.00_{-0.00}^{+6.68}$ & 1.22 &    \\
AX J181213$-$1842 &     11.895 &    273.058 &    ---/--- &    8.4/4.7 &                        &      & \\
                  &   $-$0.158 &  $-$18.707 &    5.4/4.2 &            &                        &      & \\
AX J181705$-$1607 &     14.710 &    274.272 &    ---/--- &    ---/--- &                        &      & \\
                  &      0.055 &  $-$16.132 &    6.6/4.0 &            &                        &      & \\
AX J181848$-$1527 &     15.496 &    274.704 &    4.0/5.3 &    3.7/3.9 & $5.31_{-0.89}^{+2.75}$ & 0.01 & \\
                  &      0.006 &  $-$15.463 &    ---/--- &            & $0.00_{-0.00}^{+0.57}$ & 1.14 &    \\
AX J181852$-$1559 &     15.038 &    274.719 &    ---/--- &   16.8/8.2 & $2.34_{-0.80}^{+1.37}$ & 2.20 & \\
                  &   $-$0.256 &  $-$15.991 &   15.7/8.9 &            & $7.45_{-3.91}^{+7.60}$ & 1.51 &    \\
AX J181915$-$1601 &     15.054 &    274.817 &    ---/--- &   19.5/4.3 &                        &      & SIM: V* V1962 Sgr -- Variable Star\\
                  &   $-$0.354 &  $-$16.023 &   17.5/4.6 &            &                        &      & \\
AX J181917$-$1548 &     15.251 &    274.823 &   54.9/6.8 &   52.0/5.4 &                        &      & \\
                  &   $-$0.256 &  $-$15.804 &    7.5/1.4 &            &                        &      & \\
AX J182104$-$1420 &     16.740 &    275.267 &    ---/--- &  34.5/12.8 & $2.93_{-0.46}^{+0.52}$ & 3.72 & G16.7$+$0.1\\
                  &      0.055 &  $-$14.343 &  33.4/13.6 &            & $7.22_{-1.54}^{+1.95}$ & 0.88 &    \\
AX J182216$-$1425 &     16.806 &    275.567 &   20.7/8.1 &   24.0/7.3 & $10.00_{-0.43}^{+0.00}$ & 0.25 & \\
                  &   $-$0.239 &  $-$14.424 &    ---/--- &            & $4.05_{-0.35}^{+0.33}$ & 1.81 &    \\
AX J182435$-$1311 &     18.165 &    276.149 &    7.6/4.5 &   15.4/6.0 & $8.62_{-2.05}^{+1.38}$ & 0.54 & \\
                  &   $-$0.158 &  $-$13.186 &    6.3/3.3 &            & $7.72_{-2.82}^{+2.73}$ & 0.60 &    \\
AX J182442$-$1253 &     18.443 &    276.178 &    2.0/3.3 &    4.0/4.4 &                        &      & \\
                  &   $-$0.043 &  $-$12.886 &    ---/--- &            &                        &      & \\
AX J182509$-$1253 &     18.492 &    276.291 &    ---/--- &    4.5/4.3 &                        &      & \\
                  &   $-$0.141 &  $-$12.889 &    ---/--- &            &                        &      & \\
\hline
\end{tabular}
\end{center}
\end{table}

\begin{table}
\renewcommand{\baselinestretch}{1}
\tiny
\addtocounter{table}{-1}
\begin{center}
\caption{--- Continued.}
\begin{tabular}{cccccccl}
\hline
\hline
ASCA name$^a$      & \multicolumn{2}{c}{Position$^b$[$^\circ$]} & \multicolumn{2}{c}{Flux/Significance$^c$} & \multicolumn{2}{c}{PL parameters$^d$ } & Identification$^e$\\
               & $l$ & $\alpha_{2000}$ &  0.7--2 keV & 0.7--7 keV & $\Gamma$    & Flux       & \\
               & $b$ & $\delta_{2000}$ &   2--10 keV &            & $N_{\rm H}$ & $\chi_\nu$ & \\
\hline
AX J182530$-$1144 &     19.540 &    276.377 &    ---/--- &    6.5/2.9 &                        &      & \\
                  &      0.317 &  $-$11.748 &    8.1/4.0 &            &                        &      & \\
AX J182538$-$1214 &     19.114 &    276.411 &    4.0/4.8 &    7.0/5.8 & $2.65_{-}^{-}$ & 0.14 & \\
                  &      0.055 &  $-$12.247 &    2.6/2.9 &            & $0.00_{-}^{-}$ & 2.20 &    \\
AX J182651$-$1206 &     19.376 &    276.713 &    5.2/4.4 &    6.5/3.6 &                        &      & \\
                  &   $-$0.141 &  $-$12.107 &    ---/--- &            &                        &      & \\
AX J182846$-$1116 &     20.325 &    277.194 &    6.9/4.8 &    8.9/4.5 &                        &      & \\
                  &   $-$0.174 &  $-$11.281 &    ---/--- &            &                        &      & \\
AX J183039$-$1002 &     21.635 &    277.664 &    ---/--- &    7.0/5.2 & $0.04_{-0.69}^{+0.77}$ & 2.27 & \\
                  &   $-$0.010 &  $-$10.045 &    9.8/7.2 &            & $3.07_{-2.60}^{+3.42}$ & 0.49 &    \\
AX J183114$-$0943 &     21.979 &    277.811 &    2.5/4.1 &    6.8/6.4 & $1.36_{-0.35}^{+0.40}$ & 0.33 & \\
                  &      0.006 &   $-$9.733 &    2.2/2.6 &            & $0.00_{-0.00}^{+0.29}$ & 1.35 &    \\
AX J183116$-$1008 &     21.619 &    277.818 &    ---/--- &   11.5/5.2 & $2.79_{-1.80}^{+6.93}$ & 1.11 & \\
                  &   $-$0.190 &  $-$10.143 &    8.8/4.6 &            & $8.56_{-6.73}^{+42.97}$ & 0.71 &    \\
AX J183206$-$0938 &     22.159 &    278.028 &    4.4/4.1 &   17.2/6.7 & $3.67_{-0.69}^{+0.76}$ & 1.97 & \\
                  &   $-$0.141 &   $-$9.641 &   15.5/6.6 &            & $5.48_{-1.47}^{+1.87}$ & 0.63 &    \\
AX J183206$-$0940 &     22.126 &    278.027 &    ---/--- &   13.4/5.0 &                        &      & \\
                  &   $-$0.158 &   $-$9.678 &   11.7/4.8 &            &                        &      & \\
AX J183221$-$0840 &     23.043 &    278.089 &    9.6/7.2 &  60.2/20.1 & $0.54_{-0.18}^{+0.20}$ & 11.34 & \\
                  &      0.252 &   $-$8.676 &  56.8/20.0 &            & $0.98_{-0.37}^{+0.44}$ & 1.51 &    \\
AX J183231$-$0916 &     22.535 &    278.130 &    ---/--- &    3.4/4.3 &                        &      & \\
                  &   $-$0.059 &   $-$9.270 &    1.7/2.5 &            &                        &      & \\
AX J183345$-$0828 &     23.387 &    278.440 &    ---/--- &    5.2/5.1 & $2.67_{-0.96}^{+1.64}$ & 0.43 & SIM: PSR B1830-08 -- Pulsars\\
                  &      0.039 &   $-$8.469 &    4.8/5.2 &            & $7.38_{-3.96}^{+6.24}$ & 1.90 &    \\
AX J183356$-$0822 &     23.485 &    278.486 &    ---/--- &    5.6/5.8 & $2.23_{-1.12}^{+7.77}$ & 0.48 & \\
                  &      0.039 &   $-$8.382 &    5.3/6.1 &            & $5.85_{-3.87}^{+44.32}$ & 0.81 &    \\
AX J183440$-$0801 &     23.878 &    278.669 &    ---/--- &    ---/--- &                        &      & \\
                  &      0.039 &   $-$8.033 &    2.7/4.2 &            &                        &      & \\
AX J183506$-$0806 &     23.861 &    278.778 &    ---/--- &    4.0/4.0 &                        &      & 2E1832.4$-$0812\\
                  &   $-$0.092 &   $-$8.108 &    ---/--- &            &                        &      & \\
AX J183518$-$0754 &     24.058 &    278.826 &    3.7/6.3 &  49.1/29.7 & $1.63_{-0.19}^{+0.20}$ & 5.41 & \\
                  &   $-$0.043 &   $-$7.911 &  48.4/30.5 &            & $4.96_{-0.66}^{+0.74}$ & 1.23 &    \\
AX J183528$-$0737 &     24.336 &    278.867 &    ---/--- &  26.9/18.5 & $3.23_{-0.46}^{+0.51}$ & 4.30 & \\
                  &      0.055 &   $-$7.619 &  30.6/21.0 &            & $18.72_{-3.05}^{+3.59}$ & 1.74 &    \\
AX J183607$-$0756 &     24.123 &    279.032 &    7.6/4.9 &    9.9/4.7 &                        &      & \\
                  &   $-$0.239 &   $-$7.943 &    ---/--- &            &                        &      & \\
AX J183618$-$0647 &     25.171 &    279.078 &    6.6/4.1 &   13.1/5.2 & $2.15_{-0.56}^{+1.72}$ & 0.82 & 2E1833.6$-$0649\\
                  &      0.252 &   $-$6.787 &    3.9/1.8 &            & $0.10_{-0.10}^{+0.80}$ & 0.63 &    \\
AX J183800$-$0655 &     25.236 &    279.504 &    4.9/3.8 &  55.1/14.7 & $0.47_{-0.36}^{+0.41}$ & 10.88 & 2E1835.3$-$0658\\
                  &   $-$0.190 &   $-$6.932 &  59.2/15.5 &            & $2.71_{-1.20}^{+1.62}$ & 1.73 &    \\
AX J183931$-$0544 &     26.464 &    279.880 &    ---/--- &    4.8/3.9 &                        &      & \\
                  &      0.022 &   $-$5.743 &    4.4/4.8 &            &                        &      & \\
AX J183957$-$0546 &     26.480 &    279.990 &    ---/--- &    6.0/4.6 &                        &      & \\
                  &   $-$0.092 &   $-$5.781 &    ---/--- &            &                        &      & \\
AX J184004$-$0552 &     26.415 &    280.019 &    3.4/3.2 &    6.6/4.1 &                        &      & \\
                  &   $-$0.158 &   $-$5.870 &    ---/--- &            &                        &      & \\
AX J184008$-$0543 &     26.546 &    280.035 &    ---/--- &    6.5/5.1 & $0.81_{-0.31}^{+0.59}$ & 1.00 & SIM: GSC 05125-01818 -- Star(B5)\\
                  &   $-$0.108 &   $-$5.731 &    5.3/5.2 &            & $0.15_{-0.15}^{+1.17}$ & 1.28 &    \\
AX J184024$-$0544 &     26.562 &    280.101 &    ---/--- &   12.2/7.0 & $1.75_{-0.66}^{+0.78}$ & 1.65 & \\
                  &   $-$0.174 &   $-$5.746 &    9.9/6.6 &            & $2.72_{-1.60}^{+2.07}$ & 1.37 &    \\
\hline
\end{tabular}
\end{center}
\end{table}

\begin{table}
\renewcommand{\baselinestretch}{1}
\tiny
\addtocounter{table}{-1}
\begin{center}
\caption{--- Continued.}
\begin{tabular}{cccccccl}
\hline
\hline
ASCA name$^a$      & \multicolumn{2}{c}{Position$^b$[$^\circ$]} & \multicolumn{2}{c}{Flux/Significance$^c$} & \multicolumn{2}{c}{PL parameters$^d$ } & Identification$^e$\\
               & $l$ & $\alpha_{2000}$ &  0.7--2 keV & 0.7--7 keV & $\Gamma$    & Flux       & \\
               & $b$ & $\delta_{2000}$ &   2--10 keV &            & $N_{\rm H}$ & $\chi_\nu$ & \\
\hline
AX J184121$-$0455 &     27.397 &    280.339 & 496.0/96.9 & 924.5/118.5 & $4.20_{-}^{-}$ & 28.61 & 1RXSJ184120.8$-$045612\\
                  &   $-$0.010 &   $-$4.929 & 437.6/81.5 &            & $2.53_{-}^{-}$ & 15.69  & 2E1838.6$-$0459 G27.4$+$0.0  \\
AX J184355$-$0351 &     28.641 &    280.982 &    ---/--- &    6.9/4.6 &                        &      & \\
                  &   $-$0.092 &   $-$3.860 &   40.5/3.6 &            &                        &      & \\
AX J184400$-$0355 &     28.592 &    281.004 &    ---/--- &    ---/--- &                        &      & \\
                  &   $-$0.141 &   $-$3.926 &    5.2/4.2 &            &                        &      & \\
AX J184447$-$0305 &     29.427 &    281.196 &    1.7/2.5 &    7.5/5.9 & $2.00_{-0.99}^{+1.25}$ & 0.98 & \\
                  &      0.072 &   $-$3.086 &    7.2/6.2 &            & $6.74_{-3.57}^{+5.48}$ & 1.22 &    \\
AX J184600$-$0231 &     30.066 &    281.502 &    ---/--- &    4.0/4.5 &                        &      & \\
                  &      0.055 &   $-$2.526 &    3.6/4.6 &            &                        &      & \\
AX J184610$-$0239 &     29.967 &    281.545 &    ---/--- &    3.6/4.1 &                        &      & \\
                  &   $-$0.043 &   $-$2.658 &    4.1/4.6 &            &                        &      & \\
AX J184629$-$0258 &     29.722 &    281.622 &  67.3/11.5 & 308.3/26.0 & $2.01_{-0.18}^{+0.20}$ & 27.23 & 1RXSJ184624.7$-$025854\\
                  &   $-$0.256 &   $-$2.973 & 254.8/23.8 &            & $2.15_{-0.30}^{+0.32}$ & 1.04 & 2E1843.8$-$0301 G29.7$-$0.3  \\
AX J184652$-$0240 &     30.033 &    281.721 &    3.4/4.2 &    7.8/5.9 & $4.85_{-0.61}^{+1.90}$ & 0.36 & 2E1844.1$-$0244\\
                  &   $-$0.207 &   $-$2.674 &    4.1/3.9 &            & $3.24_{-1.02}^{+1.83}$ & 0.64 &    \\
AX J184738$-$0156 &     30.770 &    281.911 &    ---/--- &   15.2/9.3 & $2.54_{-0.97}^{+1.26}$ & 2.03 & \\
                  &   $-$0.043 &   $-$1.944 &   14.3/9.6 &            & $6.72_{-2.96}^{+4.19}$ & 0.96 &    \\
AX J184741$-$0219 &     30.442 &    281.922 &    ---/--- &    6.9/4.6 &                        &      & \\
                  &   $-$0.223 &   $-$2.317 &    6.5/4.8 &            &                        &      & \\
AX J184848$-$0129 &     31.310 &    282.202 &    ---/--- &  25.0/10.8 & $3.16_{-0.71}^{+0.96}$ & 1.62 & \\
                  &   $-$0.092 &   $-$1.486 &  21.3/10.1 &            & $5.64_{-1.50}^{+1.97}$ & 1.31 &    \\
AX J184930$-$0055 &     31.883 &    282.375 & 104.9/35.4 & 146.8/40.1 & $5.44_{-}^{-}$ & 2.34 & 1RXSJ184931.1$-$005625\\
                  &      0.006 &   $-$0.931 &  42.6/20.2 &            & $2.57_{-}^{-}$ & 2.08 & G31.9$+$0.0   \\
AX J185015$-$0025 &     32.423 &    282.564 &    ---/--- &    5.4/4.0 & $3.88_{-0.92}^{+1.30}$ & 0.66 & \\
                  &      0.072 &   $-$0.420 &    8.0/6.1 &            & $9.92_{-3.09}^{+7.10}$ & 1.41 &    \\
AX J185240$+$0038 &     33.651 &    283.167 & 191.6/57.3 & 231.7/60.3 & $5.44_{-}^{-}$ & 2.40 & 1RXSJ185238.8$+$003949\\
                  &      0.022 &      0.650 &  44.9/23.8 &            & $1.54_{-}^{-}$ & 3.49 & 2E1850.0$+$0036 G33.6$+$0.1  \\
$^{*}$AX J185551$+$0129 &     34.764 &    283.966 & 1704.9/35.2 & 1955.1/25.9 &                        &      & 1RXSJ185602.3$+$011854\\
                  &   $-$0.305 &      1.491 &  426.0/6.2 &            &                        &      & 2E1853.0$+$0114(7,9)\\
AX J185608$+$0218 &     35.517 &    284.033 &    7.0/8.7 &  11.9/10.3 & $2.17_{-0.26}^{+0.31}$ & 0.40 & 1RXSJ185609.2$+$021744\\
                  &      0.006 &      2.303 &    5.4/6.4 &            & $0.00_{-0.00}^{+0.05}$ & 1.81 & SIM: GSC 00453-00136 -- Star\\
AX J185643$+$0220 &     35.615 &    284.180 &    ---/--- &    ---/--- &                        &      & \\
                  &   $-$0.108 &      2.338 &    4.7/4.2 &            &                        &      & \\
AX J185651$+$0245 &     36.008 &    284.214 &    ---/--- &    2.5/3.8 &                        &      & \\
                  &      0.055 &      2.763 &    2.6/4.3 &            &                        &      & \\
AX J185721$+$0247 &     36.090 &    284.339 &    ---/--- &    3.0/4.2 &                        &      & \\
                  &   $-$0.043 &      2.791 &    ---/--- &            &                        &      & \\
AX J185750$+$0240 &     36.041 &    284.462 &    ---/--- &    6.1/5.5 & $1.42_{-0.61}^{+0.75}$ & 0.99 & \\
                  &   $-$0.207 &      2.672 &    5.1/5.1 &            & $1.98_{-1.06}^{+1.43}$ & 1.13 &    \\
AX J185905$+$0333 &     36.974 &    284.772 &    ---/--- &   10.5/8.8 & $2.45_{-0.57}^{+0.31}$ & 1.51 & \\
                  &   $-$0.076 &      3.562 &   10.8/9.3 &            & $7.51_{-2.27}^{+3.11}$ & 0.72 &    \\
AX J190007$+$0427 &     37.890 &    285.032 &    ---/--- &    3.7/4.0 &                        &      & \\
                  &      0.104 &      4.460 &    2.9/3.3 &            &                        &      & \\
AX J190144$+$0459 &     38.545 &    285.435 &    ---/--- &    3.8/5.5 & $3.47_{-}^{-}$ & 0.32 & \\
                  &   $-$0.010 &      4.989 &    3.6/6.0 &            & $12.62_{-}^{-}$ & 2.28 &    \\
AX J190534$+$0659 &     40.755 &    286.395 &    ---/--- &    5.1/5.2 & $2.18_{-0.67}^{+0.92}$ & 0.43 & \\
                  &      0.055 &      6.984 &    3.4/4.2 &            & $0.55_{-0.50}^{+0.70}$ & 1.13 &    \\
AX J190734$+$0709 &     41.132 &    286.892 & 355.2/37.3 & 505.8/43.7 & $4.48_{-}^{-}$ & 11.39 & 1RXSJ190732.6$+$070817\\
                  &   $-$0.305 &      7.152 & 191.0/21.6 &            & $1.68_{-}^{-}$ & 8.16 & 2E1905.0$+$0704 G41.1$-$0.3  \\
\hline
\end{tabular}
\end{center}
\end{table}

\begin{table}
\renewcommand{\baselinestretch}{1}
\tiny
\addtocounter{table}{-1}
\begin{center}
\caption{--- Continued.}
\begin{tabular}{cccccccl}
\hline
\hline
ASCA name$^a$      & \multicolumn{2}{c}{Position$^b$[$^\circ$]} & \multicolumn{2}{c}{Flux/Significance$^c$} & \multicolumn{2}{c}{PL parameters$^d$ } & Identification$^e$\\
               & $l$ & $\alpha_{2000}$ &  0.7--2 keV & 0.7--7 keV & $\Gamma$    & Flux       & \\
               & $b$ & $\delta_{2000}$ &   2--10 keV &            & $N_{\rm H}$ & $\chi_\nu$ & \\
\hline
AX J190749$+$0803 &     41.967 &    286.957 &    ---/--- &    6.7/6.8 & $1.67_{-0.68}^{+1.17}$ & 0.90 & \\
                  &      0.055 &      8.059 &    7.4/7.7 &            & $6.27_{-3.10}^{+8.89}$ & 0.89 &    \\
AX J190814$+$0832 &     42.441 &    287.060 &    3.1/3.6 &    6.8/5.3 & $2.79_{-0.91}^{+2.82}$ & 0.15 & \\
                  &      0.186 &      8.541 &    3.1/3.1 &            & $0.35_{-0.35}^{+1.51}$ & 0.65 &    \\
AX J190818$+$0745 &     41.754 &    287.078 &    6.3/4.5 &   12.9/5.4 & $0.87_{-0.42}^{+0.55}$ & 1.61 & \\
                  &   $-$0.190 &      7.757 &    7.5/3.9 &            & $0.00_{-0.00}^{+0.42}$ & 0.79 &    \\
AX J191046$+$0917 &     43.391 &    287.695 &    ---/--- &  14.6/11.5 & $1.11_{-0.39}^{+0.46}$ & 2.43 & 2E1908.3$+$0911\\
                  &   $-$0.027 &      9.285 &  14.9/12.2 &            & $2.63_{-1.03}^{+1.37}$ & 0.72 &    \\
AX J191105$+$0906 &     43.276 &    287.773 & 178.9/38.4 & 756.0/75.3 & $3.60_{-}^{-}$ & 51.00 & 1RXSJ191107.4$+$090623\\
                  &   $-$0.174 &      9.115 & 604.0/66.3 &            & $4.49_{-}^{-}$ & 15.64 & 2E1908.7$+$0901 G43.3$-$0.2  \\
AX J194152$+$2251 &     58.877 &    295.471 &    ---/--- &    9.2/4.3 &                        &      & \\
                  &   $-$0.108 &     22.862 &    7.3/3.9 &            &                        &      & \\
AX J194310$+$2318 &     59.417 &    295.793 &    9.4/5.7 &   17.6/7.5 & $2.90_{-0.75}^{+1.01}$ & 0.42 & SIM: NGC 6823 -- Star in Cluster\\
                  &   $-$0.141 &     23.314 &    8.8/4.9 &            & $0.43_{-0.34}^{+0.62}$ & 0.71 &    \\
AX J194332$+$2323 &     59.532 &    295.886 &    4.7/4.6 &    6.3/4.5 &                        &      & SIM: NGC 6823 -- Star in Cluster\\
                  &   $-$0.174 &     23.398 &    ---/--- &            &                        &      & \\
AX J194622$+$2436 &     60.907 &    296.593 &    7.7/6.8 &   15.5/9.2 & $2.22_{-0.41}^{+0.49}$ & 0.58 & SIM: BWE 1944+2427 -- Star\\
                  &   $-$0.125 &     24.613 &    7.5/6.2 &            & $0.48_{-0.36}^{+0.50}$ & 1.65 &    \\
AX J194649$+$2512 &     61.480 &    296.708 &    4.7/7.7 &  12.0/12.4 & $2.15_{-0.40}^{+0.50}$ & 0.57 & SIM: EM* VES 52 -- Emission-line Star\\
                  &      0.088 &     25.215 &    7.4/9.6 &            & $0.79_{-0.48}^{+0.72}$ & 1.43 &    \\
AX J194939$+$2631 &     62.937 &    297.416 &    ---/--- &   12.4/5.4 & $2.61_{-1.40}^{+2.43}$ & 0.97 & \\
                  &      0.203 &     26.530 &   11.2/5.2 &            & $3.39_{-1.73}^{+1.99}$ & 1.51 &    \\
AX J194951$+$2534 &     62.134 &    297.465 &    5.7/3.6 & 106.3/20.9 &                        &      & \\
                  &   $-$0.321 &     25.573 & 107.5/21.6 &            &                        &      & \\
AX J195006$+$2628 &     62.937 &    297.527 &    3.0/3.0 &    7.2/4.5 &                        &      & \\
                  &      0.088 &     26.472 &    5.2/4.0 &            &                        &      & \\
\hline
\end{tabular}
\end{center}
\end{table}

\begin{table}[p]
\begin{center}
\caption{Number of the detected sources in the ASCA Galactic plane survey in each energy band.}\label{tab:numsrc}
\begin{tabular}{c|cccc}
\tableline
\tableline
Significance & \multicolumn{4}{c}{Energy band} \\
          &  0.7--2 keV & 2--10 keV & 0.7--7 keV & Total \\
\hline
$5\sigma<$ & 49(9)$^{*}$  & 95(12)$^{*}$  & 115(14)$^{*}$  & 132(20)$^{*}$ \\
$4\sigma<$ & 88(24)$^{*}$ & 115(15)$^{*}$ & 137(30)$^{*}$  & 207(44)$^{*}$ \\
\hline
\end{tabular}

\tablenotetext{}{
$^{*}$ The values in brackets represent the number of sources expected to be fake ones 
due to causes such as stray light or diffuse emission.}

\end{center}
\end{table}

\begin{table}[p]
\begin{center}
\caption{Number of catalog sources cross-identified with the ASCA Galactic sources (AGS)
in the area surveyed.
}
\label{tab:catid}
\begin{tabular}{c|c|cccc}
\hline
\hline
Catalog 	& AGS		& IPC$^a$ & RASS$^b$	& SNR$^c$ & SIM-Star$^d$\\
\hline 	          			
Identified 	& 56       	& 19	  & 22		& 16 	  & 22\\
UnID.  	      	& 107$^*$  	& 7	  & 3		& 16	  & ---\\
\hline	          			
Total         	& 163   	& 26	  & 25		& 32	  & ---\\
\hline
\end{tabular}
\tablenotetext{}{
IPC$^a$, RASS$^b$, SNR$^c$, and  SIM-Star$^d$  represent Einstein IPC X-ray Source Catalog,
ROSAT All Sky Survey Bright Source Catalogue,
Green Galactic SNR Catalogue, 
and catalogs of optical stars in SIMBAD database,
respectively.
}
\tablenotetext{}{
$^*$Number of the ASCA Galactic sources (AGS) cross-identified with 
any other catalogs.
}
\end{center}
\end{table}


\begin{table}[p]
\scriptsize
\begin{center}
\caption{Summary of ten relatively-bright X-ray sources with $F>10^{-11}$ {\flux} in the resolved sources}
\label{tab:10brtsrc}.
\begin{tabular}{cccccl}
\hline
\hline
ASCA name         & $\Gamma{}^a$                & $N_{\rm H}{}^a$                & Flux${}^a$  & V$^b$ & Identification and References\\
\hline		   
AX J153818$-$5541 & $1.96_{-0.11}^{+0.10}$  & $6.95_{-0.80}^{+1.66}$  & 21.9 & $\bigcirc$ & Unidentified\\
AX J163159$-$4752 & $0.24_{-0.16}^{+0.18}$  & $8.77_{-1.10}^{+1.21}$  & 48.5 & $\bigcirc$ & Unidentified\\
AX J163904$-$4642 & $-0.01_{-0.60}^{+0.66}$ & $12.82_{-6.88}^{+8.58}$ & 19.2 & $\bigcirc$ & Unidentified\\
AX J171804$-$3726 & $4.45_{-}^{-}$          & $5.56_{-}^{-}$          & 13.3 &            & SNR:G349.7$+$0.2 \cite{Yamauchi1998}\\
AX J183221$-$0840 & $0.54_{-0.18}^{+0.20}$  & $0.98_{-0.37}^{+0.44}$  & 11.3 & $\bigcirc$ & 1549-s pulsation \cite{Sugizaki2000}\\
AX J183800$-$0655 & $0.47_{-0.36}^{+0.41}$  & $2.71_{-1.20}^{+1.62}$  & 10.9 & $\bigcirc$ & 2E1835.3$-$0658\\
AX J184121$-$0455 & $4.20_{-}^{-}$          & $2.53_{-}^{-}$          & 28.6 &            & SNR:G27.4$+$0.0, Kes73, 1E1841-045\cite{Vasist1997}\\
AX J184629$-$0258 & $2.01_{-0.18}^{+0.20}$  & $2.15_{-0.30}^{+0.32}$  & 27.2 &            & SNR:G29.7$-$0.3, Kes75 \cite{Blanton1996}\\
AX J190734$+$0709 & $4.48_{-}^{-}$          & $1.68_{-}^{-}$          & 11.4 &            & SNR:G41.1$-$0.3, 3C397 \cite{Chen1999}\\
AX J191105$+$0906 & $3.60_{-}^{-}$          & $4.49_{-}^{-}$          & 51.0 &            & SNR:G43.3$-$0.2, W49B \cite{Fujimoto1995,Hwang2000}\\ 
\hline
\end{tabular}
\tablenotetext{a}
{Best-fit values of power-law model, extracted from Table \ref{tab:srclist}
(photon index $\Gamma$, absorption column density $N_{\rm H}$ [10$^{22}$ cm$^{-2}$], 
and flux [10$^{-12}$ ergs cm$^{-2}$ s$^{-1}$] in the 0.7--10 keV band).}
\tablenotetext{b}
{Flux variability with a confidence above 99\%.}
\end{center}

\end{table}

\begin{table}[p]
\begin{center}
\caption{Grouping criteria and the number of each spectral group.}
\label{tab:classification}
\begin{tabular}{c|cc|ccc}
\hline
\hline
Group ID & \multicolumn{2}{c|}{Condition}           & \multicolumn{3}{c}{Number} \\
          & $\Gamma$ & $\nh$ [$10^{22}${\col}]     & Total & SNR & SIM-Star \\
\hline
(a)     & $\Gamma < 1.0$      &    any             & 11 & 0 & 1\\
\hline
(b-i)   & $1.0< \Gamma< 3.0$  &  $\nh <0.8$        & 23 & 0 & 6\\
(b-ii)  &                     &  $ 0.8< \nh < 3.0$ & 10 & 0 & 1\\
(b-iii) &                     &  $ 3.0 < \nh$      & 17 & 3 & 0\\
\hline
(c-i)   & $3.0< \Gamma$ &  $ 0.8< \nh < 3.0$       & 11 & 3 & 3\\
(c-ii)  &                     &  $ 3.0 < \nh$      & 21 & 3 & 0\\
\hline
\end{tabular}

\end{center}
\end{table}

\begin{table}[p]
\scriptsize
\begin{center}
\caption{Summary of spectral fittings to grouping faint-source spectra.}
\label{tab:specfit}
\begin{tabular}{c|ccccccccc}
\hline
\hline
Group ID & Fitting Model$^a$        & \multicolumn{8}{c}{Best-fit Parameters$^b$} \\
         &                      & $N_{\rm H1}$              & $\Gamma$ or $kT_1$      & $Z_1(=Z_2)$  & $F_1{}^c$ & $N_{\rm H2}$ & $kT_2$ & $F_2{}^c(<F_1)$ & $\chi^2_\nu/\nu$ \\
\hline
(a)     & Abs$_1$$\cdot$PL 		& $0.0_{-0.0}^{+0.53}$   & $0.0_{-0.12}^{+0.17}$  & --- & 2.8 & ---  & --- & --- & 1.83/36 \\
	& Abs$_1$$\cdot$RS$_1$  	      & $5.3_{-}^{-}$    & $64_{-}^{-}$           & 1.0:fixed & 2.2 & --- & --- & --- & 3.48/36 \\
	& Abs$_1$$\cdot$RS$_1$$+$Abs$_2$$\cdot$RS$_2$ & $14.5_{-3.3}^{+4.7}$ & $30_{-15}^{+34}$ & 1.0:fixed & 1.9 & $0.12_{-0.12}^{+0.44}$ & $=kT_1$ & 0.6 & 1.28/34\\
\hline			                           			   	      		    
(b-i)   & Abs$_1$$\cdot$PL 		& $0.0_{-0.0}^{+0.03}$   & $1.93_{-0.07}^{+0.07}$ & ---       & 1.1 & --- & --- & --- & 1.62/36 \\
	& Abs$_1$$\cdot$RS$_1$  	       	& $0.0_{-}^{-}$          & $4.2_{-}^{-}$          & 1.0:fixed & 1.1 & --- & --- & --- & 2.24/36 \\
	& Abs$_1$$\cdot$RS$^*_1$  	       	& $0.0_{-0.0}^{+0.015}$  & $4.4_{-0.5}^{+0.5}$    & $0.30_{-0.23}^{+0.25}$ & 1.1 & --- & --- & --- & 1.89/35 \\
	& Abs$_1$$\cdot$(RS$^*_1$$+$RS$^*_2$) & $0.26_{-0.26}^{+0.42}$ & $5.7_{-1.0}^{+2.5}$    & $0.28_{-0.13}^{+0.22}$ & 0.9 & --- & $0.73_{-0.37}^{+0.18}$ & 0.23 & 1.21/33\\
\hline			                           	     		   	      		    
(b-ii)  & Abs$_1$$\cdot$PL 		& $1.45_{-0.50}^{+0.60}$ & $1.37_{-0.28}^{+0.31}$ & --- & 1.0 & --- & --- & --- & 1.86/36 \\
	& Abs$_1$$\cdot$RS$_1$  	       	& $1.79_{-0.39}^{+0.45}$ & $9.0_{-3.1}^{+8.9}$    & 1.0:fixed & 0.96 & --- & --- & --- & 1.65/36 \\
	& Abs$_1$$\cdot$(RS$_1$$+$RS$_2$) & $4.7_{-1.6}^{+2.9}$    & $6.7_{-3.5}^{+6.0}$    & 1.0:fixed & 0.90 & --- & $0.28_{-0.10}^{+0.17}$ & 0.08 & 1.47/34\\
\hline		     	                           	     		   	      		    
(b-iii) & Abs$_1$$\cdot$PL 		& $4.98_{-0.69}^{+0.77}$ & $1.94_{-0.21}^{+0.21}$ & --- & 2.1 & --- & --- & --- & 1.47/36 \\
\hline	                           	     		   			      		    
(c-i)   & Abs$_1$$\cdot$PL 		& $0.9_{-}^{-}$          & $4.7_{-}^{-}$          & --- & 1.7 & --- & --- & --- & 4.12/36 \\
	& Abs$_1$$\cdot$RS$_1$ 		& $1.33_{-0.05}^{+0.05}$ & $0.69_{-0.03}^{+0.06}$ & 1.0:fixed & 1.6 & --- & --- & --- & 1.74/36 \\
	& Abs$_1$$\cdot$(RS$_1$$+$RS$_2$) & $1.37_{-0.09}^{+0.08}$ & $0.58_{-0.06}^{+0.07}$ & 1.0:fixed & 1.4 & --- & $2.3_{-0.7}^{+62}$ & 0.3 & 1.34/34\\
\hline	                           	     		   			      		    
(c-ii)  & Abs$_1$$\cdot$PL 		& $3.2_{-}^{-}$  & $3.3_{-}^{-}$                  & --- & 1.3 & --- & --- & --- & 2.08/36 \\
	& Abs$_1$$\cdot$RS$_1$  	& $2.5_{-}^{-}$  & $2.1_{-}^{-}$                  & 1.0:fixed & 1.3 & --- & --- & --- & 2.40/36 \\
	& Abs$_1$$\cdot$(RS$_1$$+$RS$_2$) & $5.0_{-1.2}^{+1.6}$    & $2.2_{-0.5}^{+0.8}$    & 1.0:fixed & 1.0 & --- & $0.36_{-0.13}^{+0.26}$ & 0.3 & 1.52/34 \\
	& Abs$_1$$\cdot$(RS$^*_1$$+$RS$^*_2$) & $4.3_{-1.7}^{+1.1}$    & $3.4_{-1.0}^{+2.1}$    & $0.36_{-0.20}^{+0.30}$ & 1.0 & --- & $0.53_{-0.17}^{+0.18}$ & 0.4 & 1.35/33 \\
\hline
\end{tabular}
\tablenotetext{a}{
Abs, PL and RS represent a photoelectric absorption with a hydrogen column density, ${\nh}$ [$10^{22}${\col}], 
using Morrison \& McCammon cross-sections, a power-law with a photon index, $\Gamma$, and 
a thin-thermal plasma emission model with a temperature, $kT$ [keV], coded by Raymond \& Smith. 
Parameters in multi-component models are distinguished by suffix figures.
RS$^*$ means that the plasma abundance, $Z$, is floated.
}
\tablenotetext{b}{
All the errors represent the 90\% confidence limit. 
Dash lines ('$-$') in the errors mean that the best-fit model is unacceptable 
within a 99.9\% confidence limit ($\chi_\nu<2$).
}
\tablenotetext{c}{
$F$ is a flux of each emission model in a unit of [$10^{-12}${\flux}].
}
\end{center}
\end{table}

\begin{table}[p]
\scriptsize
\begin{center}
\caption{Summary of identification of grouped faint X-ray sources with $F=(0.5-10)\times 10^{-12}$ {\flux}.}
\label{tab:discussrc}
\begin{tabular}{c|c|l|c|l}
\hline
\hline
Source  & Grouping           & \multicolumn{1}{c|}{Properties of}      & Number density & \multicolumn{1}{c}{Identifications} \\
Group   & criteria             & \multicolumn{1}{c|}{X-ray spectrum} & \& Luminosity  & \\ 
        & $\Gamma$,            &                                     & $\rho$[pc$^{-3}$]    &\multicolumn{1}{c}{Number of cataloged sources}\\
        & $N_{\rm H}$[10$^{22}$cm$^{-2}$]  &                         & $L_{\rm X}$[{\lumi}] &\multicolumn{1}{c}{\& Candidates}\\
										       
\hline    									       
(a)     & $\Gamma<1$           & Power-law                   & $5\times 10^{-10}$                 & SNRs:0/11, Stars:1/11\\
        &                      & $\Gamma\simeq 0$            & $1.9\times 10^{34} d_{\rm 8 kpc}^2$ & HMXB pulsars\\
\hline										       
(b-i)   & $1\leq\Gamma<3$,     & Power-law                   & $3\times 10^{-7}$                  & SNRs:0/23, Stars:6/23\\
        & $N_{\rm H}<0.8$      & $\Gamma\simeq 2$            & $8.9 \times 10^{32} d_{\rm 2.6 kpc}^2$ & Quiescent LMXBs, Crab-like pulsars\\
\hline										       
(b-ii)  & $1\leq\Gamma<3$,     & Thin-tharmal plasma         & $3\times 10^{-9}$                        & SNRs:0/10, Stars:1/10\\
        & $0.8\leq N_{\rm H}<3$& Two-temp. ($kT_1\simeq$7 keV, $kT_2\simeq$0.3 keV) & $7.5 \times 10^{33} d_{\rm 8ks}^2$  & CVs\\ 
\hline										       
(b-iii) & $1\leq\Gamma<3$,     & Power-law                   & ---                                & SNRs:3/17, Stars:0/17\\
        & $3\leq N_{\rm H}$    & $\Gamma\simeq 2$            & ---                                & Extragalactic sources (AGNs):$\sim$12/17\\
\hline										       
(c-i)   & $3\leq\Gamma$,       & Thin-thermal plasma         & $3\times 10^{-9}$                   & SNRs:3/11, Stars:3/11\\
        & $N_{\rm H}<3$        & $kT\simeq$0.6 keV plus hard tail  & $1.3\times 10^{34} d_{\rm 8 kpc}^2$ & Active coronae, OB stars, Quiescent LMXBs\\
\hline										       
(c-ii)  & $3\leq\Gamma$,       & Thin-thermal plasma          & $9\times 10^{-10}$                & SNRs:3/21, Stars:0/21\\
        & $3\leq N_{\rm H}$    & Two-temp. ($kT_1\simeq$2 keV, $kT_2\simeq$0.4 keV) &  $\sim 1.6\times 10^{34} d_{\rm 10 kpc}^2$ & CVs, Active coronae\\
\hline
\end{tabular}
\end{center}
\end{table}

\end{document}